\documentstyle[aaspp4,12pt,flushrt]{article}

\newcommand\kms{km~s$^{-1}$}
\newcommand\mzw{m_{\rm Zw}}
\newcommand\br{\hbox{$B\!-\!R$}}
		
\begin{document}
\title{An Imaging and Spectroscopic Survey of\\
Galaxies within Prominent Nearby Voids I.\\
The Sample and Luminosity Distribution}

\author{Norman A. Grogin and Margaret J. Geller} 

\affil{Harvard-Smithsonian Center for Astrophysics,
60 Garden Street, Cambridge, MA 02138\\ 
E-mail: ngrogin, mgeller@cfa.harvard.edu} 

\slugcomment{To appear in the {\it Astronomical Journal}, Dec.~1999} 

\abstract{We study the optical properties of a large sample of
galaxies in low-density regions of the nearby universe.  We make a $5
h^{-1}$ Mpc-smoothed map of the galaxy density throughout the Center
for Astrophysics Redshift Survey (CfA2, \cite{gh89}) to identify
galaxies within three prominent nearby ``voids'' with diameter
$\gtrsim 30 h^{-1}$ Mpc.  We augment the CfA2 void galaxy sample with
fainter galaxies found in the same regions from the deeper Century
(\cite{cspaper}) and 15R (\cite{15rcite}) Redshift Surveys.  We 
obtain $B$ and $R$ CCD images and high signal-to-noise longslit
spectra for the resulting sample of 149 void galaxies, as well as for
an additional 131 galaxies on the periphery of these voids.

Here we describe the photometry for the sample, including B isophotal
magnitudes and \br\ colors.  For the 149 galaxies which lie in regions
below the mean survey density, the luminosity functions in $B$ and $R$
are well-fit by Schechter functions with respective parameters
($\alpha_B = -0.5\pm0.3$, $B_* = -18.9\pm0.2$) and ($\alpha_R =
-0.9\pm0.3$, $R_* = -20.4\pm0.3$).  The $B$ luminosity function (LF)
is consistent with typical survey LFs (e.g.~the Southern Sky
Redshift Survey; \cite{ssrs2}), and the $R$ LF is consistent with the
Century Survey.  The $B$ and $R$ LFs of 131 galaxies in the ``void
periphery'', regions between the mean density and twice the mean, have
similar Schechter parameters.  The CfA2 LF is inconsistent with both
samples at the $3.5\sigma$ level.

When we narrow our analysis to the 46 galaxies in regions below half
the mean density, the LF is significantly steeper:
$\alpha\sim-1.4\pm0.5$.  The typical survey LFs are inconsistent with
this subsample at the $\sim\!2\sigma$ level.  The \br\ color
distribution of galaxies in the lowest-density regions is also shifted
significantly ($\sim\!3\sigma$) blue-ward of the higher density
samples.  The most luminous red galaxies ($R \lesssim -21$) are
absent from the lowest density regions at the $2.5\sigma$ level.  }


\begin{section}{Introduction}
Wide-angle redshift surveys during the last decade have provided a
picture of galaxies distributed within large, coherent sheets, with
clusters embedded in these structures, bounding vast ($10^{5-6}$
Mpc$^3$) and well-defined ``voids'' where galaxies are largely absent.
The relationship between local density in these structures and galaxy
properties are of interest for constraining galaxy formation models.
One of the most obvious indications of environmental dependence is the
morphology-density relation (e.g., \cite{dr80,pg84}), which quantifies the
increasing fraction of ellipticals and lenticulars with increasing
local density.  In the lowest density regions, the voids, the
observational evidence of trends in morphological mix, luminosity
distribution, star formation rate, etc., is still rudimentary because
of the intrinsic scarcity of void galaxies and the difficulties in
defining an unbiased sample for study.  Here we use a broadband
imaging and spectroscopic survey of a large optically-selected sample
to compare ``void'' galaxies with their counterparts in denser
regions.

A better understanding of the properties of void galaxies is useful to
constrain proposed theories of galaxy formation and evolution.  For
example, the peaks--bias paradigm of galaxy formation in a flat, cold
dark matter universe (\cite{ds86,hsw92}) predicts that the voids
should be populated with ``failed galaxies'' identified as diffuse dwarfs
and that the $3\sigma$ massive galaxies in voids should
have extended, unevolved, low surface-brightness (LSB) disks like
Malin I (\cite{malin1}).  Unfortunately for this theory, dwarf galaxies
are observed to trace the distribution of the more luminous galaxies
(\cite{bing}), the environments of the observed Malin 1-type objects
are not globally underdense (\cite{bot93}), and H{\sc i} searches in
voids have not turned up LSB giants (\cite{sz96b,wei91,hk89}).  Two
recent efforts to measure redshifts for faint galaxies toward nearby
voids, one comprising 185 optically-selected galaxies (\cite{khe97})
and the other comprising 234 emission-line objects selected from
objective prism plates (\cite{phe97}), both failed to discover faint
galaxies filling the voids.  However, the sky coverage in both cases
was modest.

Balland, Silk, and Schaeffer (1998) recently proposed a variation on
the peaks--bias model in which collision-induced galaxy formation
drives the morphological biasing.  This new model quantitatively
recovers the cluster morphology-density relation, predicts essentially
no difference in the morphological mix from the field to the voids,
and predicts that non-cluster ellipticals must have all formed at high
redshift ($z\gtrsim2.5$).  Similarly, Lacey et al.\ (1993) have
proposed a model for the evolution of the galaxy luminosity function
(LF) in which the rate of star formation is controlled by the
frequency of tidal interactions.  Their model predicts that luminous
galaxies should not have formed in underdense regions for want of
tidal interactions to trigger star formation.

There have been few previous measurements of the galaxy LF in low
density regions, i.e., in the voids.  Park et al.~(1994) found two
indications of luminosity bias in volume-limited samples from the
Center for Astrophysics Redshift Survey (\cite{gh89}, hereafter CfA2).
For samples with $B_{\rm lim}=-19.1$ and $-18.5$, the power spectrum
for the brighter half of the sample has $\sim\!40\%$ larger amplitude,
independent of scale.  Furthermore, the lower-density regions appear
to be deficient in the brightest galaxies ($B\sim-20$) at the
$\sim\!2\sigma$ level.  El-Ad and Piran (1997) mapped out voids in the
Southern Sky Redshift Survey (\cite{ssrs2}, hereafter SSRS2),
comparable in depth to CfA2.  They identify 61\% of the survey volume
out to $80h^{-1}$~Mpc as ``voids''; this volume contains 19\% of the
fainter $B > -19$ galaxies in the sample but only 5\% of the brighter
$B \leq -19$ galaxies.  The significance of this result is difficult
to interpret, because the void-detection algorithm depends only on the
brighter galaxies.

Bromley et al.\ (1998) also investigated the environmental dependence of
the LF in their recent spectral analysis of the Las Campanas Redshift
Survey (\cite{lcrs}, hereafter LCRS).  Their density discriminant is
a friends-of-friends algorithm (\cite{fof}) to separate cluster and
group galaxies from the rest.  Their absorption-line objects have a
much shallower LF in lower-density regions ($\alpha=0.19$ versus
$-0.40$), and they observed a strong LF dependence on spectral type
coupled with a substantial change in the spectroscopic-type mix with local
density (akin to the morphology-density relation).

Most previous studies of the properties of individual void galaxies
have focused on emission line-selected and $IRAS$-selected objects in
the Bo\"otes void at $z\sim0.05$ (\cite{koss1,koss2}), and all studies
have been limited to a few dozen objects.  Broadband multicolor
imaging of 27 Bo\"otes void galaxies (\cite{cwh97}) showed that they
are brighter on average than emission-line galaxies at similar
redshift.  Moreover, a large fraction ($\approx40$\%) of this sample
shows unusual or disturbed morphology.  Weistrop et al.\ (1995)
obtained H$\alpha$ images for a subset of 12 galaxies and reported
star formation rates ranging from 3--55 $\cal M_\odot$ yr${}^{-1}$,
with the most active galaxies producing stars at almost three times
the rate found in normal field disk systems.  This finding confounds
the naive expectation for void galaxies in the Lacey et al.\ (1993)
model.

Szomoru, van Gorkom, \& Gregg (1996) surveyed $\sim\!1\%$ of the
Bo\"otes void volume in H{\sc i} with the VLA around 12
$IRAS$-selected void galaxies.  They detected these galaxies along
with 29 companions.  Szomoru et al.\ (1996) then argue that the
Bo\"otes void galaxies are mostly late-type gas-rich systems with
optical and H{\sc i} properties and local environments similar to
field galaxies of the same morphological type.  They conclude that the
void galaxies formed as normal field galaxies in local density
enhancements within the void and that the surrounding global
underdensity is irrelevant to the formation and evolution of these
galaxies.  These findings are in concert with the conclusions of
Thorstensen et al.~(1995), who examined an optically-selected sample of
27 galaxies within a nearby CfA2 void.  The fraction of
absorption-line galaxies in their sample is typical of regions away
from cluster cores, and the local morphology-density relation appeared
to hold even within the global underdensity.

Our goal is to try to resolve some of the apparent inconsistencies
among previous studies by collecting high-quality optical data for a
large sample of void galaxies with well-defined selection parameters.
We thus obtained multi-color CCD photometry and high signal-to-noise
spectroscopy for $\sim\!150$ optically-selected galaxies within
prominent nearby voids.  We work from the CfA2 Redshift Survey, which
has the wide sky coverage and dense sampling necessary to delineate
voids for $cz\lesssim10000$~\kms.  These conditions are not met
for the Bo\"otes void, making the definition of Bo\"otes void
galaxies in previous studies harder to interpret.  

Using a straightforward density
estimation technique, we identify three large ($\gtrsim
30h^{-1}$~Mpc) voids within CfA2.  In addition to the void
galaxies from CfA2, we include fainter galaxies found in the
same regions by the deeper Century Survey (\cite{cspaper}; hereafter
CS) and 15R Survey (\cite{15rcite}).  At the cost of mixing
$B$-selected and $R$-selected samples, we thereby gain extra
sensitivity at the faint end of the void luminosity distribution.

Our large sample, which covers essentially the entire volume of
three distinct voids, should afford better constraints on the morphology,
luminosity distribution, and star formation rate of void galaxies.
Moreover, our sample drawn from $B$- and $R$-selected redshift surveys
may be more broadly representative than the previous studies of
emission-line, {\sl IRAS}-selected, and H{\sc i}-selected void galaxies.

Here we introduce the sample, describe the broadband imaging survey,
and derive the void galaxy luminosity distribution and the broad-band
color distribution as a function of local density.  Grogin \& Geller
(2000, hereafter \cite{paper2}) will address the morphologies and
spectroscopic properties of the galaxies in our sample.  In
\S\ref{samplesec} we describe the selection of the void galaxy sample.
We discuss the multiple redshift surveys involved, describe the
density estimation technique for identifying voids, and show maps of
the galaxy density field.  Section \ref{obsphotsec} describes the
observations, reduction, and photometry.  In \S\ref{lffitsec} we
derive a method for fitting luminosity functions in $B$ and $R$ to our
heterogeneous galaxy sample.  We apply the method to various density
cuts through the sample and compare with typical redshift survey LFs.  We
discuss our results in \S\ref{discsec} and conclude in
\S\ref{concsec}.
\end{section}

\begin{section}{Sample Selection} \label{samplesec}
In \S\ref{czdatasec} we briefly describe the three redshift surveys,
CfA2, Century, and 15R, from which we select the void galaxies for
this study.  In \S\ref{denmethsec} we review the estimator of the
smoothed galaxy number density field within CfA2 (\cite{gg98}).  In
\S\ref{sampdefsec} we display maps of the density field around voids
in this study.

\begin{subsection}{Redshift Surveys} \label{czdatasec}
The Center for Astrophysics Redshift Survey of galaxies in the Zwicky
Catalog (\cite{cgcg}) now contains more than 14000 homogeneous redshifts
and is 98\% complete to $\mzw \leq 15.5$ (where Zwicky estimated his
catalog was complete) over the regions ($8^{\rm h} \leq \alpha_{1950}
\leq 17^{\rm h}$, $0\arcdeg \leq \delta_{1950} \leq 50\arcdeg$: CfA2
North) and ($20^{\rm h} \leq \alpha_{1950} \leq 4^{\rm h}$, $-2\fdg5
\leq \delta_{1950} \leq 50\arcdeg$: CfA2 South).  These
redshifts are all contained in the Updated Zwicky Catalog 
(\cite{uzc}, hereafter UZC), which
also includes arcsecond coordinates for all objects with $\mzw \leq
15.5$.  We use the coordinates and redshifts from the UZC as the basis
for the density field estimation of \S\ref{denmethsec}.

The Century Survey is a complete photometric and spectroscopic survey
covering the region $8^{\rm h}30^{\rm m} \leq \alpha_{1950} \leq 16^{\rm
h}20^{\rm m}$ and $29\arcdeg \leq \delta_{1950} \leq 30\arcdeg$
(0.03 steradians) to a limiting $m_R = 16.13$.  The CS catalog
was constructed from POSS E plate scans (\cite{csplate}) and
calibrated with drift-scan and pointed CCD photometry.  The best-fit
Schechter (1976) luminosity function to the 1762 CS galaxies has
$M_* = -20.73\pm0.18$ and $\alpha = -1.17\pm0.19$.  The CS $M_*$ is
consistent with the red-selected Las Campanas Redshift Survey
(\cite{lcrs}, hereafter LCRS) of $>18000$ galaxies.  The faint-end
slope of the LCRS is significantly shallower than the CS: $\alpha_{\rm
LCRS} = -0.70\pm0.05$.  This discrepancy may arise from the additional
central surface-brightness cut used in LCRS, which may preferentially
reject the faintest galaxies.  Strong dependence of $\alpha_{\rm
LCRS}$ on spectral type may also be the explanation, as suggested by
Bromley et al.~(1998).  Similarly deep blue-selected surveys such as
AUTOFIB (\cite{autofib}) and the ESO Key Project (\cite{esokp}) have
faint-end slopes indistinguishable from the CS\@.  We therefore take the CS
as our fiducial $R$ galaxy LF in comparisons with the
$R$ luminosity distribution of our void-selected samples.

The 15R Survey is an $R$-limited photometric and spectroscopic survey
of two wider declination strips: ($8^{\rm h}30^{\rm m} \leq
\alpha_{1950} \leq 16^{\rm h}30^{\rm m}$, $26\arcdeg30\arcmin \leq
\delta_{1950} \leq 32\arcdeg30\arcmin$), which almost entirely
overlaps the original CfA ``Slice of the Universe'' (\cite{dlgh86});
and a smaller region within CfA2 South ($20^{\rm h} \leq \alpha_{1950}
\leq 4^{\rm h}$, $10\arcdeg30\arcmin \leq \delta_{1950} \leq
13\arcdeg30\arcmin$).  The survey was originally intended to identify
and measure redshifts of all galaxies down to a limiting $m_R = 15.4$
in the 0.2 steradians covered by the survey.  The photometric catalog
was constructed from POSS E plate scans analogously to the CS, and
plate magnitudes in 15R North were calibrated using galaxies common to
both surveys.  The Southern 15R survey magnitudes still require
calibration: in \S\ref{cs15risosec} we use the photometry of our 15R
South void galaxies to calibrate roughly the magnitude limit for 
each of the plates in 15R South.

Longslit CCD spectra of the 15R survey galaxies were obtained with the
FAST spectrograph on the F.~L.~Whipple Observatory (FLWO) 1.5~m
Tillinghast reflector over the period 1994--1997.  The spectra were
reduced as part of the Center for Astrophysics redshift survey
pipeline and radial velocities were extracted via template
cross-correlation (\cite{km98}).  In the northern strip, the 15R
redshift survey is complete to a limiting magnitude $m_R=15.42$.  In
the south, our calibrations indicate that most surveyed plates are
complete to $m_R\approx16$; none are shallower than $m_R\approx15.4$.
\end{subsection}
\begin{subsection}{Density Estimation Method} \label{denmethsec}
To identify regions of low galaxy density, 
we first transform the point distribution of the CfA2 survey into a
continuously-defined number density field in redshift
space.  We smooth each CfA2 galaxy in redshift space with a unit-normalized
Gaussian kernel $W$ of width $\sigma = 5h^{-1}$~Mpc:
\begin{equation}
W\left({{\bf r}-{\bf r}_{\rm gal}}\right) = \left({2\pi \sigma^2}\right)^{-3/2} 
	\exp\left({\left|{{\bf r}-{\bf r}_{\rm gal}}\right|^2/2\sigma^2}\right),
\end{equation}
where ${\bf r}$ is the 3D redshift-space coordinate.
We choose a $5h^{-1}$~Mpc smoothing length to coincide
with the galaxy-galaxy correlation length (\cite{peeb}; \cite{pecvel}; 
\cite{lcrscorr}) and with the pairwise velocity dispersion in the
survey (\cite{pecvel}).

We correct the redshift survey heliocentric velocities to
the rest frame of the Local Group,
\begin{equation} \label{lsreq}
cz = cz_{\sun} + (300\,\hbox{\rm km s}^{-1}) \sin l \cos b,
\end{equation}
for a galaxy at Galactic longitude $l$ and latitude $b$.  Otherwise,
we make no attempt to remove peculiar velocity distortions (cluster
``fingers'', etc.) from the redshift survey.  We place each object at
a comoving distance $r$ appropriate for a $q_0 = 0.5$ universe with
pure Hubble flow:
\begin{equation} \label{comove}
r(z) = \left({{2c}\over{H_0}}\right) \left[{1 - (1 + z)^{-1/2}}\right].  
\end{equation}
For the low redshifts of interest here, $q_0$ has little effect on
$r(z)$.  Our smoothing kernel effectively washes out peculiar
velocities $\lesssim 500$~\kms, close to the $540\pm180$~\kms\
pairwise velocity dispersion measured by Marzke et al.~(1995) for the
combined CfA2 and SSRS2.  We underestimate spatial overdensities
associated with clusters, which are broadened in the radial direction.

Because the CfA2 Redshift Survey is flux-limited, an increasing
fraction of the galaxies at larger redshift fall below the magnitude
limit and do not appear in the survey.  In computing the density
field, we compensate for the magnitude-limited sample by assigning
each galaxy a weight $1/\psi$, where the selection function $\psi$ is
\begin{equation} \label{selfneq}
\psi(\alpha,\delta,z) = {{\displaystyle\int_{-\infty}^{M_{\rm lim}(\alpha,\delta,z)}
{\phi(M)\,dM}}
\over{\displaystyle\int_{-\infty}^{M_{\rm cut}}{\phi(M)\,dM}}}.
\end{equation}
Here $M_{\rm lim}$ is the effective absolute magnitude limit at the
galaxy position and $\phi(M)$ is the differential luminosity function.
For $M_{\rm lim}$ fainter than a fixed luminosity cutoff $M_{\rm
cut}=-16.5$ (\cite{mhg94}), we assign galaxies unit weight.  Unless
otherwise noted, all magnitudes in this section refer to Zwicky
magnitudes. Numerical values for absolute magnitudes throughout this
paper implicitly include the $h$-dependence in equation
(\ref{comove}).

\cite{mhg94} fit the CfA2 LF to a Schechter function
$\phi_{\rm SF}$ (\cite{sch76}), convolved with a Gaussian error of
$\sigma_M = 0.35$~mag (\cite{h76}) in the Zwicky magnitudes:
\setlength{\arraycolsep}{0.5mm}
\begin{eqnarray}
\phi_{\rm SF}(M) &=& \phi_* \, (0.4 \ln 10) \, 10^{0.4\, (M_* - M)\, (1 + \alpha)} 
	\exp\left[{-10^{0.4 (M_* - M)}}\right]; \nonumber \\
\phi_{\rm CfA2}(M) &=& \label{cfa2lfeq}
{1\over{\sqrt{2\pi}\sigma_M}} \int_{-\infty}^\infty{\phi_{\rm SF}(M')
\exp\left[{-(M'-M)^2/2\sigma_M^2)}\right]\,dM'}.
\end{eqnarray}
\setlength{\arraycolsep}{3mm} 
We adopt the values $\phi_* = 0.04\,({\rm Mpc}/h)^{-3}$, $M_* =
-18.8$, and $\alpha = -1.0$ (\cite{mhg94}).  These values have not
changed as a result of the revisions to the redshift catalog between
1994 and the release of the UZC (Marzke 1999).  For computational
convenience in determining $\psi$, we replace the convolution of
equation (\ref{cfa2lfeq}) with $\phi_{\rm CfA2}(M) \approx \phi_{\rm
SF}(M + 0.1\,{\rm mag})$.  This approximation recovers the true $\psi$
to better than 5\% for $cz\lesssim12000$~\kms.

For a galaxy at position ($\alpha,\delta$) and at luminosity distance
$D_L(z) = (1+z)\,r(z)$ in a survey of limiting apparent magnitude
$m_{\rm lim}$, we estimate the absolute limiting magnitude $M_{\rm
lim}$ according to
\begin{equation}
M_{\rm lim}(\alpha,\delta,z) = m_{\rm lim} - 5 
\log\left[{{(1+z)\,r(z)}\over{1h^{-1}\,
\rm{Mpc}}}\right] - 25 - \Delta m_{\rm K}(z)  
- \Delta m_{\rm ext}(\alpha,\delta). \label{mlimeq}
\end{equation}
In equation (\ref{mlimeq}), $\Delta m_{\rm K}$ is a $K$-correction,
and $\Delta m_{\rm ext}$ is a correction for Galactic extinction.  For
the density estimation technique, $m_{\rm lim}$ represents the
CfA2 flux limit in Zwicky magnitudes.  Photoelectric photometry of
Zwicky galaxies (\cite{zwsc}) suggests that Volume I of the Zwicky
catalog goes $\approx0.4$~mag fainter than the other volumes at the
CfA2 magnitude limit, $m_{\rm Zw} = 15.5$.  To correct for this Volume
I scale error, we adopt $m_{\rm lim}=15.9$ for all CfA2 North galaxies
with $\delta \leq 14\fdg5$.

Lacking precise morphological types for the majority of CfA2, we apply
a generic $K$-correction appropriate for the median type Sab
(\cite{kcorr}): $\Delta m_{\rm K}(z) = 3z$ for the Zwicky magnitudes.
To obtain the correction $\Delta m_{\rm ext}(\alpha,\delta)$ for
Galactic extinction along a particular line of sight, we first
interpolate the H{\sc i} map of Stark et al.~(1992).  We then convert
from H{\sc i} to reddening with the relation $\left<{N({\rm H}{\sc
i})/E(\bv)} \right> = 4.8 \times 10^{21}$ cm$^{-2}$~mag$^{-1}$
(\cite{zom90}).  We adopt an extinction law $\Delta m_{\rm ext} \equiv
A_B = 4.0 E(\bv)$ (\cite{zom90}).

We also employ equation (\ref{mlimeq}) for the luminosity function
fitting of \S\ref{lffitsec}.  In that case, $m_{\rm lim}$ represents
the limiting magnitude of the appropriate redshift survey (CfA2, 15R,
or Century Survey), corrected to the isophotal magnitude system used
here (cf.~\S\S\ref{zwisosec} and \ref{cs15risosec}).  We again adopt
generic $K$-corrections of $\Delta m_{\rm K}(z) = 3z$ for the CCD $B$
magnitudes and $\Delta m_{\rm K}(z) = 0.8z$ for the CCD $R$ magnitudes
(\cite{fg94}).  Because we are only interested in the 15R and CS
galaxies within the CfA2 redshift range ($z \lesssim 0.05$), the error
in the $K$-correction will be small in any case.  We use $\Delta
m_{\rm ext} \equiv A_B = 4.0 E(\bv)$ to correct for extinction in the
CCD $B$ magnitudes, and $\Delta m_{\rm ext} \equiv A_R = 2.8 E(\bv)$
to correct the CCD $R$ magnitudes (\cite{zom90}).

We compute the smoothed galaxy number density $n$ at a given point
${\bf r} \equiv (\alpha,\delta,r(z))$ by summing the contributions
from all $i$ galaxies in the CfA2 survey:
\setlength{\arraycolsep}{0.5mm}
\begin{eqnarray}
n({\bf r}) &=& \sum_i{{W({\bf r} - {\bf r}_i)}\over\psi_i}  \nonumber \\
	&=& \sum_i{{W({\bf r} - {\bf r}_i) 
\displaystyle\int_{-\infty}^{M_{\rm cut}}{\phi(M)\,dM}
}\over{\displaystyle\int_{-\infty}^{M_{\rm lim}({\bf r}_i)}
{ \phi(M)\,dM}}} \nonumber \\
&=& \bar n \sum_i{{W({\bf r} - {\bf r}_i) 
\over{\displaystyle\int_{-\infty}^{M_{\rm lim}({\bf r}_i)}{ \phi(M)\,dM}}}}.
\end{eqnarray}
\setlength{\arraycolsep}{3mm}
For the CfA2 Survey, \cite{mhg94} derive a mean density $\bar n \equiv
\int_{-\infty}^{M_{\rm cut}}{\phi(M)\,dM} = 0.07\,({\rm Mpc}/h)^{-3}$
with $M_{\rm cut}=-16.5$.  This analysis yields a continuous field of
dimensionless galaxy density contrast, $n({\bf r})/\bar n$, which we
use as the basis of our void galaxy selection.
\end{subsection}
\begin{subsection}{The Sample}\label{sampdefsec}
We restrict our study to galaxies within three of the largest
underdense regions in CfA2 ($\gtrsim 30 h^{-1}$~Mpc diameter).  This
selection minimizes the contamination by interlopers with large
peculiar velocities.  Table \ref{voidtab} lists the approximate
redshift-space boundaries of the three voids: ``NV1'' in CfA2 North,
and in CfA2 South ``SV1'' (western) and ``SV2'' (eastern).  In Figures
\ref{cfa2nfig} and \ref{cfa2sfig} we display successive $3\arcdeg$
declination slices of CfA2 North and South, respectively, which
contain NV1, SV1, and SV2.  We superpose contours of the CfA2 galaxy
density field $n$, with underdensities in linear decrements of
$0.2 \bar n$ (dotted contours) and overdensities in logarithmic
intervals (solid contours) denoting $\bar n$, $2\bar n$, $4\bar n$,
etc.  The declination thickness of each slice exceeds the
$5h^{-1}$~Mpc density smoothing length for $cz\gtrsim 10,000$~km/s; at
lower redshifts there is density-field redundancy between adjacent
slices.  We indicate the locations of CfA2 galaxies with crosses and
the subset chosen for this study with larger circles.
\placetable{voidtab} 
\placefigure{cfa2nfig} 
\placefigure{cfa2sfig}

We attempted to include all survey galaxies within the $(n/\bar n =
1)$ contour around each of the three voids.  We define these galaxies
as the ``full void sample'', hereafter FVS\@.  Because the FVS includes
$\approx150$ galaxies, we also examine the properties of two FVS
subsamples: the lowest-density void subsample (hereafter LDVS) of 46
galaxies with $(n/\bar n < 0.5)$, and the complementary higher-density
void subsample (hereafter HDVS) with $(0.5 n/\bar n < 1)$.

Our sample also includes some of the galaxies around the periphery of
the voids where $n/\bar n > 1$.  Typically the region surrounding the
voids at $1 < n/\bar n < 2$ is narrow (cf.~Fig.~\ref{cfa2nfig}),
intermediate between the voids and the higher-density walls and
clusters.  An exception is in the eastern half of our southern region
of interest (cf.~Fig.~\ref{cfa2sfig}), where this contour is comparatively
wide.  Although our void periphery sample (hereafter 
VPS) is far from complete, we have selected these galaxies based only
upon their proximity to the voids.  We thus should not have introduced
any luminosity selection bias between the FVS and the VPS\@.  We
use the VPS as a higher-density reference for the FVS and its
subsamples.

In Figure \ref{c15rfig} we show declination slices of the combined CS
and 15R North (top) and of 15R South (bottom) where they intersect the
three voids, along with superposed CfA2 isodensity contours.  We plot
the same span in right ascension as the CfA2 slices, denoting the
survey boundaries with radial dotted lines.  We indicate the 15R
galaxy locations with crosses, the CS galaxies with triangular
crosses, and the subset chosen for the current study with larger
circles.  It is striking to note the absence of 15R North galaxies
within the NV1 region.  Unfortunately we do not include the four 15R
galaxies in NV1 at $\alpha_{1950} \gtrsim 16^{\rm h}$ because the
measurement of POSS plate 329 redshifts was completed near the end of
the 15R survey, after most of the observations for this study.  We
therefore set the magnitude limit for our study in this section of 15R
North ($\alpha_{1950} > 15^{\rm h}58^{\rm m}$) to the CfA2 limiting
magnitude limit $\mzw = 15.5$ rather than the magnitude limit
$m_R=15.42$ for the rest of 15R North.  The apparent magnitude limit
of the CS allows detection of galaxies in NV1 down to absolute
magnitudes of $R\approx-18$, some three magnitudes fainter than $L_*$.
\placefigure{c15rfig}
\end{subsection}
\end{section}
\begin{section}{Observations and Photometry} \label{obsphotsec}
We acquired Johnson-Cousins $B$ and $R$ images of the 297 galaxies in
our sample (cf.~Tab.~\ref{phottab} below) over the course of several
observing runs at the FLWO 1.2~m
telescope: May/June 1995, October 1995, March 1996, June 1996,
September 1996, April 1997, June 1997, and September/October 1997.
Typical exposure times for each galaxy were 300 s in $R$ and
$2\times300$ s in $B$.  The median seeing was $\approx2\farcs0$ and
varied between $1\farcs4$ and $3\farcs3$.  In May/June 1995 we used a
thick, front-side illuminated, Ford $2048\times2048$ CCD.  We read out
the data in $2\times2$ binned mode, giving an $\approx11\arcmin$ field
with $0\farcs64$/pixel.  For all following observations we used a
thinned, back-side illuminated, antireflection-coated Loral
$2048\times2048$ CCD.  Again we had $2\times2$ binned readout to
obtain an $\approx11\arcmin$ field at $0\farcs63$/pixel.

\begin{subsection}{Reduction Steps}
We reduced our images using IRAF with the standard CCDRED tasks,
subtracting the overscan and bias, interpolating across bad columns
and pixels, and dividing by nightly combined dome or twilight sky flat
fields to correct for pixel-to-pixel sensitivity variations.  The dark
counts on both chips were negligible; thus we did not subtract dark
frames in the reduction.  We passed the flatfielded images through the
COSMICRAYS task to remove cosmic ray hits above $\sim\!60$ counts or a
7.8\% flux ratio.  We then fit a world coordinate system (WCS) to each
frame by matching stars from the US Naval Observatory UJ1.0 Catalog
(\cite{uac}) against stars in each field that we extracted with the
SExtractor program (\cite{sex}) and placed onto a global tangent
projection.  The typical rms deviation in the matched UJ1.0 objects
was $0\farcs3$.  We stacked the two $B$ frames for each galaxy using
offsets determined from the fitted WCS and re-fit a WCS to the
combined image.

For images taken under photometric conditions, we calibrated the
photometry with $B$ and $R$ images of several photometric standard
fields (\cite{landolt}) taken at varying air-mass throughout the
night.  To obtain our photometric solution, we used the PHOTCAL task
to fit to an instrumental zero point term, an air-mass term, and a
\br\ color term to the instrumental (large aperture) magnitudes of the
standard stars as obtained with the PHOT task.  The scatter in a
photometric solution fit is typically 0.015--0.020 mag.  Because of
poor weather during much of our observing time in 1996 and 1997,
roughly half our our images required follow-up photometric
calibration.

For images taken under nonphotometric conditions, we calibrated the
photometry with follow-up "snapshot" images of the same fields taken
during photometric conditions with the Loral CCD on the FLWO 1.2 m
from 1996--1998.  We used the same procedure to reduce these
snapshots, typically 120 s exposures, as for the longer exposures,
including WCS fitting.  We then ran SExtractor on both the
uncalibrated and calibration frames, extracting stars common to both
images by WCS position-matching.  This procedure yielded $\gtrsim30$
calibration stars per frame, enabling us to the recover the magnitude
zero-points of the nonphotometric images to $\lesssim0.02$ mag.  This
uncertainty is commensurate with the typical scatter in the
photometric solutions.

We used a modified version of the GALPHOT surface photometry package
(\cite{galphot}) to obtain galaxy isophotal magnitudes from our
flatfielded images.  We first estimated and subtracted the sky
background around the target galaxies by interactive marking of sky
boxes on the images.  Because our galaxies only span
$\lesssim2\arcmin$ within $10\arcmin$ images, we typically marked
$\sim\!10,000$ sky pixels around each galaxy for local sky
subtraction.  All the galaxies in our sample are comparatively bright
($m_B \gtrsim 17.5$), high surface-brightness objects for which the
isophotal magnitude uncertainty due to sky subtraction is at worst on
par with the 0.02--0.03 mag uncertainties from the photometric
calibration.

Next we interactively masked foreground stars near the galaxies with
the IMEDIT task.  In the few rare cases when a star appeared too close
to a galaxy center for simple masking, we used the DAOPHOT package to
model the image point spread function (PSF) and to interactively fit
and subtract a scaled PSF from the star's position.  We then masked
any obvious residual in the PSF-subtracted stellar core with IMEDIT.

We determined the galaxy surface brightness profiles with the IRAF
isophotal analysis package ISOPHOTE, part of the Space Telescope
Science Data Analysis System.  The package's contour fitting task
ELLIPSE takes an initial guess for an isophotal ellipse, then steps
logarithmically in major axis.  At each step it finds the optimal
isophotal ellipse center, ellipticity, and positional angle.  Masked
pixels are ignored by ELLIPSE.  Because the ELLIPSE algorithm averages
pixels within an elliptical annulus, it is capable of fitting
isophotes out to a surface brightness well below the sky noise.  Far
enough from the galaxy center, the fitting algorithm will ultimately
fail to converge, and ELLIPSE enters a non-fitting mode which fixes
larger ellipses to be similar to the largest convergent isophote.  We
generally ran ELLIPSE non-interactively, but in cases where a peculiar
galaxy surface brightness profile sent the task into non-fitting mode
prematurely, we stepped through the isophote fitting interactively.

Rather than fitting $R$ isophotal ellipses to the galaxy $R$ images,
we determined the galaxies' $R$ aperture magnitudes through overlaid
$B$ image isophotes.  We transferred the $B$ isophotal ellipses using
the images' fitted WCS, thereby compensating for variations in image
scale and orientation.  As a final step, we color-corrected the
resulting $B$ and $R$ surface-brightness profiles at each isophote
with the color term from the photometric solution.  We do not correct
for internal extinction.  Our limiting $B$ isophote was governed by
the early images taken with the thick Ford CCD, which had less sensitivity
in $B$ than the Loral CCD.  These images could not be fit by ELLIPSE
beyond $\mu_B \sim 26$ mag arcsec${}^{-2}$; we adopt this limit for
the entire sample.  The uncertainty in the $\mu_B=26$ mag
arcsec${}^{-2}$ isophote is $\lesssim0.15$ mag arcsec${}^{-2}$ on the
thick chip images and $\lesssim0.05$ mag arcsec${}^{-2}$ on the thin
chip.  Although this procedure gave us the detailed surface brightness
profiles of each galaxy, we defer that analysis and focus here solely
on the isophotal magnitudes and colors.
\end{subsection}

\begin{subsection}{Magnitudes and Colors} \label{magcolsec}
Table \ref{phottab} lists the isophotal $B$ magnitudes ($b_{B26}$) and
corresponding $R$ aperture magnitudes ($r_{B26}$) for the galaxies in
our study.  The uncertainty in these magnitudes is conservatively
$\approx0.05$ mag.  We empirically verified this error from photometry
of galaxies imaged in more than one observing run.  We segregate the
table by survey: CfA2 (Zwicky catalog) galaxies followed by 15R
galaxies followed by Century Survey galaxies.  We note that some of
the galaxies are common to more than one of these overlapping surveys.
For each galaxy we also provide arcsecond B1950 coordinates, the
redshift corrected to the local standard of rest
(cf.~eq.~[\ref{lsreq}]), and our estimate of the $5h^{\rm
-1}$-smoothed galaxy density $(n/\bar n)$ at the galaxy location.  The
typical error in this large-scale density estimator is $\lesssim0.1$ for
$cz \lesssim 10000$~\kms\ (\cite{gg98}).
\placetable{phottab}

In Figure \ref{tridencolfig} we plot the absolute magnitudes (as
determined with eq.~[\ref{mlimeq}]) $B_{B26}$ (top) and $R_{B26}$
(bottom) versus the $(\br)_{B26}$ colors for galaxies in the VPS
(upper panel), the HDVS (middle panel), and the LDVS (lower panel).
For notational convenience, we shall refer to the absolute magnitudes
henceforth as $B$ and $R$.  We also superpose onto Figure
\ref{tridencolfig}b the histograms of the various samples' $\br$
distribution (solid lines).  Clearly there is a shift toward bluer
galaxies in the LDVS, although the VPS and HDVS samples have similar
color distributions.  A Kolmogorov-Smirnov (K-S) test between the VPS
and LDVS colors gives only a 0.3\% probability of their being drawn
from the same underlying distribution.  The corresponding probability
for the VPS and HDVS colors is 55\%, but only 3.2\% between the HDVS
and LDVS colors.  
\placefigure{tridencolfig}

One concern in interpreting the color distributions is that the
distribution of absolute magnitudes may differ from sample to sample.
Thus we must investigate color-magnitude correlation as a possible
source of the LDVS color shift in Figure~\ref{tridencolfig}.  To model
the effect of a color-magnitude correlation, we select galaxies from
the VPS according to the LDVS $R$ luminosity distribution and then
compare the resulting color distribution of the galaxies selected from
the VPS (lower panel, dotted line) with the LDVS colors (lower panel,
solid line).  The clear difference between the solid and dotted
histograms (K-S probability of 0.08\%) demonstrates that the blueward
shift of the LDVS is {\sl not} attributable to a difference in the
absolute magnitude distribution for the sample galaxies.

One may also be concerned by a possible systematic color difference
between the $B$-selected and $R$-selected galaxies in these samples,
i.e.~an $R$-limited sample should include redder objects near the
magnitude limit than a $B$-limited sample.  In our study, the
$R$-selected galaxies are from deeper surveys (15R and CS) than the
$B$-selected galaxies from CfA2, and thus disproportionately populate
the faint end of our $R$ magnitude range.  We note from Figure
\ref{tridencolfig}b that the faintest $R$ galaxies are also slightly
bluer in the mean: $\left<{\br}\right>=1.11\pm0.25$~mag for
$R>-20$~mag compared with $\left<{\br}\right>=1.32\pm0.25$~mag for
$R<-20$~mag.  Because the intrinsically less luminous galaxies tend to
be $R$-selected and these galaxies are not particularly red, we do not
see the anticipated red-ward shift of the $R$-selected galaxy colors.
A K-S test of the \br\ colors for the $B$-selected and $R$-selected
galaxies cannot distinguish between the two distributions ($P_{\rm KS}
= 72$\%).  We therefore employ the overall color distribution,
regardless of selection filter, in our derivation of the $B$ and $R$
LFs.

For an external check on our color distributions, we compare with the
\br\ CCD colors of 193 galaxies from the Nearby Field Galaxy Survey of
Jansen et al.\ (1999).  Their survey also used the FLWO 1.2 m, and
with an observing setup identical to ours.  Their distribution of
field galaxy colors is indistinguishable from the HDVS ($P_{\rm KS} =
71$\%), and is similarly skewed redward of the LDVS at the $\approx\!2\sigma$
confidence level ($P_{\rm KS} = 5.6$\%).
\end{subsection}
\begin{subsection}{Calibration of Redshift Survey Limiting Magnitudes}
To analyze the luminosity dyistribution of the void galaxies, we need
to convert the various redshift survey limiting magnitudes into the
isophotal system used here.  This task is simplest for CfA2 and the
Century Survey, where the entire survey is characterized by a single
magnitude limit.  For the northern 15R Survey, calibrated against the
CS, we assume that the whole strip is again described by a single
magnitude limit.  We use galaxies in this survey to calibrate the 15R
South magnitudes on a plate-by-plate basis and obtain the limiting
$r_{B26}$ magnitude on each of the plates.  Section \ref{zwisosec}
describes our calibration of the void galaxy Zwicky magnitudes.  We
examine the Century Survey and 15R magnitude calibrations in
\S\ref{cs15risosec}.

\begin{subsubsection}{Calibration of Void Galaxy CGCG Magnitudes}
\label{zwisosec}There have been suggestions that Zwicky deviated 
from a Pogson scale when assigning magnitudes to his faintest ($\mzw
\gtrsim 15$) galaxies (\cite{bs82,gh84}).  Such a scale error would
have a significant impact on luminosity functions derived for the CfA2
survey (cf.~\cite{mhg94}).  The largest previous investigation of the
faint CGCG magnitudes with CCD photometry was by Bothun \& Cornell
(1990), who studied 107 cluster spirals of which 66 have $\mzw
\geq 15.0$.  They found that in the mean, $\mzw$ corresponds well to
$b_{B26}$ with a scatter of 0.31 mag, comparable to the scatter at
brighter $\mzw$ (\cite{h76}).  Although Bothun \& Cornell observed
that the Zwicky magnitudes were not closely isophotal, they found a
linear relation between the magnitude systems with a slope
$db_{B26}/d\mzw = 1.05\pm0.05$, consistent with zero scale error.

From Table \ref{phottab} we have 230 isophotal $B$ magnitudes of
Zwicky catalog galaxies down to the estimated catalog completeness
limit of $\mzw \leq 15.5$.  Of that number, 165 have $\mzw \geq
15.0$ and furnish an excellent dataset with which to gauge the
accuracy of Zwicky's magnitude estimates near the CfA2 survey limit.
Our sample is complementary to that of Bothun \& Cornell because we
look at CGCG galaxies {\sl outside} of clusters, comprising both early
and late types.  In Figure \ref{zwcalfig} we plot $b_{B26}$ versus
$\mzw$ for the $\mzw \leq 15.5$ galaxies (crosses) along with an
additional 8 Zwicky galaxies with $\mzw = 15.6$--15.7 (open squares)
from 15R and CS\@.  The solid line in the figure is $b_{B26}=\mzw$; the
dotted line is a linear least-squares fit to our data (with one outlier at
$>3\sigma$ removed), which has slope $1.09\pm0.06$ and an offset 
of $+0.10\pm0.03$ mag from $b_{B26}=\mzw$ at $\mzw=15.5$.  
\placefigure{zwcalfig}

Our slope is consistent with the Bothun \& Cornell (1990) measurement,
and represents the detection of a modest (0.09 mag mag${}^{-1}$) scale
error in the faint Zwicky magnitudes at a $1.7\sigma$ significance.
The scatter in the linear fit is 0.32 mag, remarkably close to the
scatter found by Bothun \& Cornell.  We do not include the $\mzw >
15.5$ galaxies in the fit because these galaxies are not in CfA2.  We
are chiefly interested in converting the CfA survey limit
($\mzw=15.5$) to the $b_{B26}$ scale.  Even with only 8 galaxies at
$\mzw > 15.5$, we can see from Figure \ref{zwcalfig} that Zwicky, just
as he claimed, included substantially fainter objects in his last two
magnitude bins.

With the redshifts for these Zwicky galaxies (Tab.~\ref{phottab}), we
can also investigate the Zwicky magnitude error ($b_{B26}-\mzw$) as a
function of {\sl absolute} Zwicky magnitude $M_{\rm Zw}$ (Figure
\ref{zwerramagfig}).  The Spearman rank correlation between the
magnitude error and $M_{\rm Zw}$ is a negligible 0.02.  A linear
least-squares fit to the points in Figure \ref{zwerramagfig} (with
$2\sigma$-clipping) yields a Zwicky absolute magnitude scale error of
$0.06\pm0.04$ mag mag${}^{-1}$ with a scatter of 0.28~mag.  Thus we
conclude that the Zwicky's magnitude estimation was not biased by
the intrinsic brightness of the galaxy over the range $-20.5 \lesssim M_{\rm Zw}
\lesssim -18$.
\placefigure{zwerramagfig}

\cite{mhg94} found a highly significant discrepancy in the maximum-likelihood
LF parameters $\alpha$ and $M_*$ between CfA2 North and CfA2 South
subsamples, with the CfA2 North $M_*$ some $0.26$ mag fainter.  In
addition to the possibility that the shape of the
$cz\lesssim10,000$~\kms\ LF really does vary between the two Galactic
caps, the authors suggest a number of potential systematic differences
between the North and South Zwicky magnitudes that could reproduce the
discrepancy (cf.~Fig.~6 of \cite{mhg94}): scale error offset, zero-point
offset, varying faint-end incompleteness, and variation in the
estimated Zwicky magnitude scatter.  For example, they suggest that a
differential scale error of 0.1 mag mag${}^{-1}$ coupled with modest
extra incompleteness of $\sim\!15\%$ by $\mzw=15.5$ would bring the
CfA2 North and South LFs into agreement.

Our sample is sufficiently large to constrain most of the proposed
Zwicky magnitude discrepancies between northern and southern Galactic
caps.  We repeat the linear fit between $b_{B26}$ and $\mzw$ on
subsamples of 84 CfA2 North galaxies and 146 CfA2 South galaxies.  We
summarize the results in Table \ref{isonstab}.  Although the
differential slope and offset in the linear fits to the North and
South $\mzw$ have relatively large errors, they have roughly equal and
opposite effects on the North/South LF discrepancy (\cite{mhg94}).  It is
curious that the CfA2 North galaxies in our study have a noticeably
smaller scatter than in CfA2 South (0.24 mag versus 0.35 mag).  Huchra
(1976) did not see this north/south difference in scatter with a
somewhat smaller sample of photoelectric photometry.  Bothun and
Cornell (1990) did not address the issue.  If the reduced
dispersion is representative of the entire CfA2 North, this
difference would drive the North and South $M_*$ estimates closer, but
only by $\sim\!0.1$ mag.
\placetable{isonstab}
\end{subsubsection}
\begin{subsubsection}{Calibration of Century Survey and 15R Magnitudes}
\label{cs15risosec}
For the 13 Century Survey galaxies in our study, we find a linear fit
between $r_{B26}$ and $m_{\rm CS}$ with slope $1.13\pm0.07$ mag
mag${}^{-1}$ and offset of $+0.08\pm0.07$ mag at the CS limiting
magnitude $m_{\rm lim, CS} = 16.13$.  The scatter in the fit is 0.18
mag, near the $\sim\!0.25$ mag scatter estimated by Geller et
al.~(1997) between the CS plate magnitudes and CCD calibrations.
Table \ref{cs15rcaltab} summarizes these results, along with the 15R
calibrations below.
\placetable{cs15rcaltab}

15R North is calibrated from the Century Survey, thus we simply use
the linear fit to the CS magnitudes to find the 15R North $r_{B26,\rm
lim}$.  The offset at the 15R North magnitude limit $m_{\rm lim, N15R}
= 15.42$ is $-0.01\pm0.05$ mag.  The scatter in the 15R North
magnitudes about the Century Survey fit is 0.25 mag, not appreciably
worse than the CS scatter.

The situation for 15R South is less straightforward because the survey
is uncalibrated except for the measurements here.  Seven of the 12
POSS plates containing 15R void galaxies in this study have four or
more members (cf.~Tab.~\ref{cs15rcaltab}).  In these cases we fit a
linear model for $r_{B26}$ as a function of $r_{\rm 15R}$.  We find
that the scatter about these plate-specific calibrations is typically
low, $\sim\!0.1$ mag, but the number of galaxies involved is far too
small to comment on the accuracy of the plate scan magnitudes.  

Five other 15R South plates contain only one or two galaxies from our
sample, giving us minimal constraints on those plates' limiting
magnitudes.  For these plates we adopt a linear relation with fiducial
slope 0.75 mag mag${}^{-1}$ and passing through the single calibration
datum or the mean of the two data as appropriate.  Where there are two
calibrators per plate, we see that the ``scatter'' of those points
about the fiducial linear model is similar to the $N\geq4$ plates.
Although the $r_{B26, \rm lim}$ estimates for these plates are highly
uncertain, there are only eight galaxies in our sample of $\sim\!300$
which are affected.

Figure \ref{sidebysidefig} summarizes the results of our 15R
calibrations.  On the left we plot $r_{B26}$ versus $r_{\rm 15R}$ for
all the 15R galaxies in our sample, with the (dotted) line $r_{B26} =
r_{\rm 15R}$ for comparison.  We represent galaxies from different
plates with unique symbols as indicated in the figure.  On the right
we plot $r_{B26}$ versus the ``corrected'' 15R magnitudes $r_{\rm
15R'}$ using the plate-specific linear models for 15R South galaxies
and the CS linear model for the 15R North galaxies (from plates 324
and 325).  The linear corrections are sufficient to bring the
discrepant $r_{\rm 15R}$ into good agreement with $r_{B26} = r_{\rm
15R'}$ (dotted line, right panel).
\placefigure{sidebysidefig}
\end{subsubsection}
\end{subsection}
\end{section}
\begin{section}{Luminosity Function Fitting} \label{lffitsec}
Here we use our two-color photometry to analyze the
void galaxy luminosity function in both $B$ and $R$.  Because we have 
drawn the sample from both $B$-selected and $R$-selected redshift surveys,
however, the procedure for fitting a luminosity function is more
involved than for a sample selected from a single magnitude-limited survey.  
We describe the
technique in \S\ref{lftecsec} and the results in \S\ref{lfressec}.

\begin{subsection}{Technique} \label{lftecsec}
We follow the method of Sandage, Tammann, and Yahil (1979, hereafter
STY), who solve for an optimal parametric luminosity function $\phi$
by varying the parameters to maximize the likelihood of the observed
luminosity distribution.  With a sample of N galaxies located at
$r_i$, the likelihood ${\cal L}$ is given by the product of the
probability $p$ that the observed absolute magnitude $M_i$ of each is
drawn from $\phi$:
\begin{equation} \label{origprobeq}
{\cal L} = \prod_{i=1}^N{p(M_i, r_i | \phi)} = \prod_{i=1}^N{{\phi(M_i)}\over
{\int_{-\infty}^{M_{\rm lim}(r_i)}{\phi(M)\,dM}}},
\end{equation}     
where $M_{\rm lim}$ is given by equation (\ref{mlimeq}).  The STY technique
is particularly well-suited to our void sample because it is
independent of the survey geometry, easily accommodates variations in
survey magnitude limit, and is minimally biased by nonuniform density
fields (Efstathiou, Ellis, \& Peterson 1988, hereafter \cite{eep}).

The STY likelihood of equation (\ref{origprobeq}) assumes that the survey
limiting magnitude at each galaxy location, $M_{\rm lim}(r_i)$, is
always in the same passband as the LF we are estimating.  Our sample
does not satisfy this assumption: our heterogeneous collection of void
galaxies is drawn from surveys that are $B$-limited (CfA2) as well as
$R$-limited (15R and CS).  Rather than attempting to fit a bivariate
LF in $B$ and $R$ to our modest sample, we use the color
distribution of the sample to transform a limiting absolute magnitude
in filter Y into an approximate distribution of limiting magnitudes in
filter X as necessary.  We then perform an appropriate summation in
the denominator of equation (\ref{origprobeq}) over this distribution of
limiting magnitudes:
\begin{equation} \label{fuzzcoleq}
\sum_j^{n_j}{f\left[{(M_X\!-\!M_Y)_j}\right] \times
\int_{-\infty}^{M_{Y,{\rm lim}}(r_i) + (M_X\!-\!M_Y)_j}{\phi(M_X)\,dM_X}},
\end{equation}
where $f\left[{(M_X\!-\!M_Y)_j}\right]$ is the fraction of sample
galaxies with color $(M_X\!-\!M_Y)_j$.  

In practice, we set $f[(\br)_j]$ equal to the sample's (\br) histogram
in 0.1 mag intervals (comparable to our color uncertainty for 5\%
photometry).  As a purely notational convenience
for the following derivations, we introduce the terms $B_{\rm
lim}^{\rm eff}$ and $R_{\rm lim}^{\rm eff}$ as shorthand for the
summation in (eq.~[\ref{fuzzcoleq}]): \setlength{\arraycolsep}{0.5mm}
\begin{eqnarray*} 
\int_{-\infty}^{B_{\rm lim}^{\rm eff}(r_i)}{\phi_B(B)\,dB} &\equiv&
\sum_j^{n_j}
{f\left[{(B\!-\!R)_j}\right] 
\times \int_{-\infty}^{R_{\rm lim}(r_i) + 
(B\!-\!R)_j}{\phi_B(B)\,dB}}, \\
\int_{-\infty}^{R_{\rm lim}^{\rm eff}(r_i)}{\phi_R(R)\,dR} &\equiv&
\sum_j^{n_j}
{f\left[{(B\!-\!R)_j}\right] 
\times \int_{-\infty}^{B_{\rm lim}(r_i) - 
(B\!-\!R)_j}{\phi_R(R)\,dR}}.
\end{eqnarray*}

This treatment assumes that within a given sample, there is negligible
color-magnitude correlation in the data over the range of limiting
absolute magnitude.  One may worry that this assumption is violated by
the data, as we have already seen in \S\ref{magcolsec} that the color
distribution is correlated with density environment.  In fact there is
little color-magnitude correlation in $B$ for any of the samples
(Tab.~\ref{colmagcorrtab}).  Although Table \ref{colmagcorrtab} shows
that $R$ is more strongly correlated with \br, the correlation largely
vanishes for $R\gtrsim-20$, a range which includes all $R$-selected
limiting magnitudes.  Despite the color variation with density, we may
safely assume that the colors and magnitudes remain uncorrelated, at
least near the absolute magnitude limits, for any particular density
subsample.  \placetable{colmagcorrtab} \setlength{\arraycolsep}{3mm}

Inserting the cross-color magnitude limit correction
(eq.~[\ref{fuzzcoleq}]) into the likelihood equation
(eq.~[\ref{origprobeq}]) and taking the logarithm, we arrive at the
expression for the log-likelihood of the $B$ luminosity function
$\phi_B$: \setlength{\arraycolsep}{0.5mm}
\begin{eqnarray} \label{bmleq}
\ln({\cal L}_B) &=& \sum_i^N{\ln[\phi_B(B_i)]} - 
\left\{{\sum_i^{N_B}{\ln\int_{-\infty}^{B_{\rm lim}(r_i)}{\phi_B(B)\,dB}}
}\right\}_{B-\rm selected} \\
&& 
{} - \left\{{\sum_i^{N_R}{\ln\int_{-\infty}^{B_{\rm lim}^
{\rm eff}(r_i)}{\phi_B(B)\,dB}}
}\right\}_{R-\rm selected},  \nonumber
\end{eqnarray}
where the numbers of $B$-selected and $R$-selected galaxies in the
sample are respectively $N_B$ and $N_R$ ($= N - N_B$).  Similarly,
the log-likelihood of the $R$ luminosity function $\phi_R$ is given
by:
\begin{eqnarray} \label{rmleq}
\ln({\cal L}_R) &=& \sum_i^N{\ln[\phi_R(R_i)]} - 
\left\{{\sum_i^{N_R}{\ln\int_{-\infty}^{R_{\rm lim}(r_i)}{\phi_R(R)\,dR}}
}\right\}_{R-\rm selected} \\
&&
{} - \left\{{\sum_i^{N_B}{\ln\int_{-\infty}^{R_{\rm lim}^
{\rm eff}(r_i)}{\phi_R(R)\,dR}}
}\right\}_{B-\rm selected}.  \nonumber
\end{eqnarray}
The quantity $-2\ln({\cal L})$ for a parametric luminosity function of
$n_p$ fitted parameters is distributed about its minimum as a $\chi^2$ with
$(N-n_p)$ degrees of freedom.  

Although our photometry is accurate to $\approx0.05$ mag, the
magnitude limits appearing in equations (\ref{bmleq}) and (\ref{rmleq})
are derived from redshift survey magnitudes that have a much larger
scatter.  We correct for the resulting Malmquist bias by convolving
the integrals in equations (\ref{bmleq}) and (\ref{rmleq}) with Gaussians
in the limiting magnitudes of respective width $\sigma_B = 0.35$ mag
(appropriate for CfA2) and $\sigma_R = 0.25$ mag (appropriate for 15R
and CS).

We test our STY implementation with Monte Carlo realizations of the
sample magnitudes.  We preserve the sample's heterogeneous selection
criteria and draw colors from the sample's \br\ distribution.  We
test both $B$ and $R$ input luminosity functions, using Schechter
functions with parameters ($\alpha_B = -1.0$, $B_* = -18.8$) and
($\alpha_R = -1.17$, $R_* = -20.73$).  In 1000 simulations of the FVS,
we recover the input LF in both passbands to 0.03 in $\alpha$
and 0.02 mag in $M_*$ in the mean.  The Monte Carlo parameter
dispersion is consistent with the parameter confidence intervals
predicted for the actual data (\S\ref{lfressec}).

Although the STY technique can find the optimal parametric LF and its
parameter uncertainties, it does not provide information about the
goodness of fit.  The goodness of fit is typically estimated by
comparison of a nonparametric LF maximum likelihood with the
parametric LF maximum likelihood (\cite{e71}).  The most commonly used
nonparametric LF is from the stepwise maximum-likelihood (SWML) method
of \cite{eep} and consists of a series of steps $\phi_k$ at regular
luminosity intervals $M_k$ separated by $\Delta M$.  

The stepwise maximum likelihood found by independently varying the
step heights $\phi_k$ is then compared, not to the STY maximum
likelihood, but to the stepwise likelihood of the expected steps from
$\phi_{\rm STY}$ (cf.~eq.~[2.15] of \cite{eep}):
\begin{equation} \label{compstyswmleq}
\phi_{k,\rm STY} \approx {{\int^{M_k+\Delta M/2}_{M_k-\Delta M/2}
\phi_{\rm STY}^210^{-0.6M}\,dM}\Bigg/{\int^{M_k+\Delta M/2}_{M_k-\Delta M/2}
\phi_{\rm STY}10^{-0.6M}\,dM}},
\end{equation}
where the approximation assumes that the survey volume in which
galaxies of luminosity $L$ are seen above the magnitude limit
scales as $L^{3/2}\propto10^{-0.6M}$.  This assumption is reasonable
for the typical sample of galaxies from a single redshift survey to a
given flux limit in a single passband within a simple spatial
geometry.  It is not accurate for our samples, which are restricted to
disjoint irregular volumes at varying distances and drawn from
multiple redshift surveys with varying sky coverage and magnitude
limits in two filters.

Because of the difficulty in obtaining $\phi_{k,\rm STY}$ for our
sample, we do not use SWML to evaluate the STY goodness of fit.
Instead we directly compare the observed luminosity cumulative
distribution function (CDF) of the data with the luminosity CDF $C(M |
r_i,\phi)$ predicted for the sample from the parametric LF\@.  For the
case of a sample selected in a single filter, the predicted luminosity
CDF has the form
\begin{equation} \label{predcdfeq}
C(M | r_i, \phi) = {1\over N}\sum_i^N{\min\left({{{\int_{-\infty}^M{\phi(M')\,dM'}}
\over{{\int_{-\infty}^{M_{\rm lim}(r_i)}{\phi(M')\,dM'}}}},1}\right)}
\end{equation}
By preserving the observed redshift distribution, this model-predicted
luminosity CDF is independent of variations in the galaxy density.  We
introduce a cross-color correction to the magnitude limit
(eq.~[\ref{fuzzcoleq}]) in the denominator of (eq.~[\ref{predcdfeq}]),
analogous to our treatment of the likelihood equation
(eq.~[\ref{origprobeq}]).  We may then express the model-predicted
luminosity CDF for a $B$ luminosity function as
\begin{eqnarray} \label{fuzzcdfeq}
C(B | r_i, \phi_B) &=& \left\{{{1\over N_B}\sum_i^{N_B}{\min\left({
{{\int_{-\infty}^B{\phi_B(B')\,dB'}}
\over{\int_{-\infty}^{B_{\rm lim}(r_i)}{\phi_B(B')\,dB'}}},1}\right)}}\right\}
_{B-\rm selected}\\
&& {} + \left\{{{1\over N_R}\sum_i^{N_R}{\min\left({
{{\int_{-\infty}^B{\phi_B(B')\,dB'}}
\over{\int_{-\infty}^{B_{\rm lim}^{\rm eff}(r_i)}{\phi_B(B')\,dB'}}},1}\right)}}\right\}
_{R-\rm selected}, \nonumber
\end{eqnarray}
where the $\min()$ operation on the ratio of integrals is to be
performed {\sl within} the summation implied by $B_{\rm lim}^{\rm
eff}$.  Swapping all instances of $B$ and $R$ in
equation (\ref{fuzzcdfeq}) yields the corresponding model-predicted
luminosity CDF in $R$.  As in our treatment of the STY technique, we
account for the Malmquist bias of the redshift surveys by convolving
the denominator integrals of $C()$ above with Gaussians in the
limiting magnitude.  \setlength{\arraycolsep}{3mm}

We compare an observed luminosity CDF with the model-predicted CDF
$C(M | r_i, \phi)$ using the Kolmogorov-Smirnov (K-S) test.  From the
resultant K-S $D$-statistic, we may obtain the probability $P_{\rm
KS}(\phi)$ of the null hypothesis that the sample was drawn from
$\phi$.  Because our computation of $C()$ incorporates the observed
color distribution via $M_{\rm lim}^{\rm eff}$, we cannot use the
standard approximation for $P_{\rm KS}$ in terms of $D$
(cf.~\cite{nr}).  This limitation pertains even when comparing the
observed luminosity CDF against the predictions from survey LFs (e.g.,
$\phi_{\rm SSRS2}$) with parameters not fit to our data.  We
explicitly compute the probability of the null hypothesis by
generating the distribution of the K-S $D$-statistic between $C()$ and
Monte Carlo realizations of the sample's luminosity CDF.  Our
realizations satisfy the same absolute magnitude limits as the sample,
with magnitudes drawn from the model LF and colors drawn from the
sample's color distribution.
\end{subsection}
\begin{subsection}{Results} \label{lfressec}
We carry out the LF analysis of \S\ref{lftecsec} identically for the
FVS ($n/\bar n < 1$), the VPS ($1 < n/\bar n < 2$), the LDVS ($n/\bar
n < 0.5$), and the HDVS ($0.5 < n/\bar n < 1$).  For each sample, we
show separate figures for the $B$ and $R$ LF determination.  The upper
plot of each LF figure (Figs.~\ref{fullblffig}a--\ref{hilorlffig}a)
shows the joint-probability intervals (solid contours) of the
Schechter function parameters $\alpha$ and $M_*$.  We also display the
likelihood intervals (dashed contours) which project onto the
individual $n$-$\sigma$ bounds of $\alpha$ and $M_*$.  For comparison
with the maximum-likelihood parameters (plus symbol), we indicate the
Schechter function parameters of various survey LFs.  For $B$ we mark
the CfA2 LF ($\alpha_B=-1.0$, $B_* = -18.8$; \cite{mhg94}) with a
square, the SSRS2 LF ($\alpha_B=-1.12$, $B_* = -19.43$;
\cite{lfssrs2}) with a diamond, and the Stromlo-APM Redshift Survey LF
($\alpha_{B_J}=-0.97$, $B_{J,*} = -19.50$; \cite{apm}) with a cross.
For $R$ we mark the Century Survey LF ($\alpha_R=-1.17$, $R_* =
-20.73$; \cite{cspaper}) with a triangle and the LCRS LF
($\alpha_R=-0.70$, $R_* = -20.64$; \cite{lcrslf}) with an asterisk.

The lower plot of each LF figure
(Figs.~\ref{fullblffig}b--\ref{hilorlffig}b) indicates the goodness of
fit for various Schechter functions to the given sample's luminosity
CDF (jagged solid line).  We overplot the model-predicted $C(M |
r_i,\phi)$ for our best-fit Schechter function (dotted curve) as well
as the Schechter functions for various survey LFs.  For $B$ we show the
predictions for CfA2 (long-dashed curve) and SSRS2 (short-dashed
curve).  For $R$ we show the prediction for the Century Survey (dashed
curve).  In all cases we give the K-S probability (from Monte Carlo
simulation) of the null hypothesis between the Schechter LF and the
data.  We summarize the results in Table \ref{lfrestab}.
\placetable{lfrestab}

Figures \ref{fullblffig}a and \ref{fullrlffig}a show our results for
the STY Schechter function analysis of the 149 galaxies comprising the
FVS\@.  The SSRS2 LF is brighter and steeper than the best-fit $B$ LF,
but in the direction poorly constrained by the data ($\sim\!1.5\sigma$
exclusion).  The CfA2 LF deviates less from the maximum-likelihood
values, but in a better-constrained direction ($>3.5\sigma$
exclusion).  For the $R$ LF, we find best-fit parameters close to the
Century Survey LF, which lies within the 68\% likelihood contour.
Figures \ref{fullblffig}b and \ref{fullrlffig}b show the goodness of
fit of various Schechter functions to the FVS\@.  We find a high K-S
probability of the data being drawn from the either the best-fit $B$
Schechter function or the SSRS2 LF\@.  The CfA2 Schechter function,
however, significantly underestimates the fraction of bright galaxies
actually observed and results in a $P_{\rm KS}(\phi_{\rm CfA2}) =
0.07\%$.  The best-fit $R$ Schechter function predicts a luminosity
CDF which is indistinguishable from the FVS $R$ magnitudes, as does
the Century Survey LF\@.  We find little evidence for an FVS
luminosity distribution significantly fainter than typical survey LFs.
\placefigure{fullblffig} \placefigure{fullrlffig}

Figures \ref{hiblffig}a and \ref{hirlffig}a show the STY Schechter
function analysis of the 131 galaxies comprising the VPS\@.  The
maximum-likelihood Schechter function parameters are highly consistent
with the FVS (cf.~Tab.~\ref{lfrestab}) and similarly close to SSRS2
and CS\@.  Figures \ref{hiblffig}b and \ref{hirlffig}b show the
goodness of fit of various Schechter functions to our VPS\@.  The K-S
probabilities are close to those for the FVS: the VPS luminosity
distribution is indistinguishable from the predictions of the optimal
Schechter functions in each filter or from the survey LFs from SSRS2
and CS ($P_{\rm KS}\approx60\%$).  These results, which suggest that
the VPS is in fact representative of the overall galaxy LF, appear to
argue against significant LF variation between a general redshift
survey and void galaxies.  Again the CfA2 Schechter function is a very
poor fit to the data.  We discuss this issue in \S\ref{discsec}.
\placefigure{hiblffig} \placefigure{hirlffig}

It is only when we look to the galaxies in the very lowest density
regions that we see a deviation from typical survey LFs (SSRS2
and Century).  Figures \ref{loloblffig}a and \ref{lolorlffig}a show
the STY analysis of the 46 galaxies comprising the LDVS\@.  Although the
parameter uncertainties are much larger in this smaller sample, we
find a significantly steeper $B$ LF of $\alpha_B = -1.4\pm0.6$.  This
subsample only excludes the SSRS2 LF at $\lesssim2\sigma$, while the
CfA2 LF is well within the $1\sigma$ likelihood contour.  The
maximum-likelihood $R$ LF has a similarly steep $\alpha_R =
-1.4\pm0.5$ and excludes the Century Survey LF at $\approx2\sigma$.
\placefigure{loloblffig}
\placefigure{lolorlffig}

Figures \ref{loloblffig}b and \ref{lolorlffig}b show the goodness of
fit of various Schechter functions to the LDVS\@.  The K-S probability
of the data being consistent with the best-fit $B$ Schechter function
is reasonable ($<1\sigma$ discrepancy) although smaller than for the
other samples.  The probability of the CfA2 LF being consistent with
the LDVS is comparably high, while SSRS2 LF is mildly discrepant
($P_{\rm KS} = 16\%$).  We see from Figure \ref{loloblffig}b that the
SSRS2 LF overpredicts the observed CDF at essentially all magnitudes.
All three Schechter LFs predict a steeper luminosity CDF around $B_*$
than we observe.  The maximum-likelihood Schechter function in $R$
predicts a luminosity CDF consistent with the data ($P_{\rm KS} =
66\%$), while the Century Survey LF overpredicts the fraction of
bright galaxies.  In this respect, the CS LF is more discrepant
with the $R$ magnitudes ($P_{\rm KS} = 3.5\%$) than SSRS2 in $B$.

In light of the steeper LDVS, it is not surprising that we see a
shallower LF for the 103 galaxies comprising the complementary HDVS
(Figs.~\ref{hiloblffig}a and \ref{hilorlffig}a).  As we found for the
FVS, the HDVS discrepancy with SSRS2 and CS is in the
poorly-constrained direction, while the CfA2 LF is strongly excluded
($>4\sigma$).  Although the observed luminosity CDF is fit well by the
$B$ and $R$ maximum-likelihood Schechter functions
(Figs.~\ref{hiloblffig}b and \ref{hilorlffig}b), the survey LFs
underpredict the observed number of $\sim\!M_*$ galaxies at the
$\sim\!2\sigma$ level.
\placefigure{hiloblffig}
\placefigure{hilorlffig}

An interesting feature of Figure \ref{lolorlffig}b is the sharp cutoff
of the LDVS CDF for the brightest red galaxies ($R < -21$).  The
survey LF predicts that $\approx10\%$ of the sample should be brighter
than the brightest observed $R$, a $>2.5\sigma$ discrepancy.  We
conclude from Figure \ref{lfsummfig} that there is evidence for a
steepening in the faint-end slope of the void galaxy LF at the lowest
densities and a relative deficit of red objects
(cf.~Fig.~\ref{tridencolfig}), particularly at the bright end.
\placefigure{lfsummfig}
\end{subsection}
\end{section}
\begin{section}{Discussion} \label{discsec}
The luminosity distribution of 149 galaxies within underdense ($n<\bar
n$) regions of CfA2 is very similar to the predictions of typical
survey LFs from SSRS2 ($\alpha_B=-1.12$; $B_*=-19.43$) and the Century
Survey ($\alpha_R=-1.17$; $R_* = -20.73$).  The 131 galaxies in our
study surrounding the voids at $\bar n<n<2\bar n$ also show a
luminosity distribution consistent with these survey LFs.  These two
results (cf.~Tab.~\ref{lfrestab}), as well as the similarity in colors
between the HDVS and the VPS (cf.~Fig.~\ref{tridencolfig}), suggest
that the influences of environment upon galaxy formation and
evolution that shape a survey LF also pertain in regions of at least
moderate global underdensity.

Oddly, the CfA2 LF as fit by \cite{mhg94} is a poor match to the $B$
magnitudes of our higher-density samples, which are largely drawn from
CfA2 but agree instead with the SSRS2 LF\@.  The discrepancy of
($\Delta B_*\approx0.6$ mag) between the LFs of these disjoint
wide-angle surveys to similar depth ($m_{B(0)} \approx 15.5$) has
recently received attention from Marzke et al.\ (1998).  In fitting
type-dependent LFs to SSRS2, they observe a deficit of bright CfA2
galaxies across all morphological types and suspect some systematic
error in the Zwicky magnitudes.  Our CCD photometry of 230 CGCG
galaxies places the best current limits on such errors.

Our CfA2 North magnitudes do have a lower scatter (0.24 mag) than the
0.35 mag assumed by \cite{mhg94} when fitting the CfA2 LF, although
this effect would only brighten the North LF by $\sim\!0.1$ mag.  We
see a modest faint-end CGCG scale error of $0.09\pm0.06$ mag
mag${}^{-1}$ and zero-point offset of $\lesssim0.10\pm0.03$ with an
overall scatter of 0.32~mag.  Our scale error and zero-point offset
have roughly equal and opposite effects on the derived LF parameters.
In a recent preprint Gazta\~naga \& Dalton (1999) report that their
CCD photometry of 204 Zwicky galaxies yields a large scale error of
$\gtrsim 0.3$ mag mag${}^{-1}$, highly discrepant with our findings
and those of Bothun \& Cornell (1990).  Possible explanations of the
difference may include the larger scatter of their CCD photometry or
their method of compensating for Malmquist bias in their fitting.

We have established that the Zwicky magnitude error does not
correlate with {\sl absolute} magnitude.  The deficit of bright CfA2
galaxies noted by Marzke et al.\ (1998) is not trivially explained by
Zwicky haing systematically overestimated the magnitudes of intrinsically
bright objects.
\end{section}
\begin{section}{Conclusions} \label{concsec}
Using a large-scale ($5 h^{-1}$~Mpc) density estimator applied to the
CfA2 redshift survey, we construct an optically-selected,
magnitude-limited sample of galaxies in and around three prominent
nearby voids.  With CCD photometry for these galaxies in $B$ and $R$,
we assess the luminosity and color distributions of void galaxies with
a much larger sample than previous studies.  We also have less
selection bias against early-type galaxies than most previous void
galaxy studies, which chose objects based on strong H{\sc i},
infrared, or line emission.  A goal of this study is to compare the
data against model predictions which, for instance, suggest void
galaxies should be underluminous relative to the field due to lack
of tidal interactions (\cite{lac93}).

The luminosity and color distributions for regions with $n \leq 0.5
\bar n$ (the LVDS) differ significantly from those for denser
regions. The shift toward blue galaxies in the LVDS is particularly
pronounced compared with our highest density sample (the VPS) at $1 <
n/\bar n \leq 2$, with a K-S probability of 0.6\% that the
samples' \br\ colors are drawn from the same underlying distribution.
It is noteworthy that these two samples are well separated in density;
the uncertainty in the density estimator is $\lesssim0.1$ at the
distance of the three voids in this study.  We rule out a difference
in absolute magnitude distribution as the cause of the color shift.

Both the $B$ and $R$ LFs are significantly steeper
($\alpha\approx-1.4$) in the lowest-density regions.  Despite our
optically-selected sample having less bias against the inclusion of
early-type void galaxies than previous studies based on {\sl IRAS},
H{\sc i}, or emission-line identifications, we observe that the
brightest red galaxies ($R \lesssim -21$) are missing from the LDVS at
$2.5\sigma$ relative to predictions from the field LF
(cf.~Fig.~\ref{lolorlffig}b).  The deviations of color and LF suggest
that the processes which account for luminous ellipticals are
ultimately suppressed at a sufficiently low global density threshold
($\sim\!0.5\bar n$).  Perhaps such galaxies can only form in regions of high 
local density enhancement (via mergers or otherwise) and require too great
a density contrast in regions of extreme global underdensity.

Our observed shift in galaxy properties at the lowest densities has
some precedent in recent theoretical models, although the differences
are more subtle than the models predict.  For example, the
tidally-triggered galaxy formation model of Lacey et al.~(1993)
produces too few luminous red galaxies and a present-day LF somewhat
steeper than the field.  On the other hand, underdense regions in
their model do not contain {\sl any} luminous galaxies.  Even in the
LDVS we find many galaxies with $M \gtrsim M_*$
(cf.~Figs.~\ref{loloblffig}b and \ref{lolorlffig}b), in agreement with
previous studies of the Bo\"otes void (\cite{sz96b,cwh97}).  A more
recent simulation by Kauffmann et al.~(1999) with semi-analytic galaxy
formation does predict blue galaxies in the voids, but fails to
produce the red galaxies we observe in the LDVS\@.

Because the centers of voids are so empty of galaxies, it is difficult
to increase the significance of the LDVS results further without
deeper, wide-angle redshift surveys like the Sloan Digital Sky Survey
(\cite{sdss}) and the 2dF Galaxy Redshift Survey
(\cite{2df}).  Including the galaxies within the lowest density
portions of other large CfA2 and SSRS2 voids could increase the
low density sample to $\sim\!100$ galaxies.

\acknowledgments 
We thank the several observers who took photometric calibration
snapshots for us, including P. Barmby, E. Barton, D. Koranyi,
L. Macri, K. Stanek, and A. Vikhlinin.  We also thank D. Koranyi,
M. Kurtz, and R. Marzke for useful discussions, R. Jansen for his
field galaxy photometry in advance of publication, and J. Kleyna for
the WCS-fitting software.  This research was supported in part by the
Smithsonian Institution.
\end{section}

\clearpage
\begin{table}[bp]
\caption[]{ \label{voidtab}
Approximate Bounds of Target Voids}
\medskip
\begin{tabular}{lr@{ $\leq\alpha_{1950}\leq$ }lr@{ $\leq\delta_{1950}\leq$ }l
r@{ $\leq cz \leq$ }l}\tableline \tableline
Label & \multicolumn{2}{c}{Right Ascension} & \multicolumn{2}{c}{Declination} 
& \multicolumn{2}{c}{Velocity\tablenotemark{a}\ \ [km/s]} \\ \tableline
NV1 & $13^{\rm h}45^{\rm m}$ & $16^{\rm h}45^{\rm m}$ &
$22\arcdeg$ & $44\arcdeg$ & 5300 & 8200 \\
SV1 & $21^{\rm h}$ & $23^{\rm h}35^{\rm m}$ &
$-2\fdg5$ & $20\arcdeg$ & 4500 & 7500 \\
SV2 & $23^{\rm h}30^{\rm m}$ & $2^{\rm h}$ &
$-1\arcdeg$ & $20\arcdeg$ & 5900 & 10300 \\
\tableline
\end{tabular}
\tablenotetext{a}{Velocity bounds are Galactocentric.}
\end{table}
\dummytable
\begin{deluxetable}{lcccccc}
\tablecaption{\label{phottab}
Coordinates and Magnitudes of Galaxies in this Study}
\tablewidth{0pt}
\tablenum{2}
\tablehead{
& R.A.
& Decl.
& \colhead{$cz$}
& \colhead{Density\tablenotemark{a}}
& \colhead{$b_{B26}$\tablenotemark{b}}
& \colhead{$r_{B26}$\tablenotemark{b}}\\
\colhead{Name} 
& (B1950.0)
& (B1950.0)
& \colhead{(km/s)}
& \colhead{$(n/\bar n)$}
& \colhead{(mag)}
& \colhead{(mag)}
}
\startdata
\multicolumn{7}{c}{\it CfA2 Survey Galaxies} \nl
IC 5378       & 00 00 03.98 & $+$16 21 56.9 &
 6554 & 1.19 & 14.59 & 13.20 \nl
00012$+$1555  & 00 01 10.32 & $+$15 54 30.2 &
 6636 & 1.14 & 15.79 & 14.75 \nl
00017$+$1030  & 00 01 39.12 & $+$10 30 42.5 &
 8134 & 0.91 & 15.13 & 14.00 \nl
NGC 7825      & 00 02 32.74 & $+$04 55 31.1 &
 8220 & 1.45 & 15.34 & 13.76 \nl
00055$+$0926  & 00 05 32.74 & $+$09 26 22.6 &
 6606 & 1.18 & 14.80 & 13.28 \nl
00059$+$0956  & 00 05 54.05 & $+$09 55 37.9 &
 6608 & 1.16 & 15.52 & 13.97 \nl
\multicolumn{7}{l}{\nodata}\\
\multicolumn{7}{c}{\it 15R Survey Galaxies} \\
464.015505    & 00 01 02.21 & $+$10 19 30.4 &
 8085 & 0.91 & 15.64 & 14.07 \nl
464.040541    & 00 05 19.08 & $+$11 34 57.7 &
 6685 & 1.06 & 16.43 & 15.19 \nl
464.069313    & 00 11 08.35 & $+$12 55 00.1 &
 8059 & 0.77 & 16.62 & 15.40 \nl
465.011685    & 00 26 35.78 & $+$10 19 10.9 &
 9887 & 1.78 & 16.23 & 15.19 \nl
467.045084    & 01 11 34.99 & $+$12 24 11.5 &
 5894 & 1.18 & 16.00 & 15.05 \nl
467.017701    & 01 19 00.91 & $+$10 36 51.8 &
 9989 & 1.99 & 16.18 & 14.61 \nl
\multicolumn{7}{l}{\nodata}\\
\multicolumn{7}{c}{\it Century Survey Galaxies}\nl
c20.CJ        & 13 33 32.59 & $+$29 28 09.1 &
 6414 & 1.38 & 17.41 & 16.46 \nl
c14.FB        & 13 56 20.90 & $+$29 37 54.1 &
 5748 & 0.65 & 16.85 & 16.04 \nl
c14.HQ        & 14 02 20.71 & $+$29 24 50.0 &
 7624 & 1.51 & 16.41 & 15.37 \nl
c14.JA        & 14 06 04.70 & $+$29 15 11.9 &
 7412 & 1.01 & 17.07 & 15.79 \nl
c14.JK        & 14 07 58.01 & $+$29 40 08.0 &
 6574 & 0.38 & 16.74 & 15.75 \nl
e1390.CX      & 14 44 36.60 & $+$29 23 06.0 &
 8316 & 1.21 & 16.36 & 15.43 \nl
\multicolumn{7}{l}{\nodata}\\
\enddata
\footnotesize
\tablecomments{\baselineskip 12pt
Right ascension in hours, minutes, and seconds of time.  Declination
in degrees, minutes, and seconds of arc.  Velocities are Galactocentric.
Table \ref{phottab} is available in its entirety by request from the authors.
A portion is shown here for guidance regarding its form and content.
}
\tablenotetext{a}{Density uncertainty $\lesssim0.1$ (Grogin \& Geller 1998)}
\tablenotetext{b}{Magnitude uncertainty $\approx0.05$ mag (cf.~\S3)}
\end{deluxetable}
\begin{table}[bp]
\caption[]{ \label{colmagcorrtab}
Color-Magnitude Correlation of Various Samples
}
\medskip
\begin{tabular}{lrrr}\tableline\tableline
\hfill Sample: & VPS & HDVS & LDVS \\
Correlation\tablenotemark{a} & & & \\ \tableline
$r_s(B,\br)$\dotfill& $-0.04$ & $-0.17$ & $0.03$ \\
$r_s(R,\br)$\dotfill& $-0.33$ & $-0.52$ & $-0.27$\\ 
$r_s(R>-20,\br)$ & $-0.06$ & $-0.24$ & $-0.09$\\
\tableline
\end{tabular}
\tablenotetext{a}{Spearman rank-order correlation.}
\end{table}
\begin{table}[bp]
\caption[]{ \label{isonstab}
Linear Fits Between CfA2 Isophotal and Zwicky Magnitudes}
\medskip
\begin{tabular}{lcr@{ $\pm$ }lr@{ $\pm$ }lc}\tableline \tableline
& $N$ & \multicolumn{2}{c}{Slope} & 
\multicolumn{2}{c}{$\mzw=15.5$} & RMS \\
Sample & & \multicolumn{2}{c}{[mag mag${}^{-1}$]} 
& \multicolumn{2}{c}{Offset [mag]} & [mag]\\ \tableline 
North\dotfill & 84 & 1.133 & 0.075 & 0.048 & 0.040 & 0.24\\
South\dotfill & 146 & 1.069 & 0.073 & 0.137 & 0.040 & 0.35\\
Combined\dotfill & 230 & 1.092 & 0.055 & 0.104 & 0.030 & 0.32\\
\tableline
\end{tabular}
\tablecomments{Slope is $d(b_{B26})/d(\mzw)$ of the linear fit;
offset is ($b_{B26}-\mzw$) of the fit at $\mzw=15.5$.}
\end{table}
\begin{table}[bp]
\caption[]{ \label{cs15rcaltab}
Calibration of Century Survey and 15R Magnitude Limits}
\medskip
\begin{tabular}{lccr@{ $\pm$ }lr@{ $\pm$ }lcc}\tableline \tableline
& $N$ & $r_{\rm lim}$ & \multicolumn{2}{c}{$d(r_{B26})/d(r)$} & 
\multicolumn{2}{c}{$r_{B26,\rm lim}$} & RMS \\
Region & & [mag] & \multicolumn{2}{c}{[mag mag${}^{-1}$]} 
& \multicolumn{2}{c}{[mag]} & [mag]\\ \tableline 
Century & 13 & 16.13 & 1.128 & 0.067 & 16.21 & 0.07 & 0.18\\
15R North & 5 & 15.42\tablenotemark{a} & 1.128 & \nodata & 15.41 & 0.05 & 0.25\\
15R South\\\
\hfill Plate 464 & 9 & 15.3 & 0.760 & 0.053 & 15.78 & 0.08 & 0.12\\
\hfill Plate 465 & 5 & 14.5 & 0.669 & 0.233 & 15.39 & 0.37 & 0.40\\
\hfill Plate 466 & 1 & 16.1 & 0.75 & \nodata & 16.16 & \nodata & \nodata\\
\hfill Plate 467 & 8 & 14.9 & 0.795 & 0.035 & 15.63 & 0.05 & 0.07\\
\hfill Plate 468 & 4 & 15.2 & 0.756 & 0.038 & 15.75 & 0.07 & 0.07\\
\hfill Plate 469 & 2 & 15.4 & 0.75 & \nodata & 15.81 & \nodata & 0.09\\
\hfill Plate 518 & 7 & 15.9 & 0.925 & 0.033 & 16.02 & 0.05 & 0.06\\
\hfill Plate 519 & 1 & 16.0 & 0.75 & \nodata & 15.97 & \nodata & \nodata\\
\hfill Plate 520 & 2 & 15.8 & 0.75 & \nodata & 16.14 & \nodata & 0.11\\
\hfill Plate 521 & 2 & 15.0 & 0.75 & \nodata & 16.43 & \nodata & 0.07\\
\hfill Plate 522 & 9 & 15.6 & 0.967 & 0.041 & 15.94 & 0.06 & 0.08\\
\hfill Plate 523 & 15 & 14.6 & 0.318 & 0.058 & 15.95 & 0.08 & 0.17\\
\tableline
\end{tabular}
\tablenotetext{a}{The 15R North $r_{\lim}$ is already calibrated to
the Century Survey; we estimate its $r_{B26,\rm lim}$ and RMS with the
linear fit to the Century Survey galaxies.}
\end{table}
\begin{table}[bp]
\caption[]{ \label{lfrestab}
Summary of Luminosity Function Analyses}
\medskip
\begin{tabular}{lccccccc}\tableline \tableline
Sample\tablenotemark{a} & $N$ & $\alpha$ & $M_*$ [mag] & $P_{\rm KS}(\phi_{\rm BF})$ &
$P_{\rm KS}(\phi_{\rm CfA2})$ & $P_{\rm KS}(\phi_{\rm SSRS2})$ &
$P_{\rm KS}(\phi_{\rm CS})$\\ \tableline 
FVS, $B$ & 149 & $-0.53\pm0.3$ & $-18.91\pm0.2$ & 0.795 & 0.0007 & 0.594 & \\
FVS, $R$ & 149 & $-0.92\pm0.25$ & $-20.43\pm0.25$ & 0.547 & & & 0.661\\
VPS, $B$ & 131 & $-0.75\pm0.3$ & $-19.02\pm0.2$ & 0.659 & 0.0006 & 0.583 & \\
VPS, $R$ & 131 & $-0.92\pm0.3$ & $-20.41\pm0.25$ & 0.619 & & & 0.644\\
LDVS, $B$ & 46 & $-1.42\pm0.6$ & $-19.08^{+0.5}_{-0.6}$ & 0.375 & 0.469 & 0.159 & \\
LDVS, $R$ & 46 & $-1.44\pm0.5$ & $-20.35^{+0.5}_{-0.7}$ & 0.660 & & & 0.035\\
HDVS, $B$ & 103 & $-0.09\pm0.4$ & $-18.79\pm0.2$ & 0.962 & 0.0001 & 0.145 &\\
HDVS, $R$ & 103 & $-0.54\pm0.3$ & $-20.31^{+0.2}_{-0.3}$ & 0.871 & & & 0.258\\ 
\tableline
\end{tabular}
\tablenotetext{a}{The various samples are defined in \S\ref{sampdefsec}.}
\end{table}
\clearpage
\pagestyle{empty}
\begin{figure}[bp]
\caption{\footnotesize Eight successive $3\arcdeg$ declination slices
through CfA2 North delineating the northern void.  
CfA2 galaxies are plotted with crosses; galaxies included
in this study are circled.  We overplot $5h^{-1}$ Mpc-smoothed number
density contours as determined from CfA2.  Underdensities in $0.2\bar
n$ decrements are marked with dotted contours; overdensities in
logarithmic intervals of $\bar n$, $2\bar n$, $4\bar n$, etc., are
marked with solid contours. 
\label{cfa2nfig}}
\plotfiddle{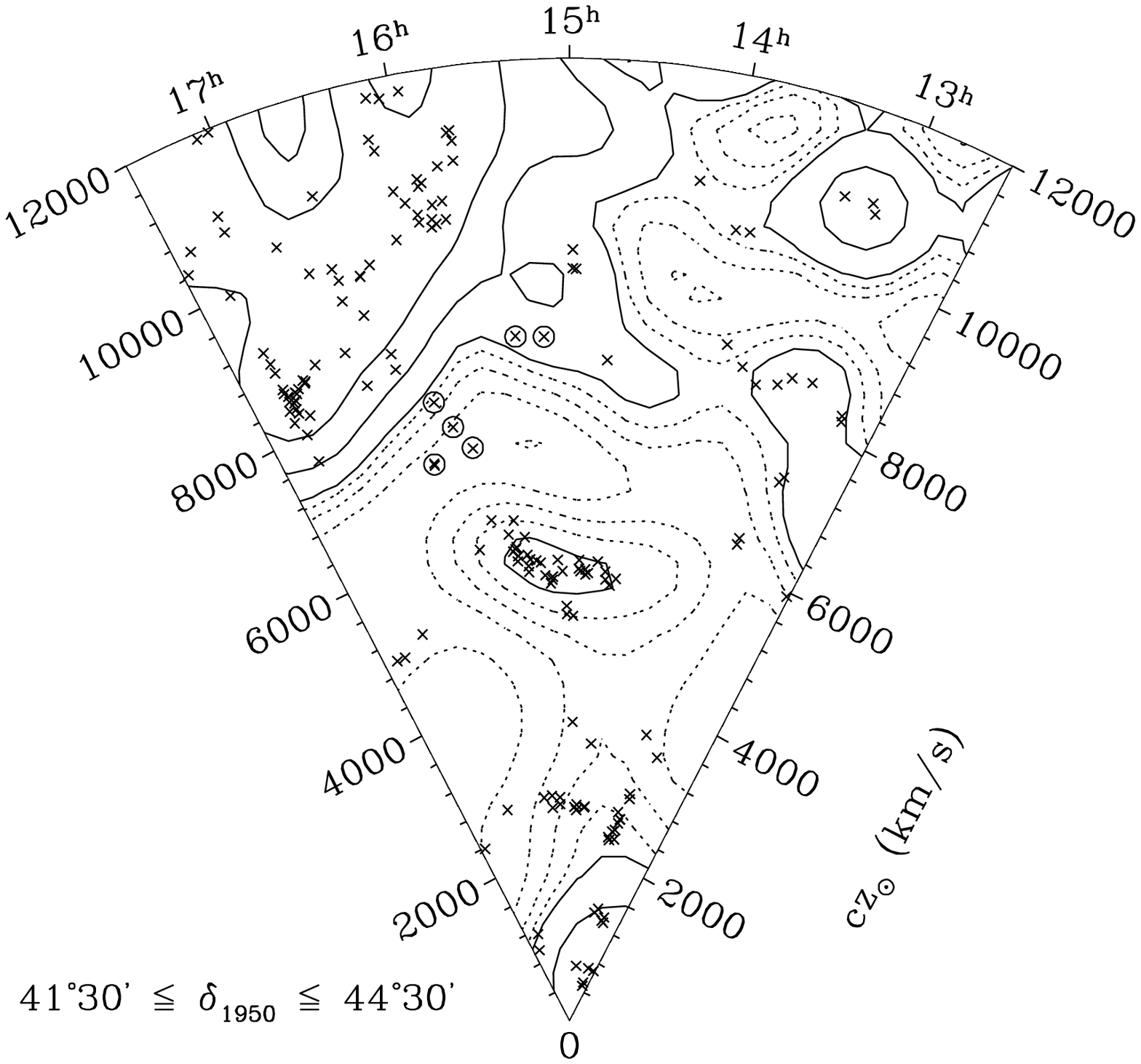}{4in}{0}{50}{50}{-280}{-50}
\plotfiddle{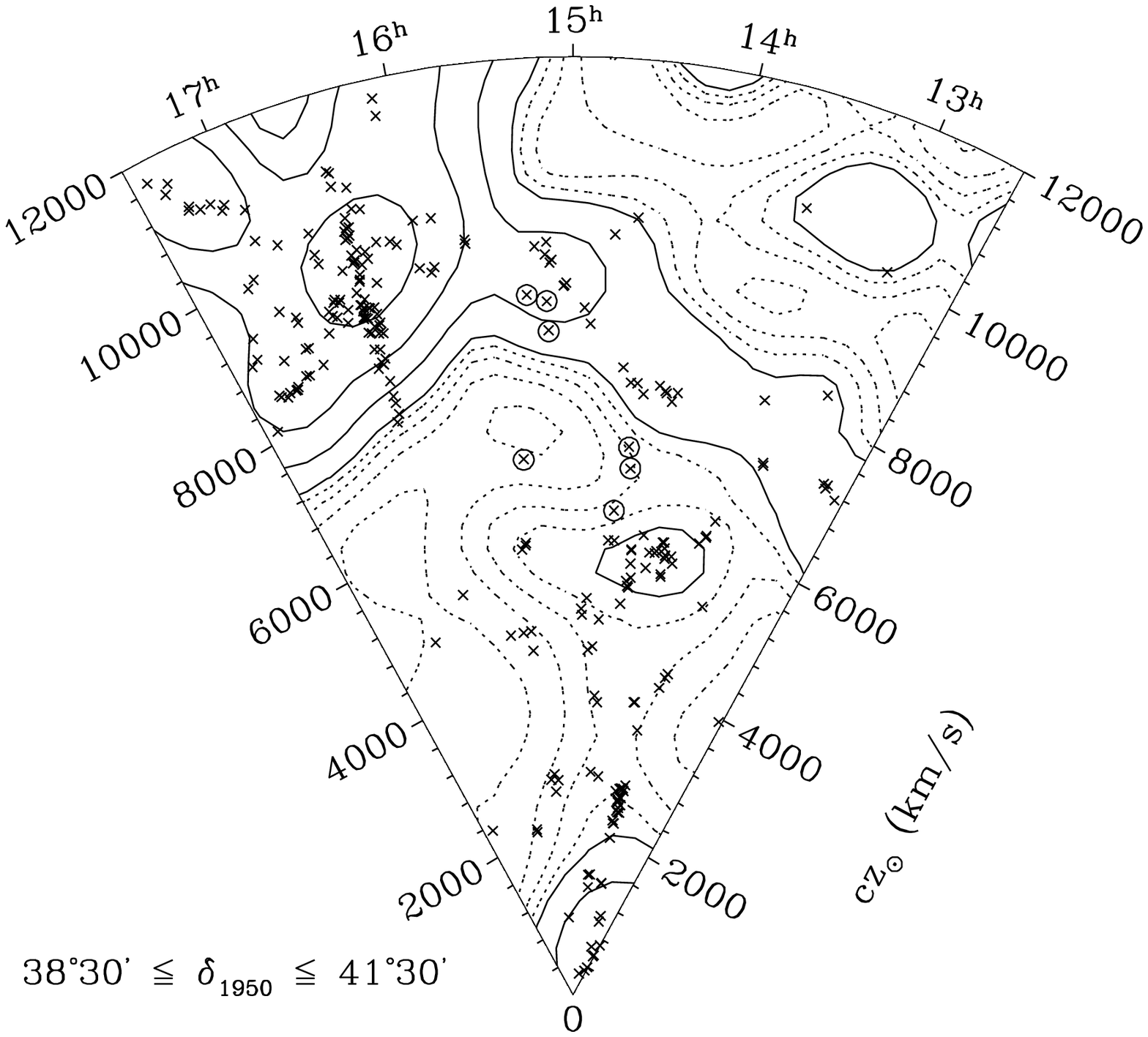}{4in}{0}{50}{50}{-10}{+260}
\plotfiddle{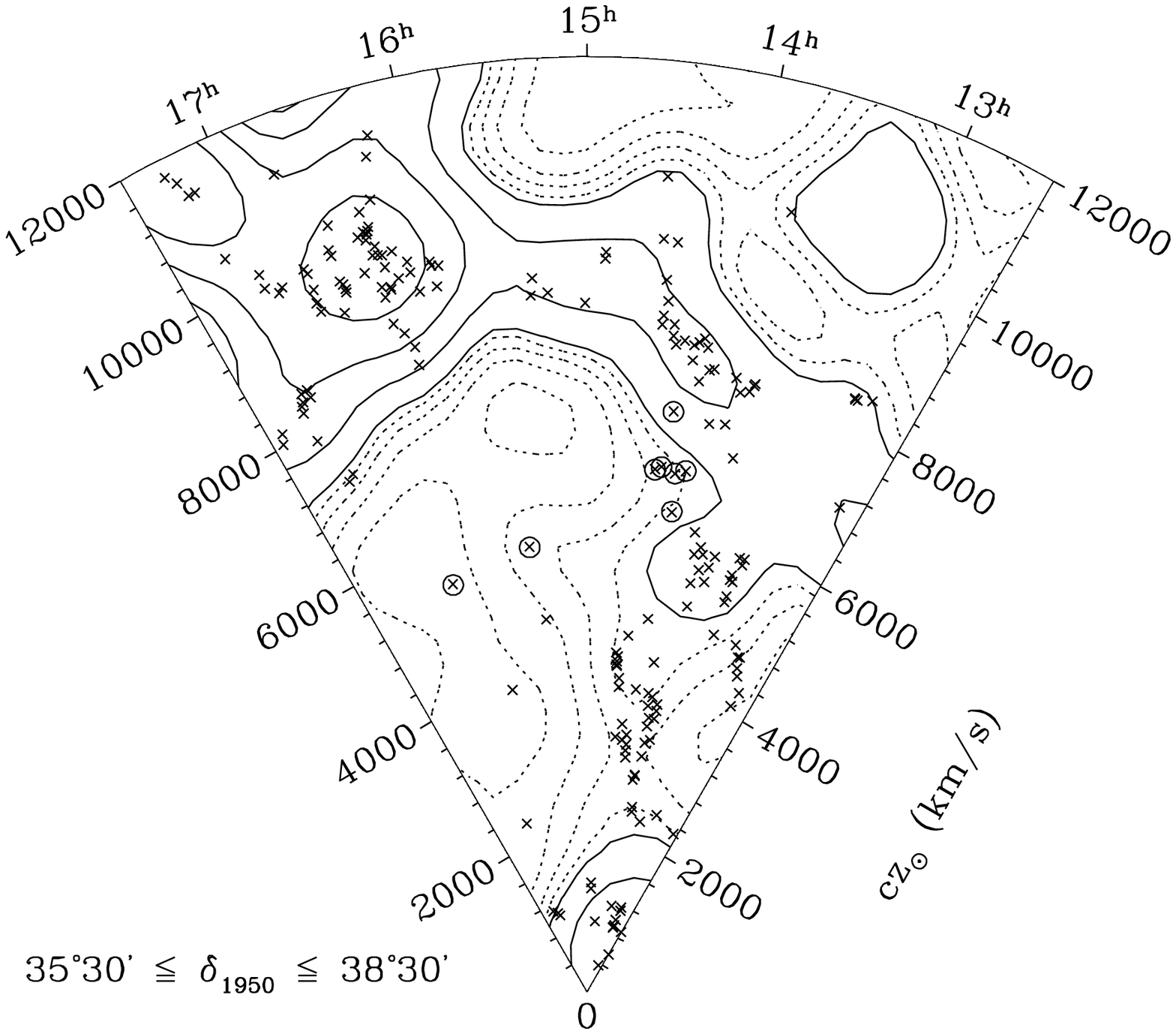}{4in}{0}{50}{50}{-280}{+310}
\plotfiddle{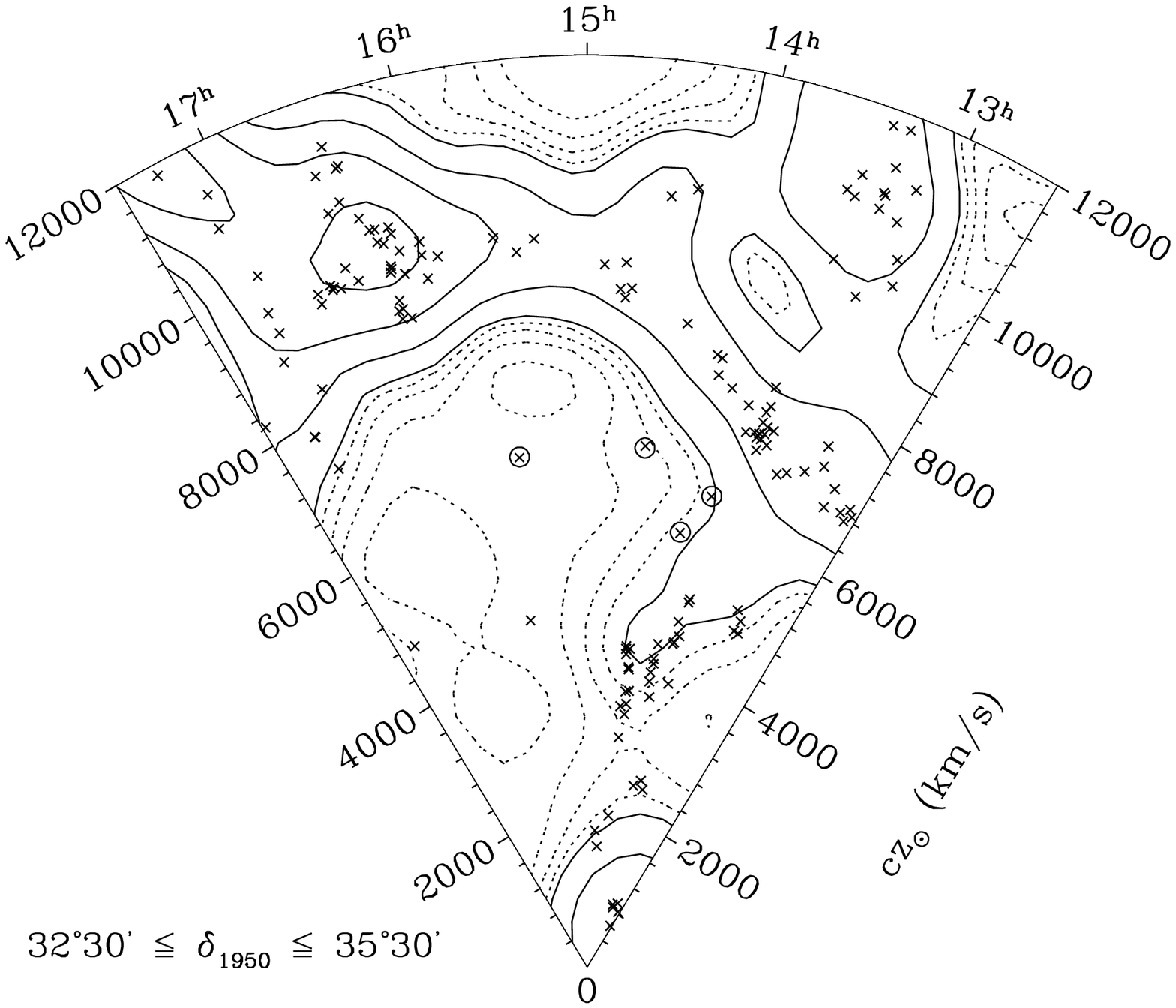}{4in}{0}{50}{50}{-10}{+620}
\end{figure}
\clearpage
\begin{figure}
\figurenum{\ref{cfa2nfig}}
\caption{\footnotesize Cont'd.}
\plotfiddle{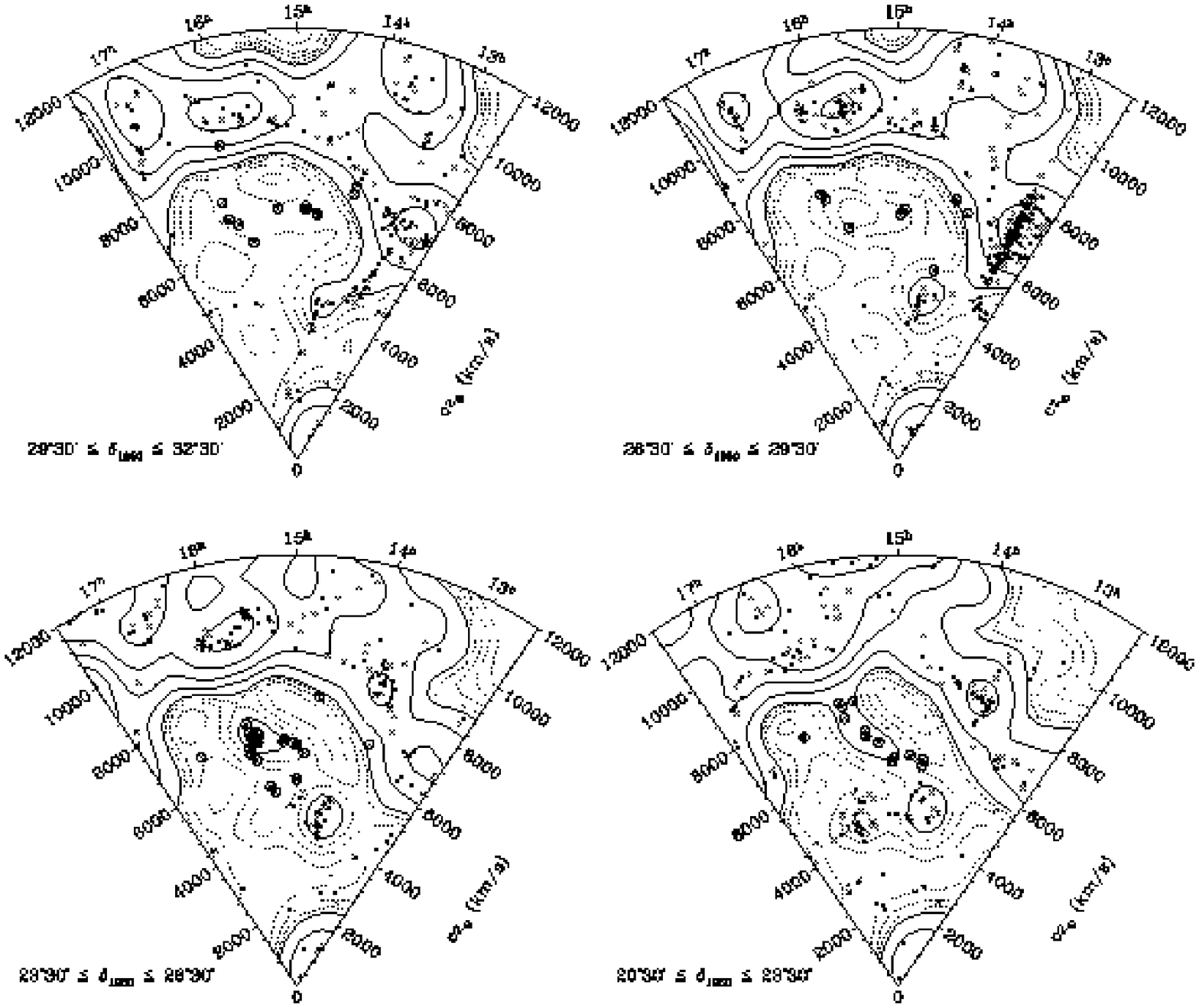}{7in}{0}{100}{100}{-290}{-300}
\end{figure}
\clearpage
\begin{figure}[bp]
\vskip -0.75in
\caption{\footnotesize Eight successive $3\arcdeg$ declination slices
through CfA2 South delineating the two southern voids: 
void SV1 is to the right and void SV2 is to
the left.  CfA2 galaxies are plotted with crosses; galaxies included
in this study are circled.  We overplot $5h^{-1}$ Mpc-smoothed number
density contours as determined from CfA2.  Underdensities in $0.2\bar
n$ decrements are marked with dotted contours; overdensities in
logarithmic intervals of $\bar n$, $2\bar n$, $4\bar n$, etc., are
marked with solid contours.
\label{cfa2sfig}}
\plotfiddle{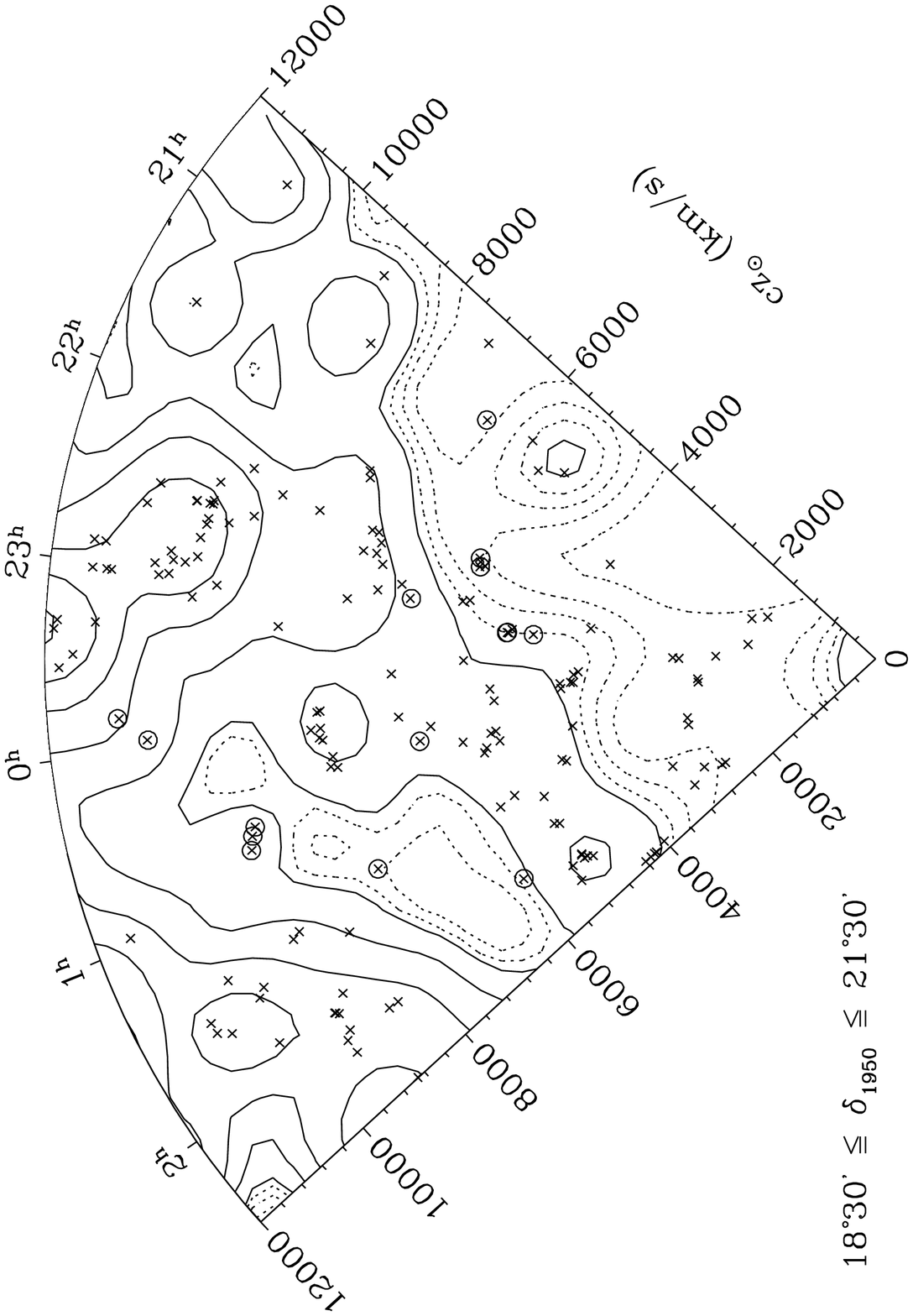}{4in}{270}{60}{60}{-220}{+325}
\plotfiddle{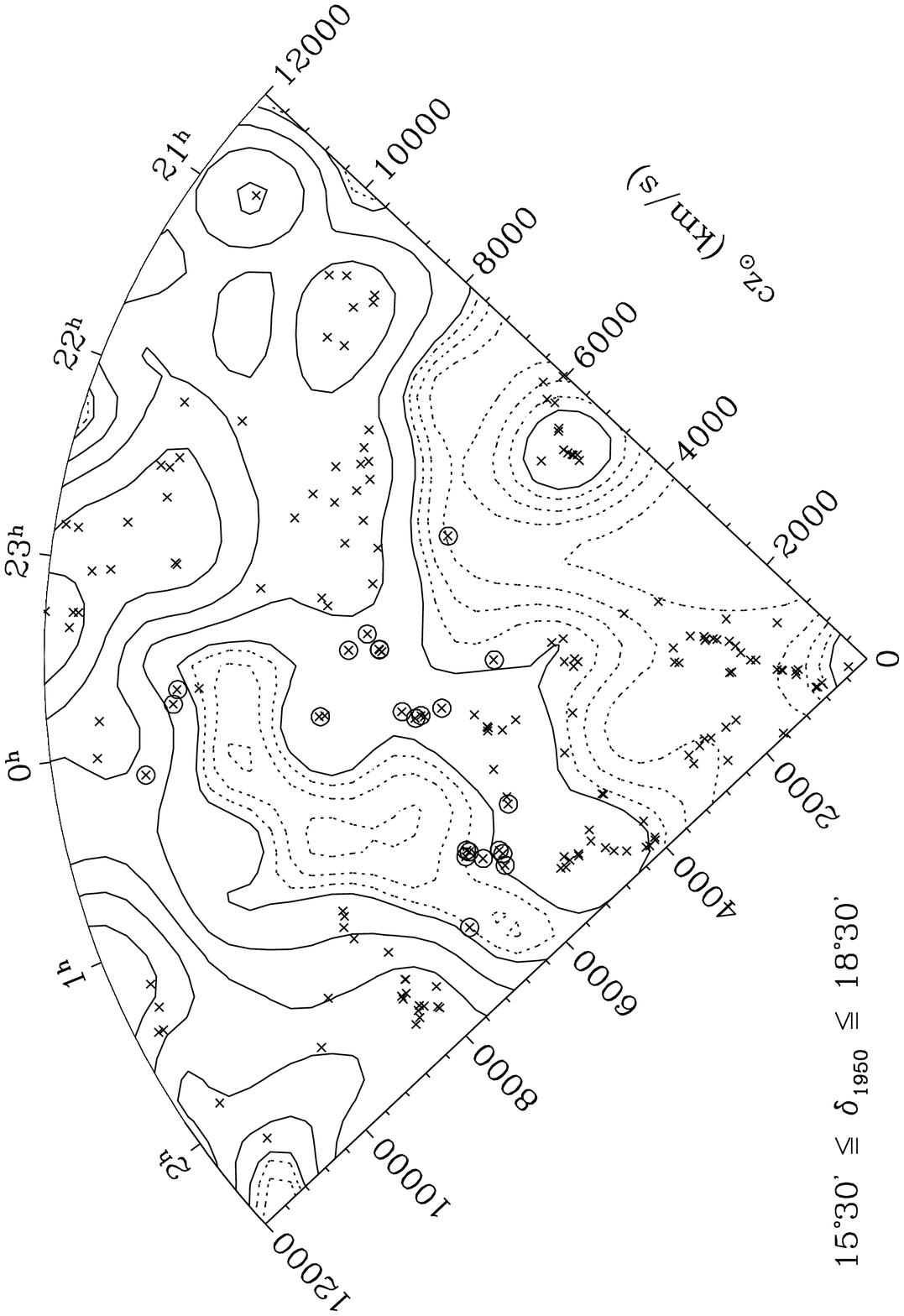}{4in}{270}{60}{60}{-220}{+335}
\end{figure}
\clearpage
\begin{figure}
\figurenum{\ref{cfa2sfig}}
\vskip -1in
\caption{\footnotesize Cont'd.}
\plotfiddle{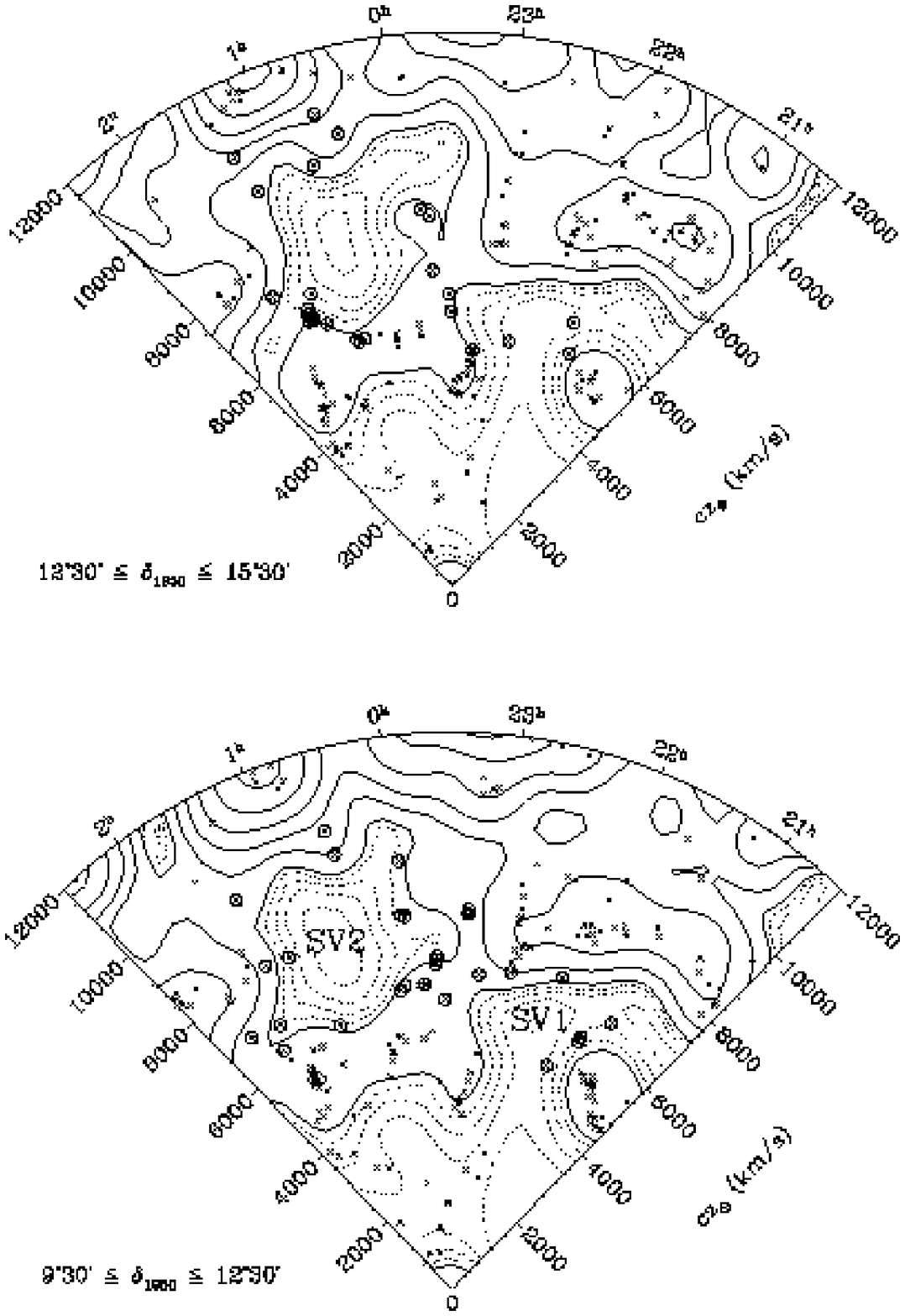}{7in}{0}{100}{100}{-220}{-280}
\end{figure}
\clearpage
\begin{figure}
\figurenum{\ref{cfa2sfig}}
\vskip -1in
\caption{\footnotesize Cont'd.}
\plotfiddle{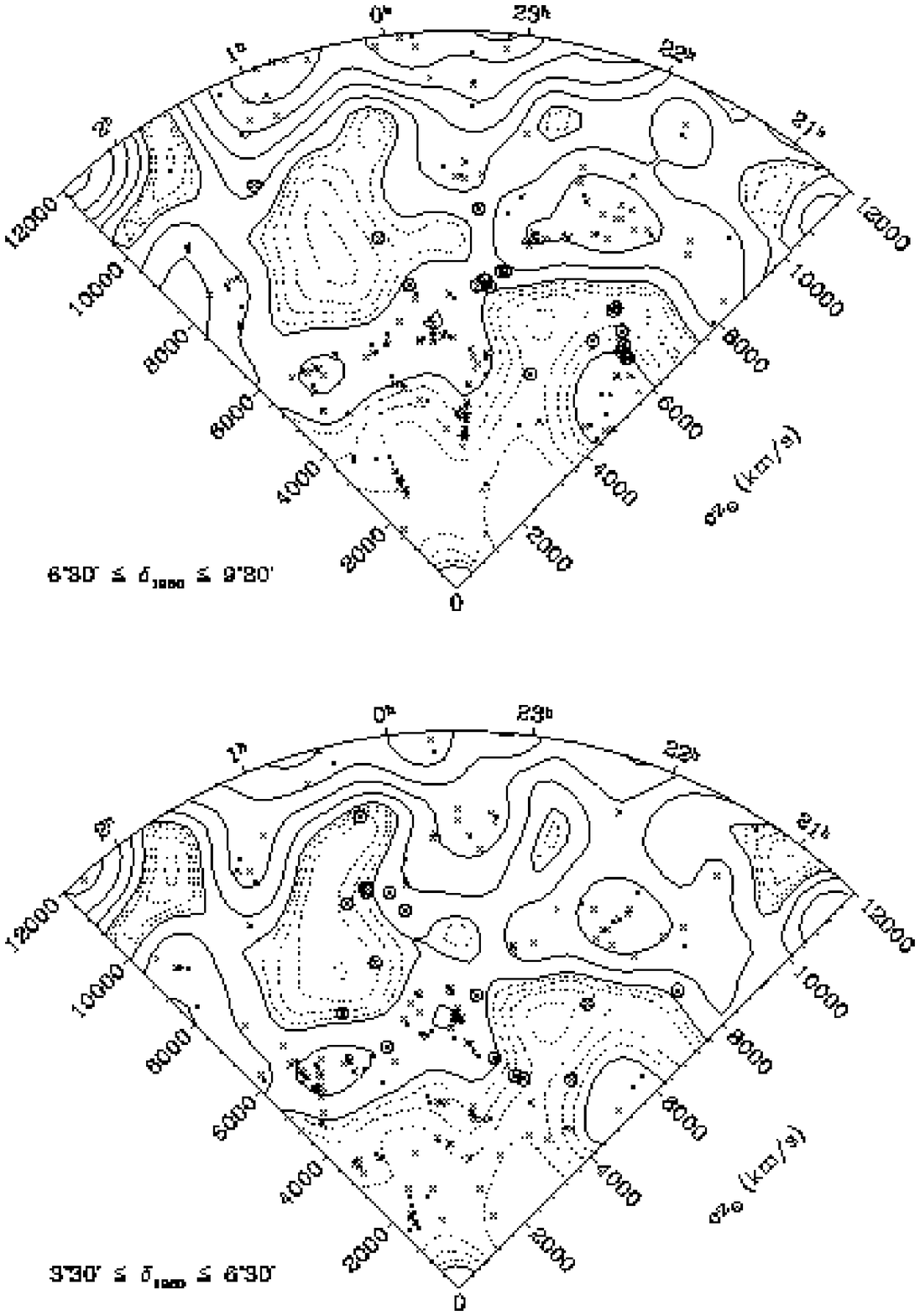}{7in}{0}{100}{100}{-220}{-280}
\end{figure}
\clearpage
\begin{figure}
\figurenum{\ref{cfa2sfig}}
\vskip -1in
\caption{\footnotesize Cont'd.}
\plotfiddle{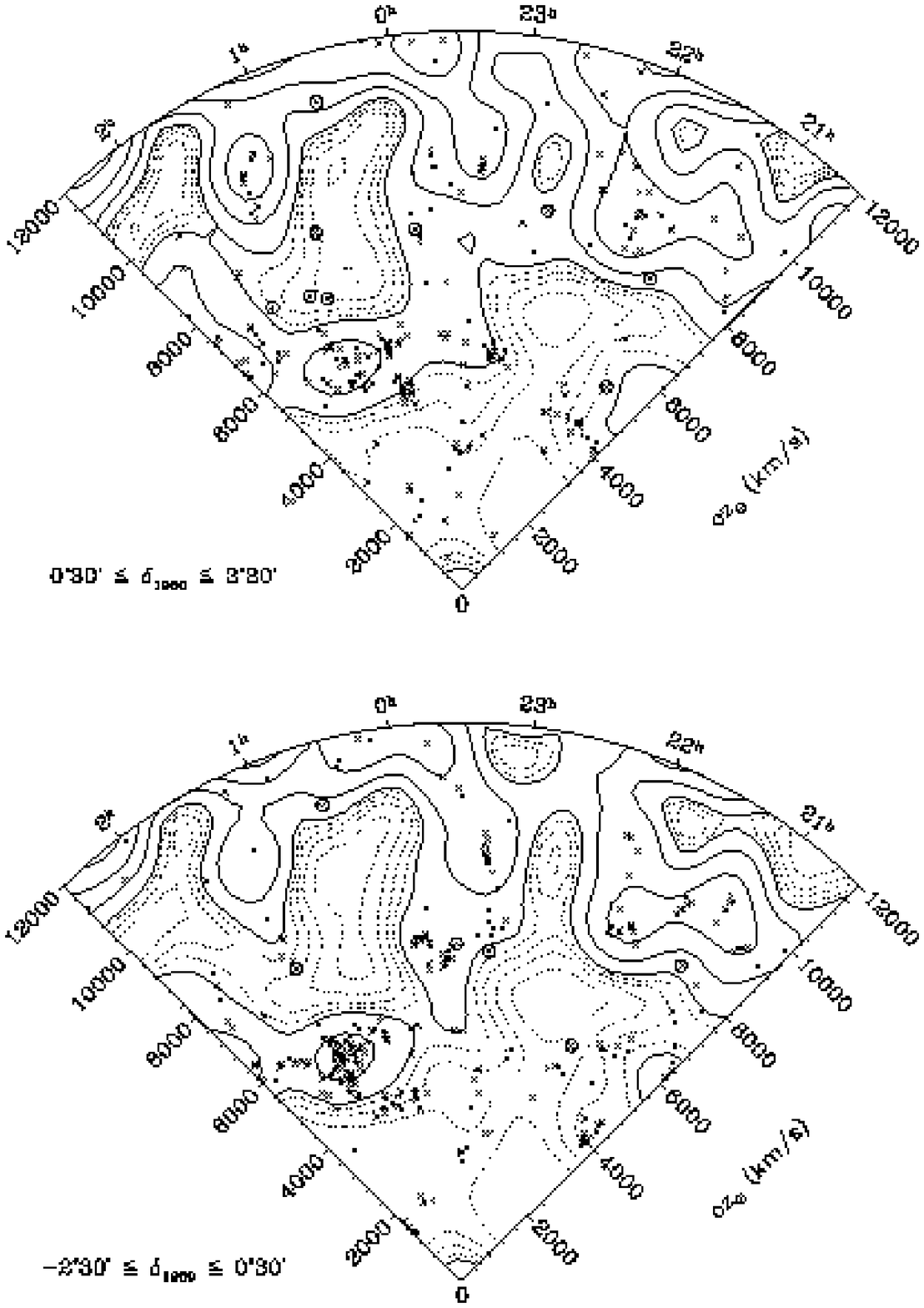}{7in}{0}{100}{100}{-220}{-280}
\end{figure}
\clearpage
\begin{figure}[bp]
\vskip -0.5in
\caption{\footnotesize At the top are the southern (left) and northern
(right) halves of 15R North (crosses) and the Century Survey
(triangular crosses) in the region of NV1.  At the bottom is 15R South
(crosses) in the region of SV1 and SV2.  Galaxies included in this
study are circled.  The dotted radial lines indicate R.A.~limits of
the surveys.  We overplot $5h^{-1}$ Mpc-smoothed number density
contours from CfA2 (cf.~Fig.~\ref{cfa2nfig}).  The uncircled void
galaxies at $\gtrsim16^{\rm h}$ in 15R North are from POSS plate 329
(cf.~\S\ref{sampdefsec}).
\label{c15rfig}}
\plotfiddle{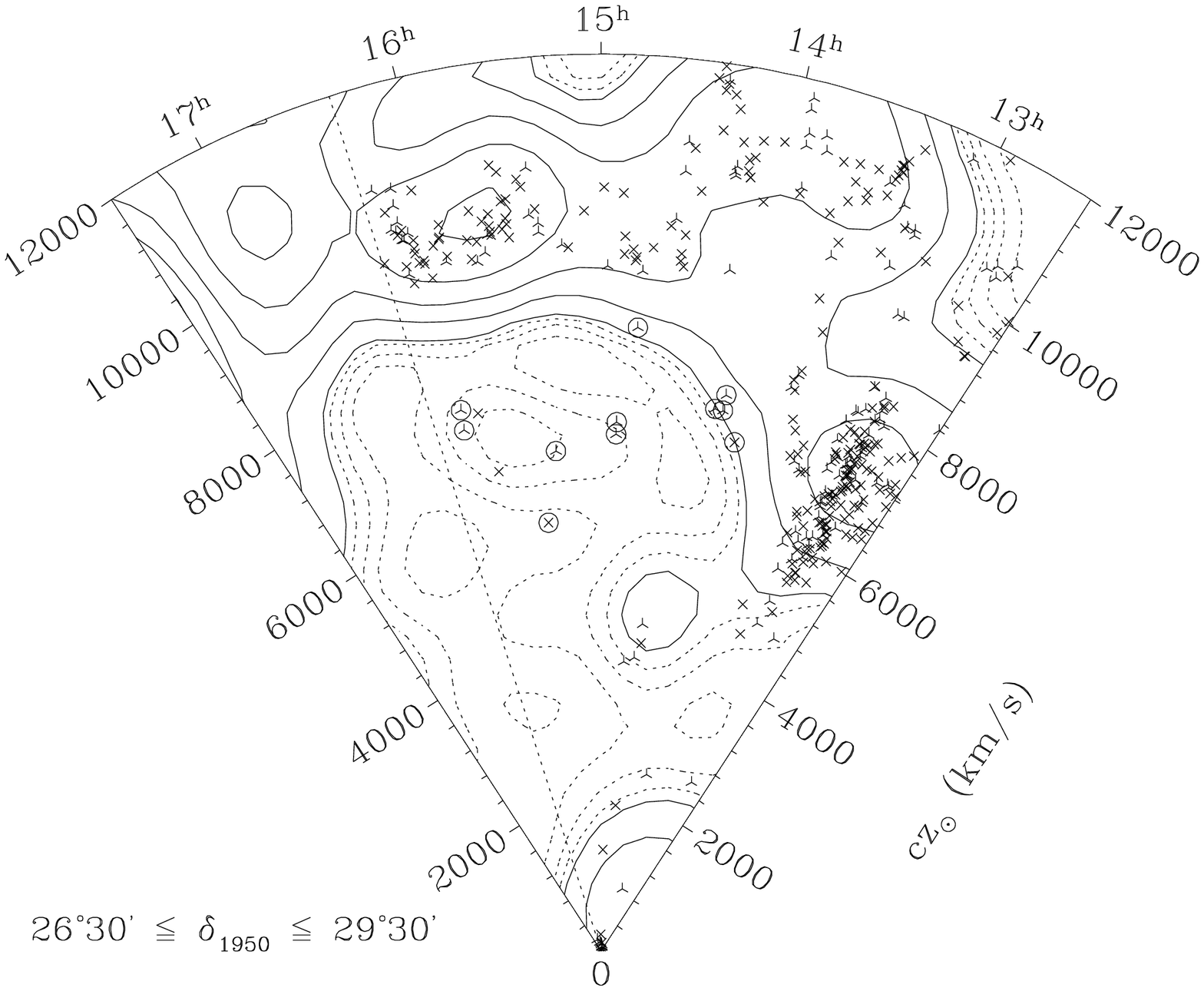}{4in}{0}{50}{50}{-290}{-50}
\plotfiddle{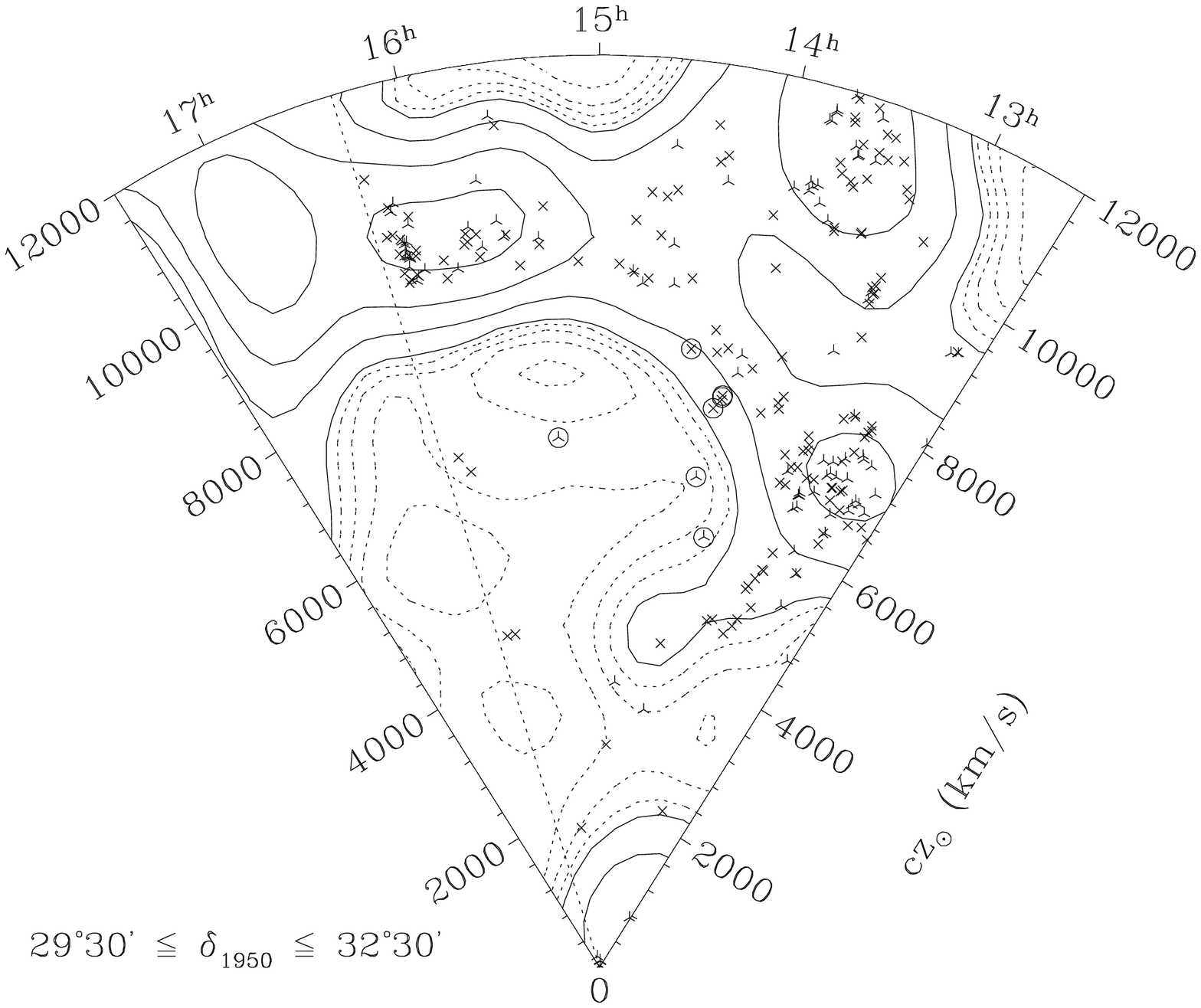}{4in}{0}{50}{50}{-10}{+262}
\plotfiddle{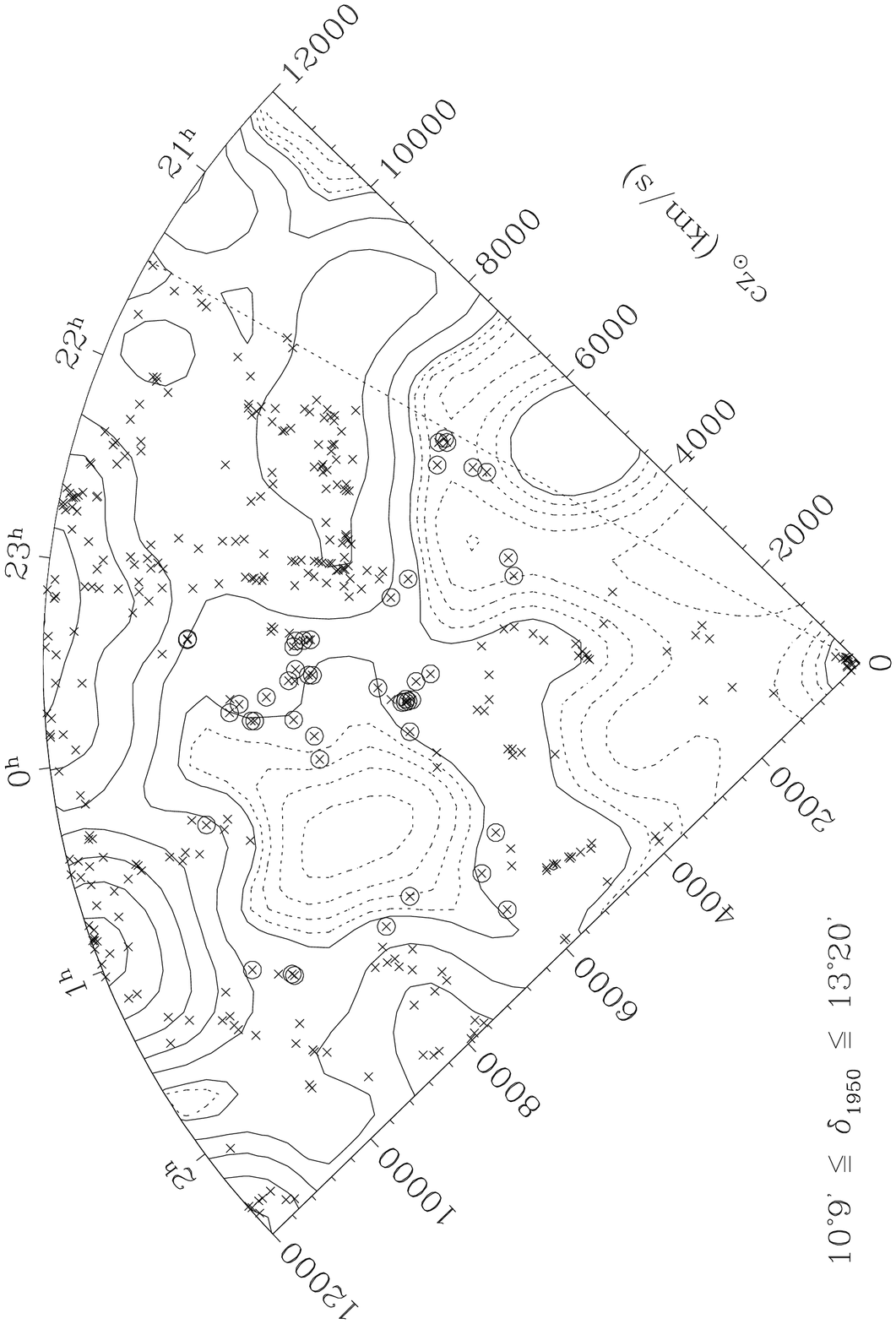}{5in}{270}{60}{60}{-250}{+750}
\end{figure}
\clearpage
\begin{figure}[bp]
\vskip -1.1in
\plotfiddle{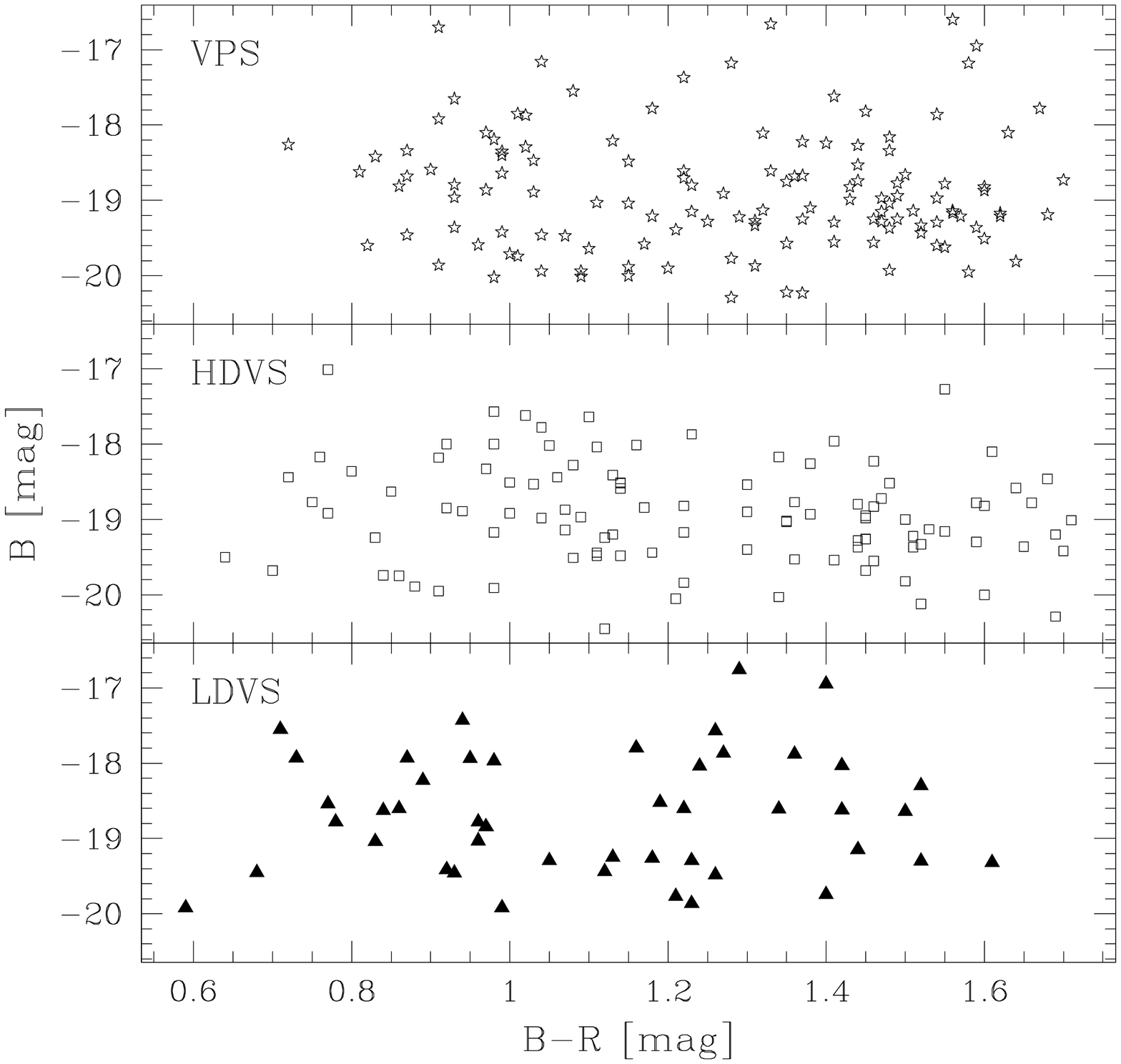}{4.5in}{0}{60}{60}{-197}{-72}
\plotfiddle{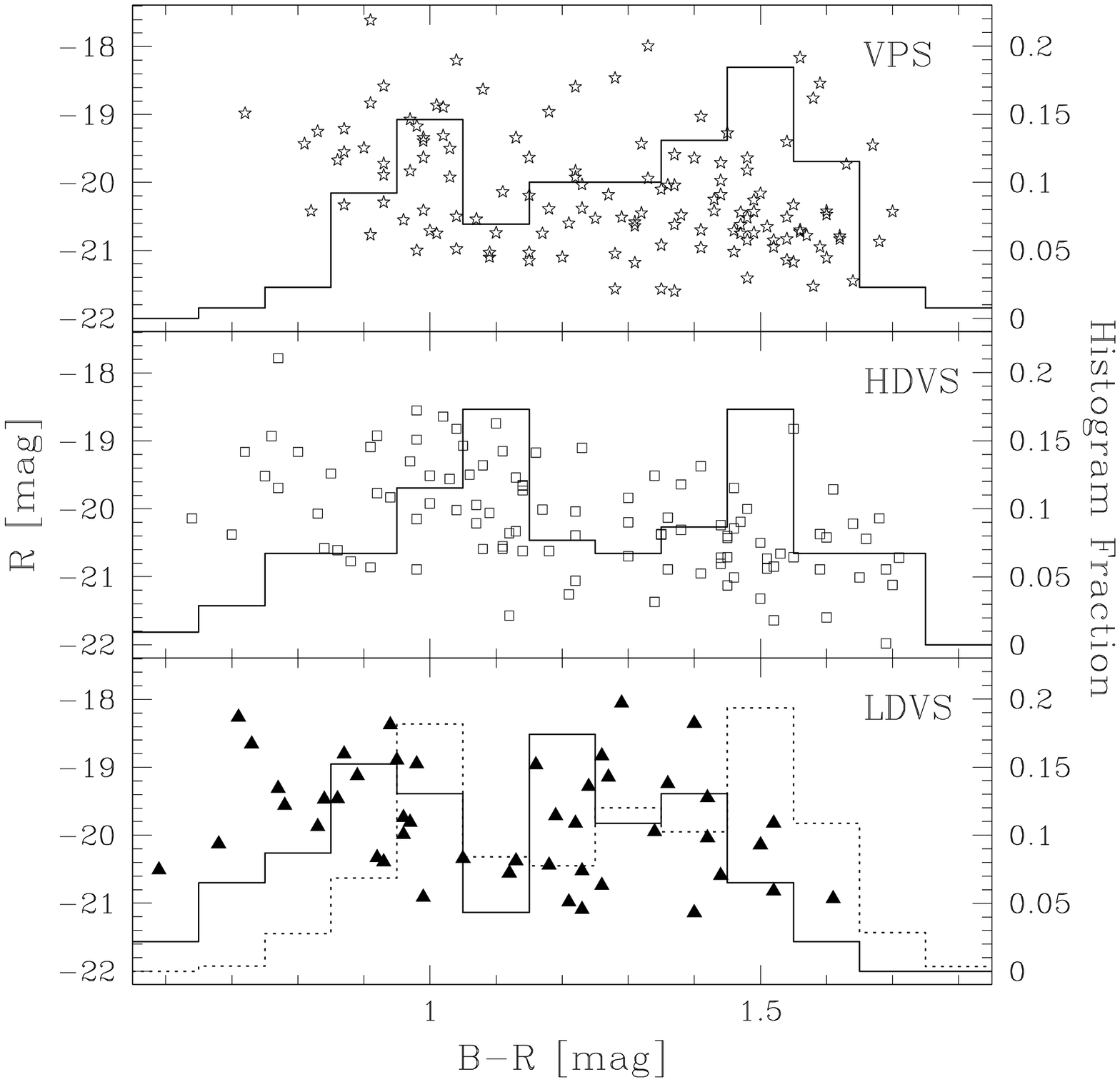}{4.5in}{0}{60}{60}{-185}{-72}
\vskip -0.2in
\caption{\footnotesize Color-magnitude diagrams in $B$ (top)
and $R$ (bottom) separated by global density environment: the VPS, at
$0.5\bar n < n < \bar n$ (open stars); the HDVS, at $0.5\bar n < n <
\bar n$ (open squares); and the LDVS, at $n< 0.5\bar n$ (filled triangles).  In the
$R$ plot, we also show the histogram of $\br$ (solid line) to
highlight the blue-ward color shift of the LDVS\@.  The dotted histogram
shows the color distribution obtained from sampling the VPS colors
with the LDVS $R$ magnitudes.
\label{tridencolfig}}
\end{figure}
\clearpage
\begin{figure}[bp]
\plotone{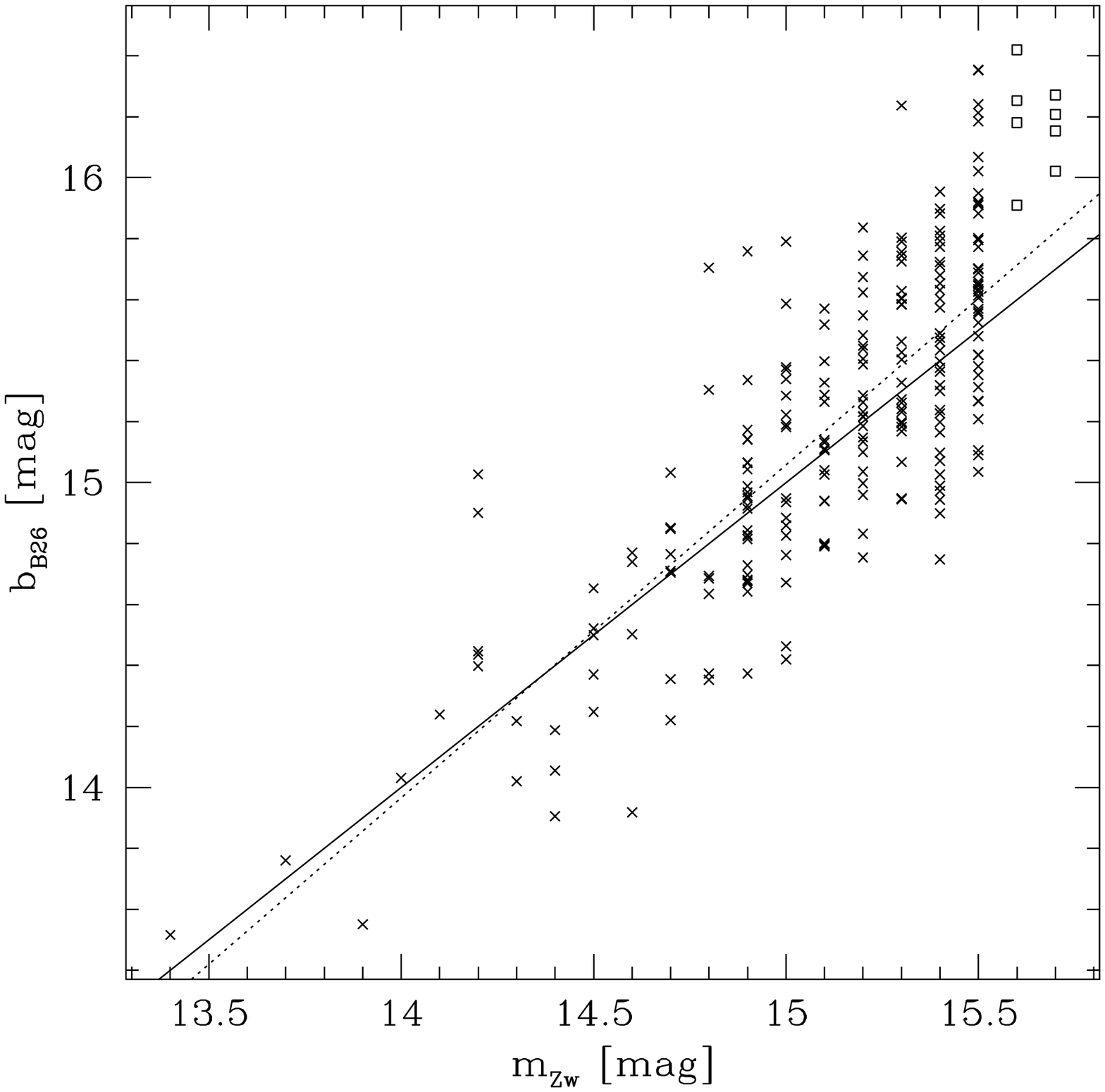}
\caption{\footnotesize Comparison between isophotal $b_{B26}$ and
CGCG magnitude $\mzw$ for 230 galaxies with $\mzw \leq 15.5$ (crosses)
and 8 galaxies with $\mzw = 15.6$--15.7 (open squares).  The solid
line shows $b_{B26}=\mzw$; the dotted line is the linear fit to the
$\mzw \leq 15.5$ galaxies.
\label{zwcalfig}}
\end{figure}
\clearpage
\begin{figure}[bp]
{
\plotone{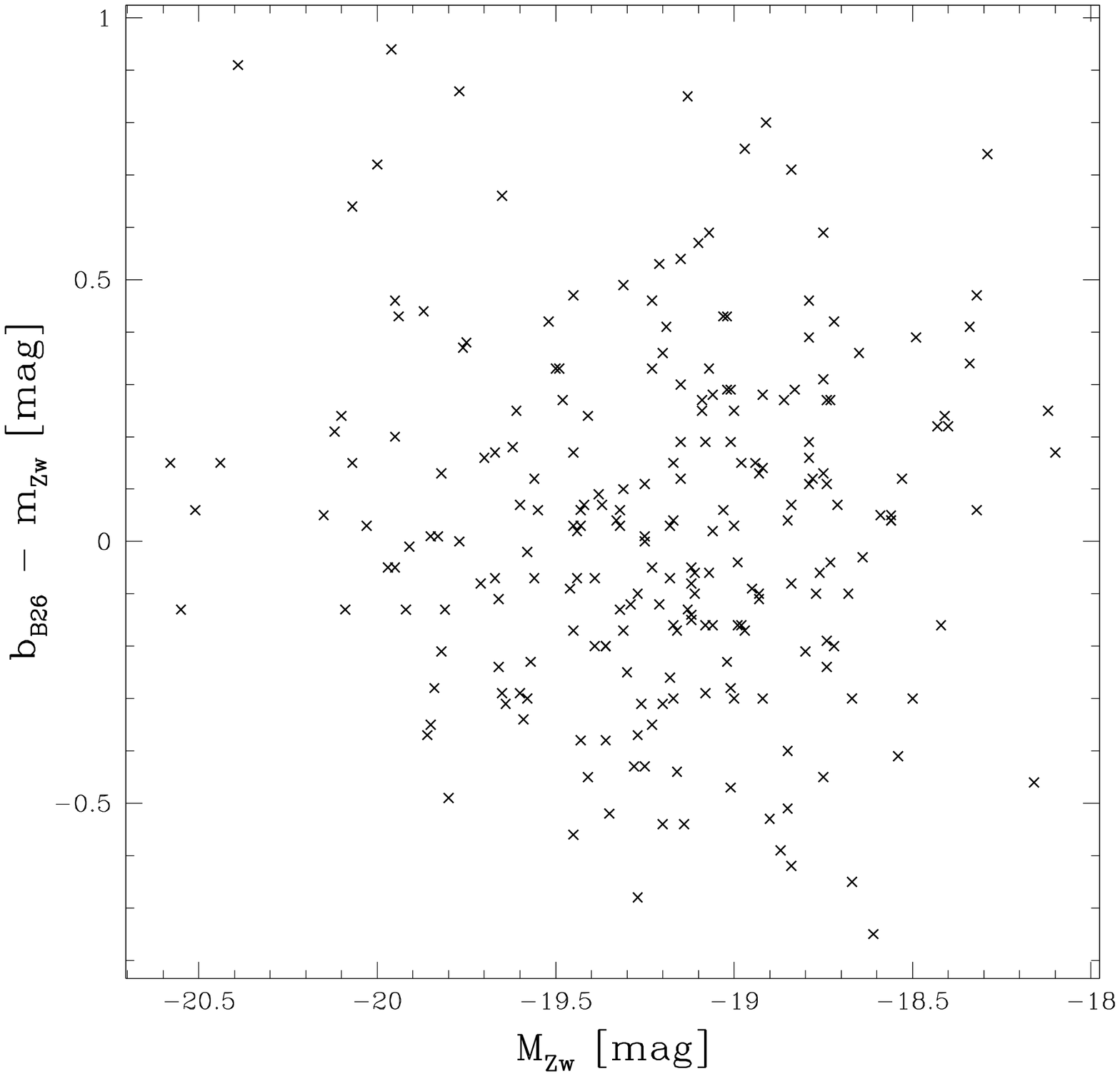}
\caption[zwerramag.ps]{\footnotesize Zwicky magnitude error as a function of
the Zwicky absolute magnitude $M_{\rm Zw}$ for the 230 galaxies in
this study with $\mzw \leq 15.5$.
\label{zwerramagfig}}
}
\end{figure}
\clearpage
\begin{figure}[bp]
\plotfiddle{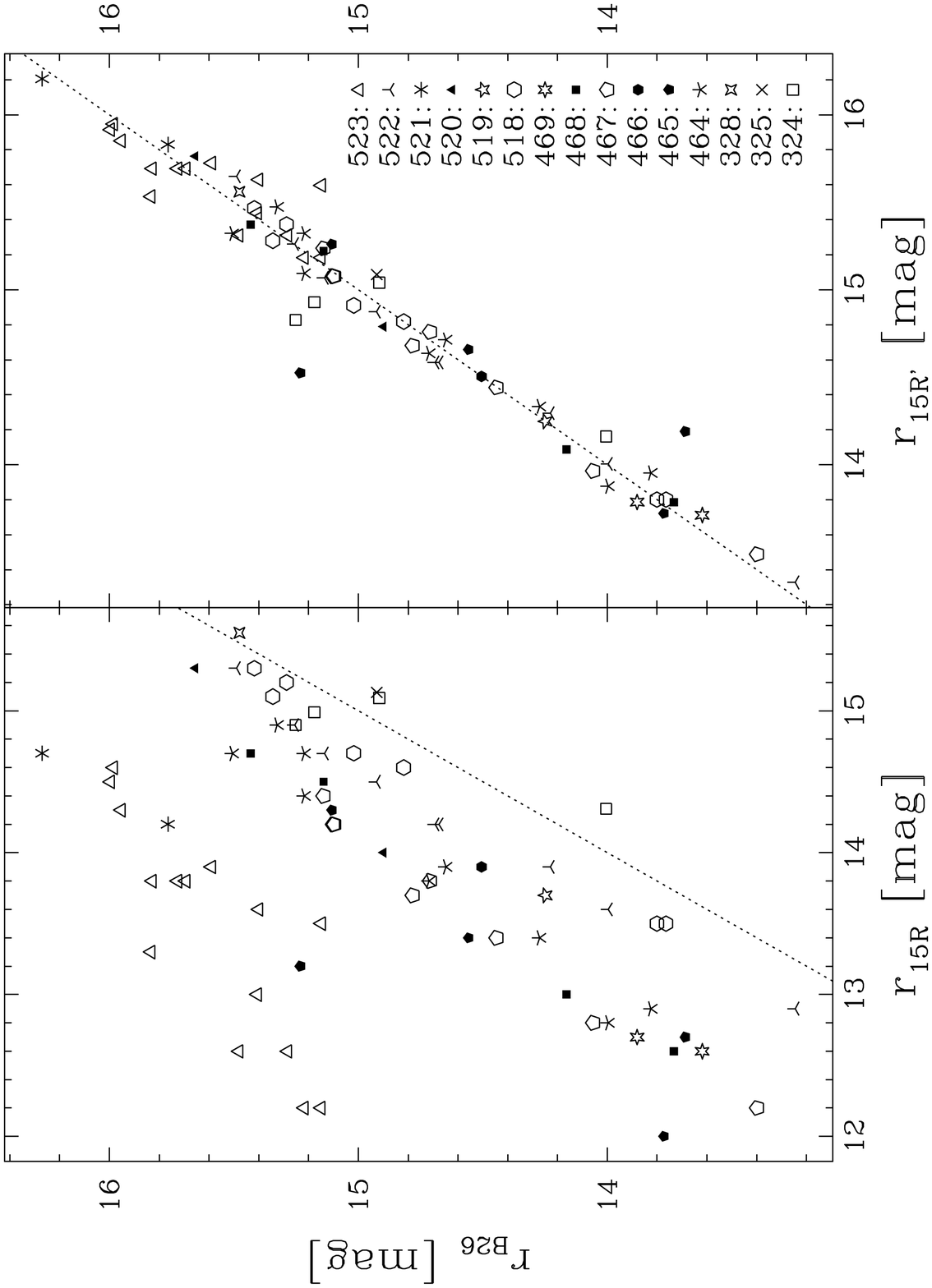}{5in}{270}{65}{65}{-250}{+400}
\caption{\footnotesize Calibration of the 15R catalog
(plate-scanned) magnitudes with our isophotal photometry.  The left
panel plots the $r_{B26}$ magnitudes of the 15R galaxies in our sample
against their catalog magnitudes $r_{\rm 15R}$.  Objects are assigned
a plate-specific symbol, given by the legend to the lower right.  In
the right panel, we plot the $r_{B26}$ magnitudes against
``corrected'' 15R magnitudes $r_{\rm 15R'}$ where we have subtracted
out a plate-specific linear fit (cf.~Tab.~\ref{cs15rcaltab}).  For
the 15R North magnitudes (plates 324, 325, and 328) we calibrate with the
linear fit to the Century Survey.
\label{sidebysidefig}}
\end{figure}
\clearpage
\begin{figure}[bp]
{
\vskip -6in
\mbox{a) \plotfiddle{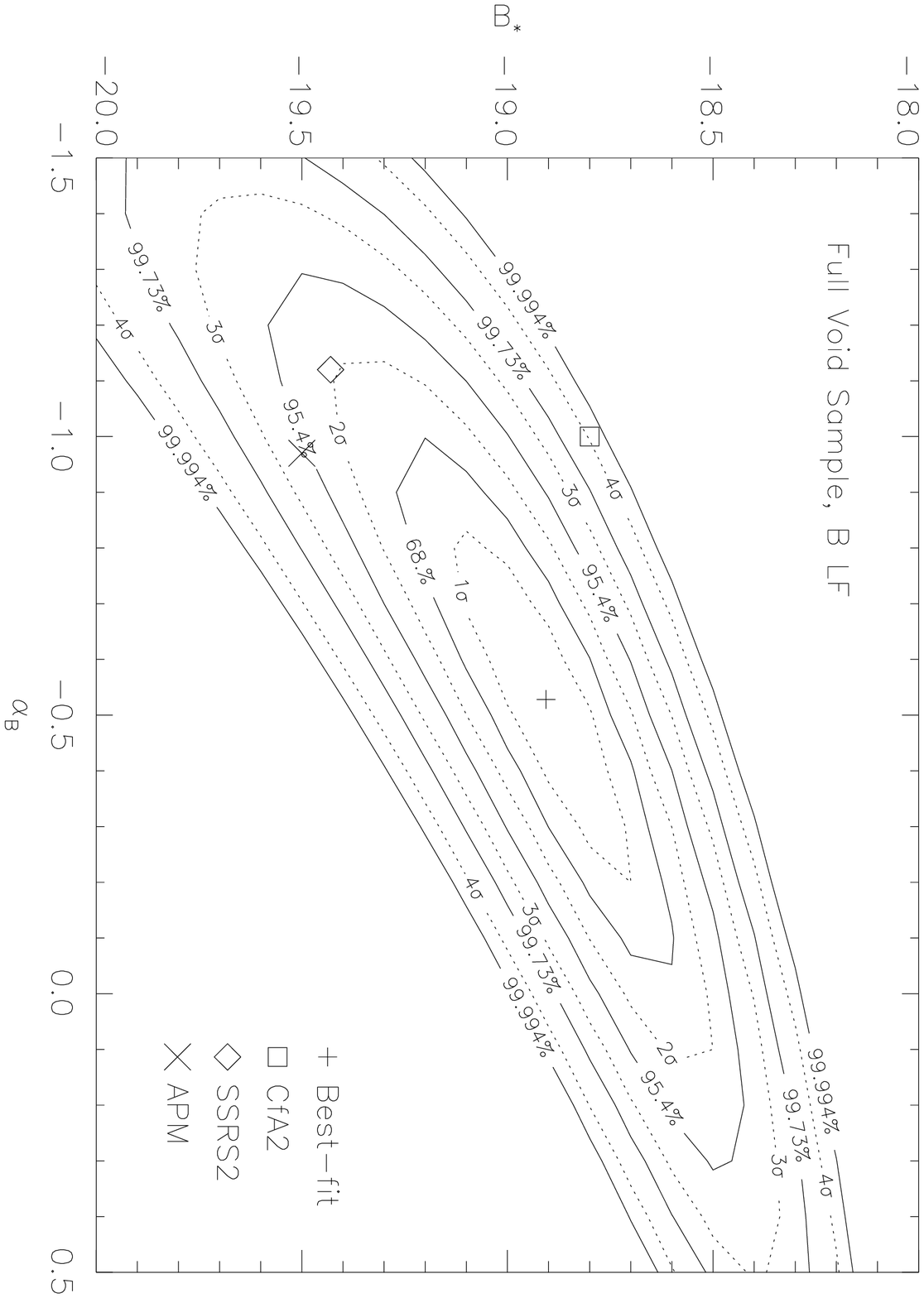}{5in}{90}{60}{60}{0}{-310}}
\vskip -0.9in
\mbox{b) \plotfiddle{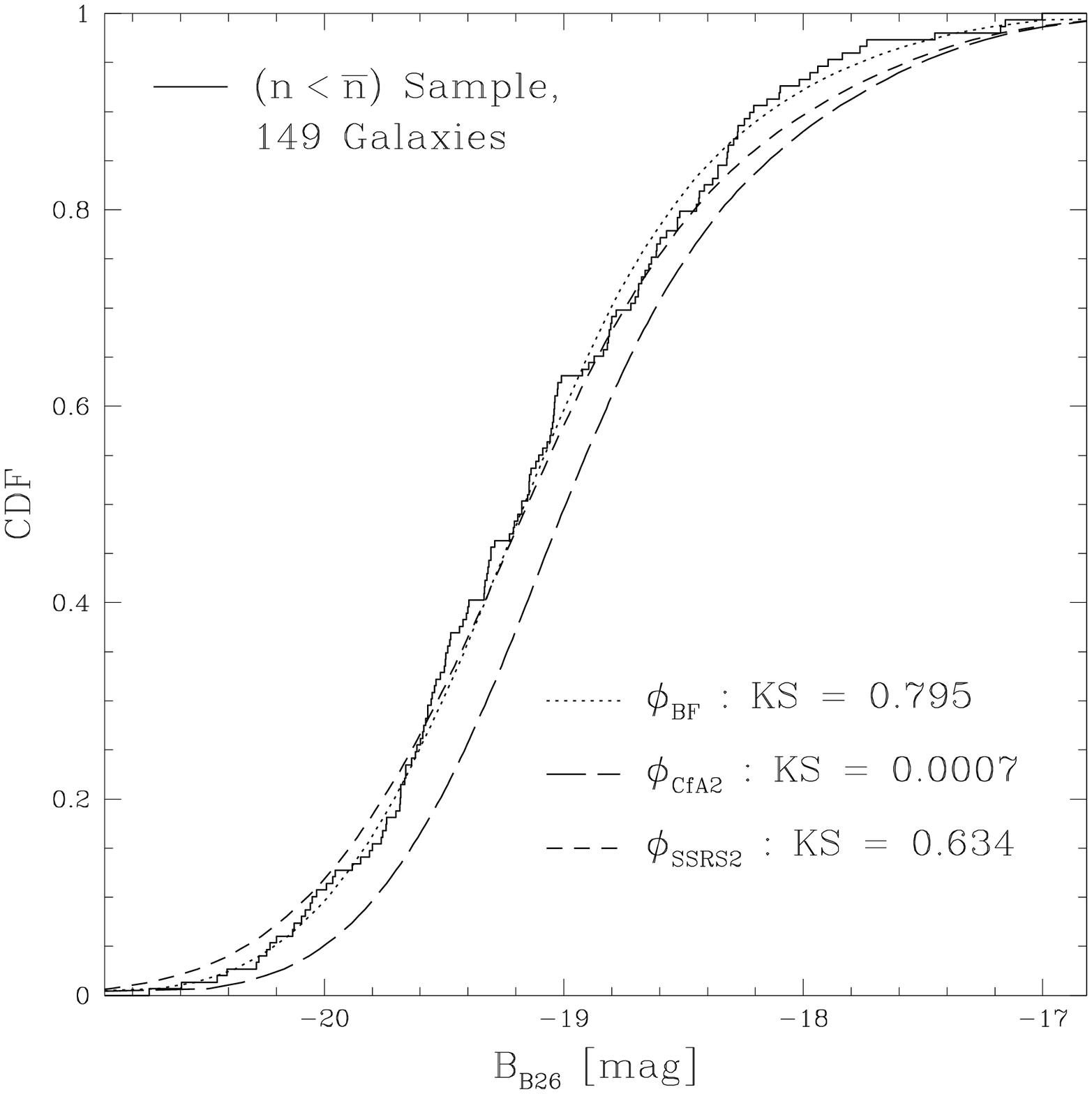}{5in}{0}{60}{60}{-430}{-405}}
\vskip 4.3in
\caption{\footnotesize a) Likelihood contours from our STY
analysis of the full void sample LF in $B$.  Solid contours denote the
joint probability distribution of Schechter function parameters
$\alpha_B$ and $B_*$.  Dotted contours project onto the $n$-$\sigma$
confidence intervals for each parameter individually.  We indicate the
location of the best-fit LF (plus symbol), the CfA2 LF (square
symbol), the SSRS2 LF (diamond symbol), and the Stromlo-APM LF (cross
symbol). b) The cumulative distribution function of $B$ absolute
magnitudes observed in the full void sample (jagged solid line).  We
also show the predictions for the Schechter functions corresponding to
CfA2 (long-dashed curve), SSRS2 (short-dashed curve), and the maximum
likelihood (dotted curve).  We note the KS probabilities that the
observed absolute magnitudes were drawn from the respective Schechter
functions.
\label{fullblffig}}
}
\end{figure}
\clearpage
\begin{figure}[bp]
{
\vskip -6in
\mbox{a) \plotfiddle{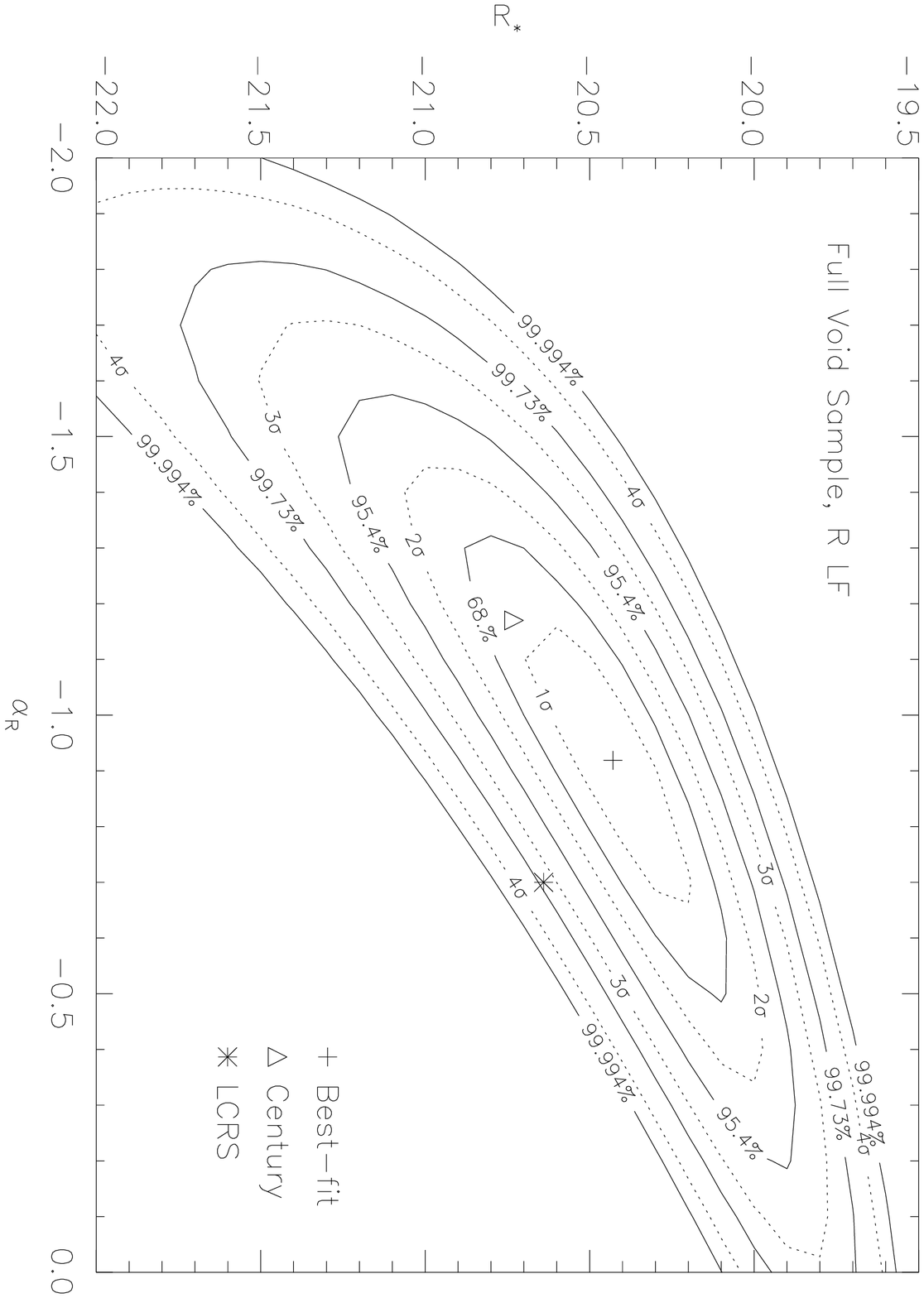}{5in}{90}{60}{60}{0}{-310}}
\vskip -0.9in
\mbox{b) \plotfiddle{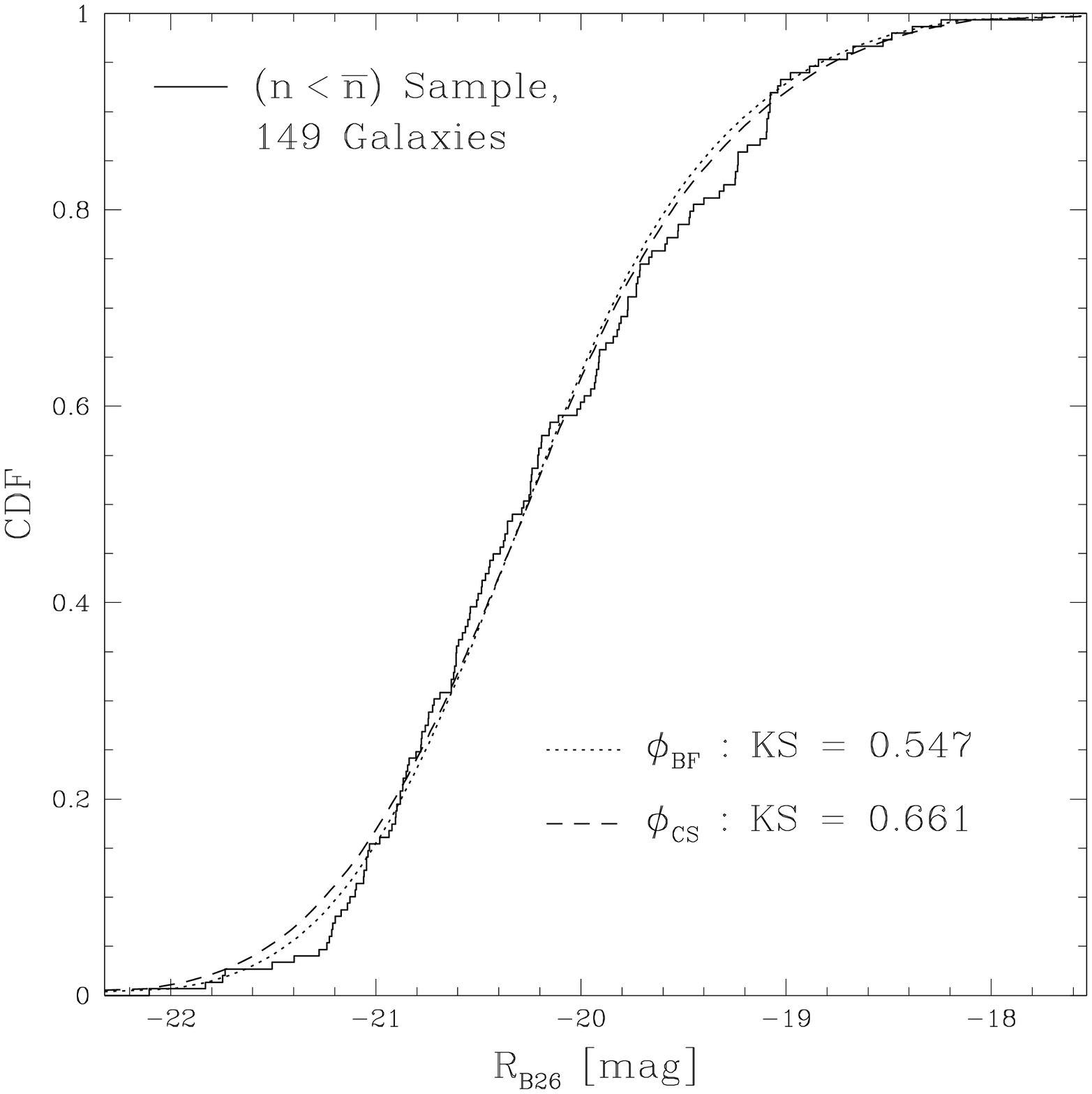}{5in}{0}{60}{60}{-430}{-405}}
\vskip 4.3in
\caption{\footnotesize a) Likelihood contours from our STY
analysis of the full void sample LF in $R$.  Solid contours denote the
joint probability distribution of Schechter function parameters
$\alpha_R$ and $R_*$.  Dotted contours project onto the $n$-$\sigma$
confidence intervals for each parameter individually.  We indicate the
location of the best-fit LF (plus symbol), the Century Survey LF
(triangle symbol), and the LCRS LF (asterisk symbol).  b) The
cumulative distribution function of $R$ absolute magnitudes observed
in the full void sample (jagged solid line).  We also show the
predictions for the Century Survey (dashed curve) and the maximum
likelihood (dotted curve).  We note the K-S probabilities that the
observed absolute magnitudes were drawn from the respective Schechter
functions.
\label{fullrlffig}}
}
\end{figure}
\clearpage
\begin{figure}[bp]
{
\vskip -6in
\mbox{a) \plotfiddle{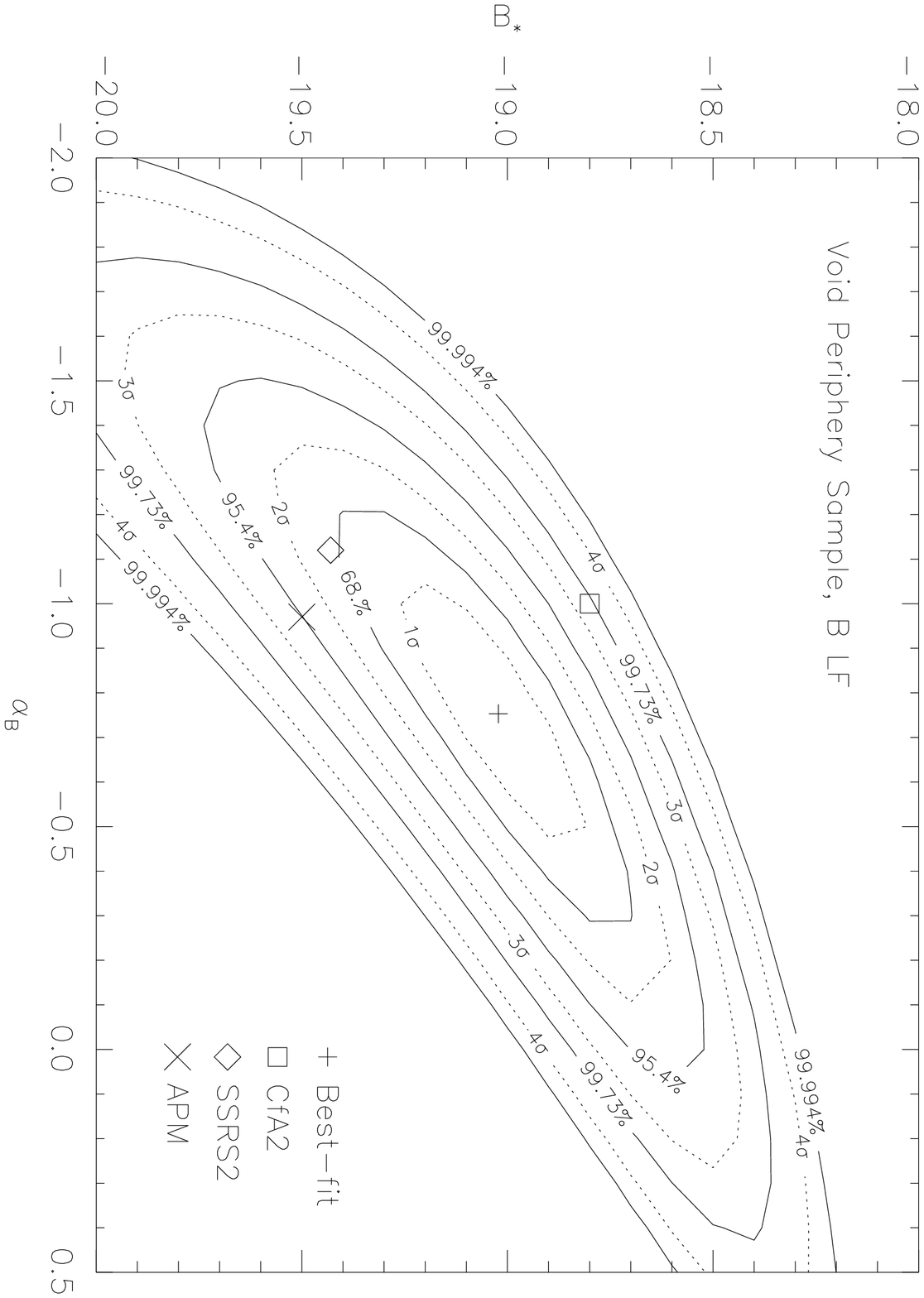}{5in}{90}{60}{60}{0}{-310}}
\vskip -0.9in
\mbox{b) \plotfiddle{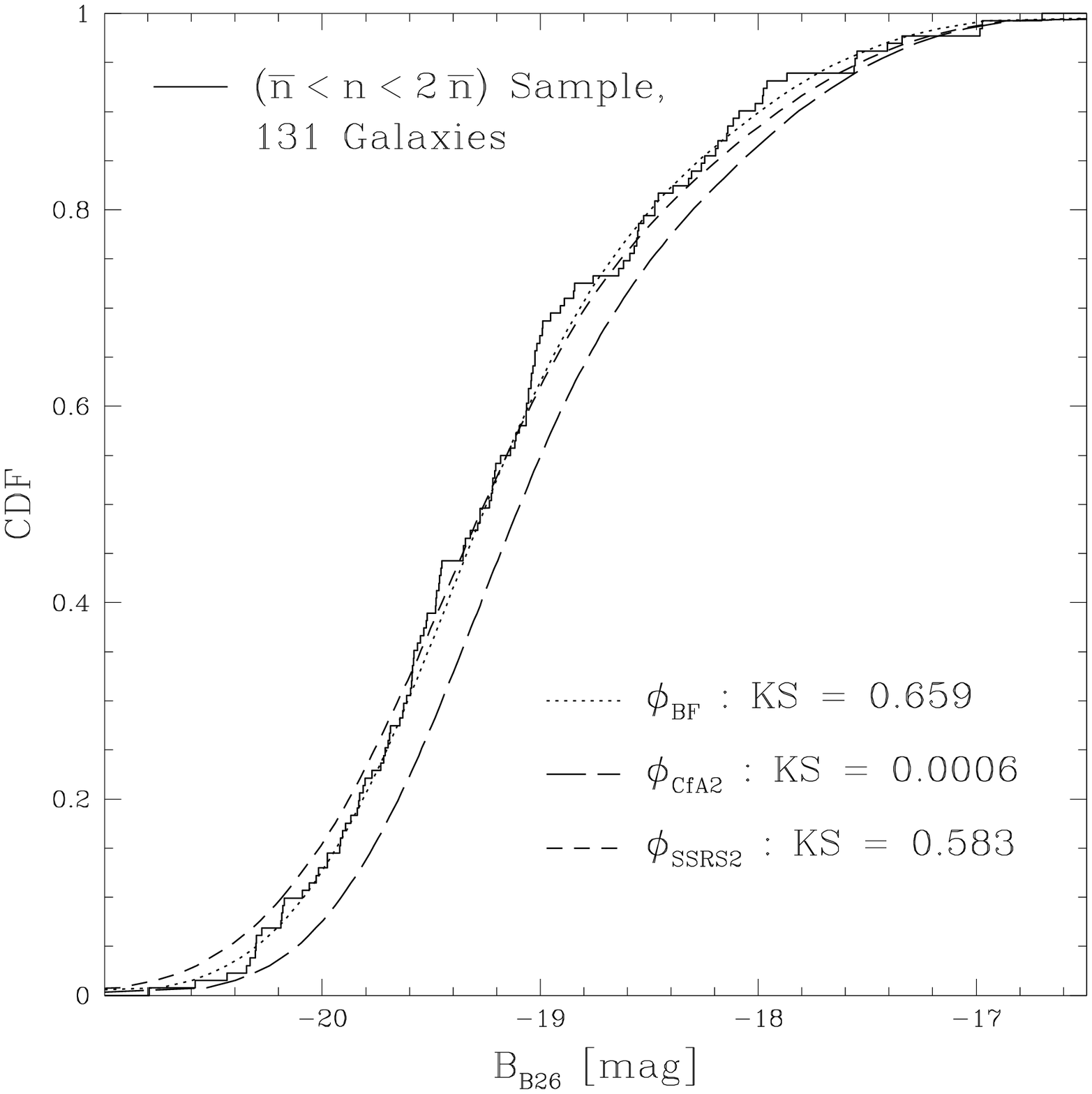}{5in}{0}{60}{60}{-430}{-405}}
\vskip 4.3in
\caption{\footnotesize  As Figure \ref{fullblffig}, but with the
$B$ magnitudes in the void periphery sample.
\label{hiblffig}}
}
\end{figure}
\clearpage
\begin{figure}[bp]
{
\vskip -6in
\mbox{a) \plotfiddle{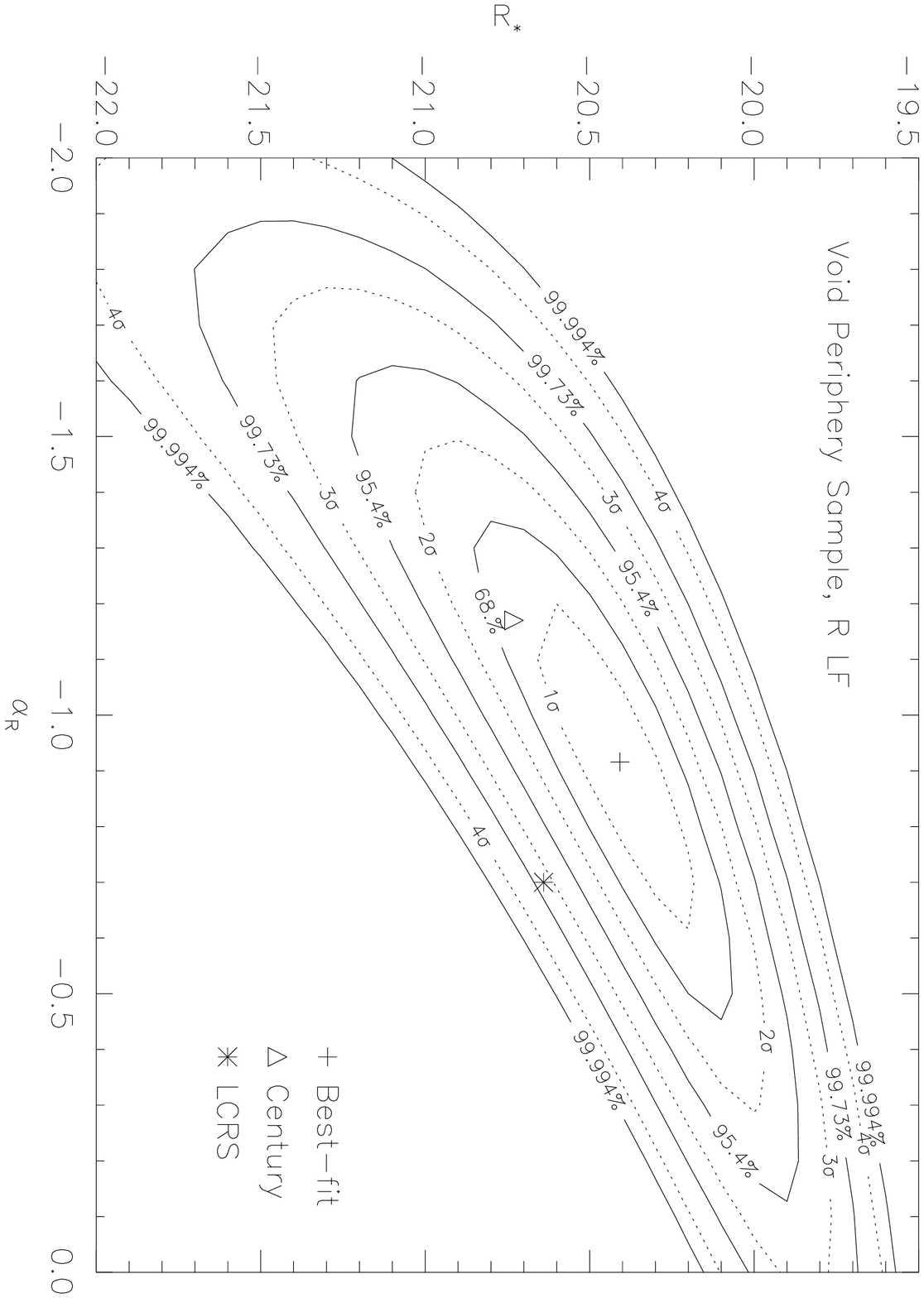}{5in}{90}{60}{60}{0}{-310}}
\vskip -0.9in
\mbox{b) \plotfiddle{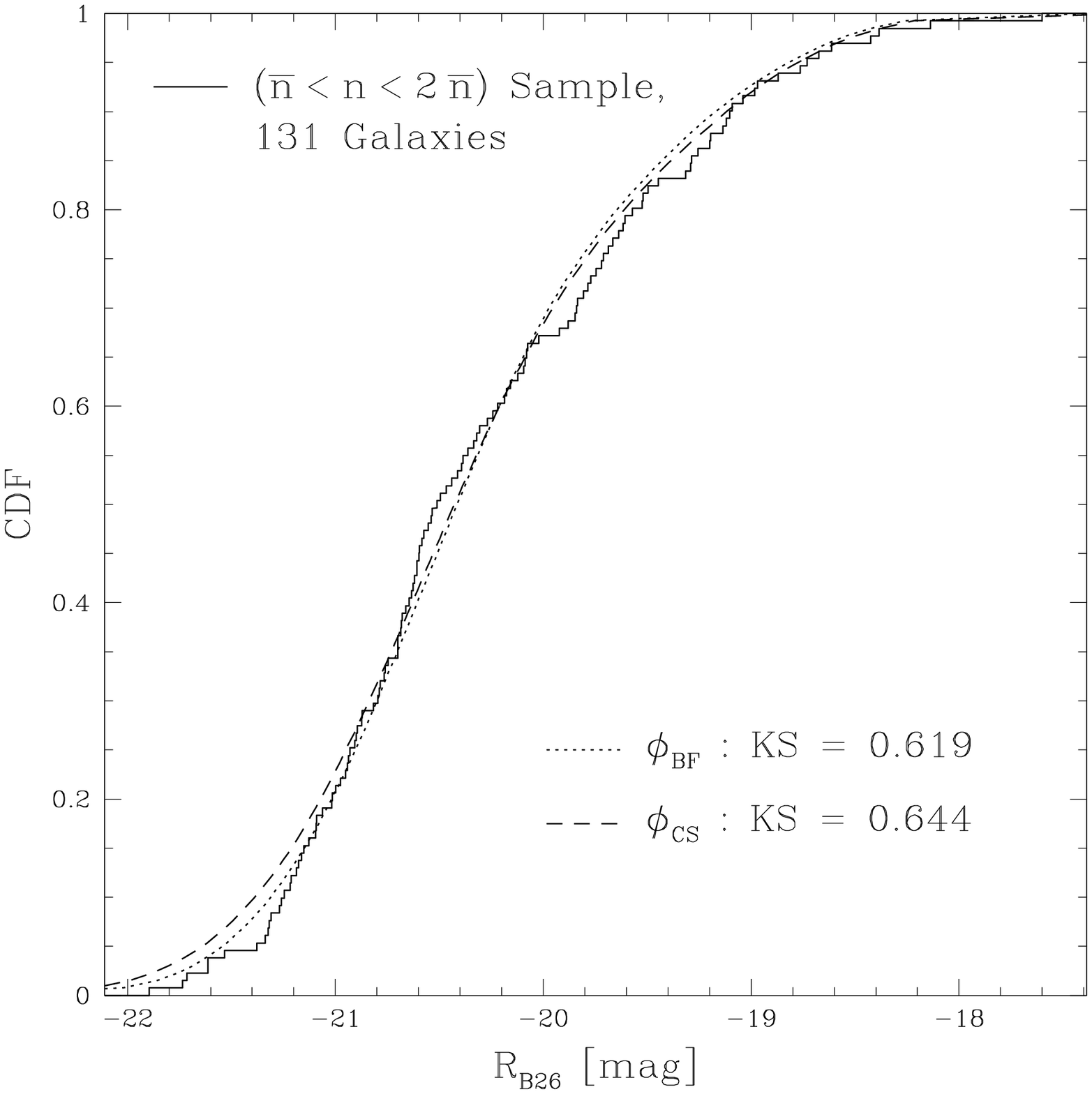}{5in}{0}{60}{60}{-430}{-405}}
\vskip 4.3in
\caption{\footnotesize  As Figure \ref{fullrlffig}, but with the
$R$ magnitudes in the void periphery sample.
\label{hirlffig}}
}
\end{figure}
\clearpage
\begin{figure}[bp]
{
\vskip -6in
\mbox{a) \plotfiddle{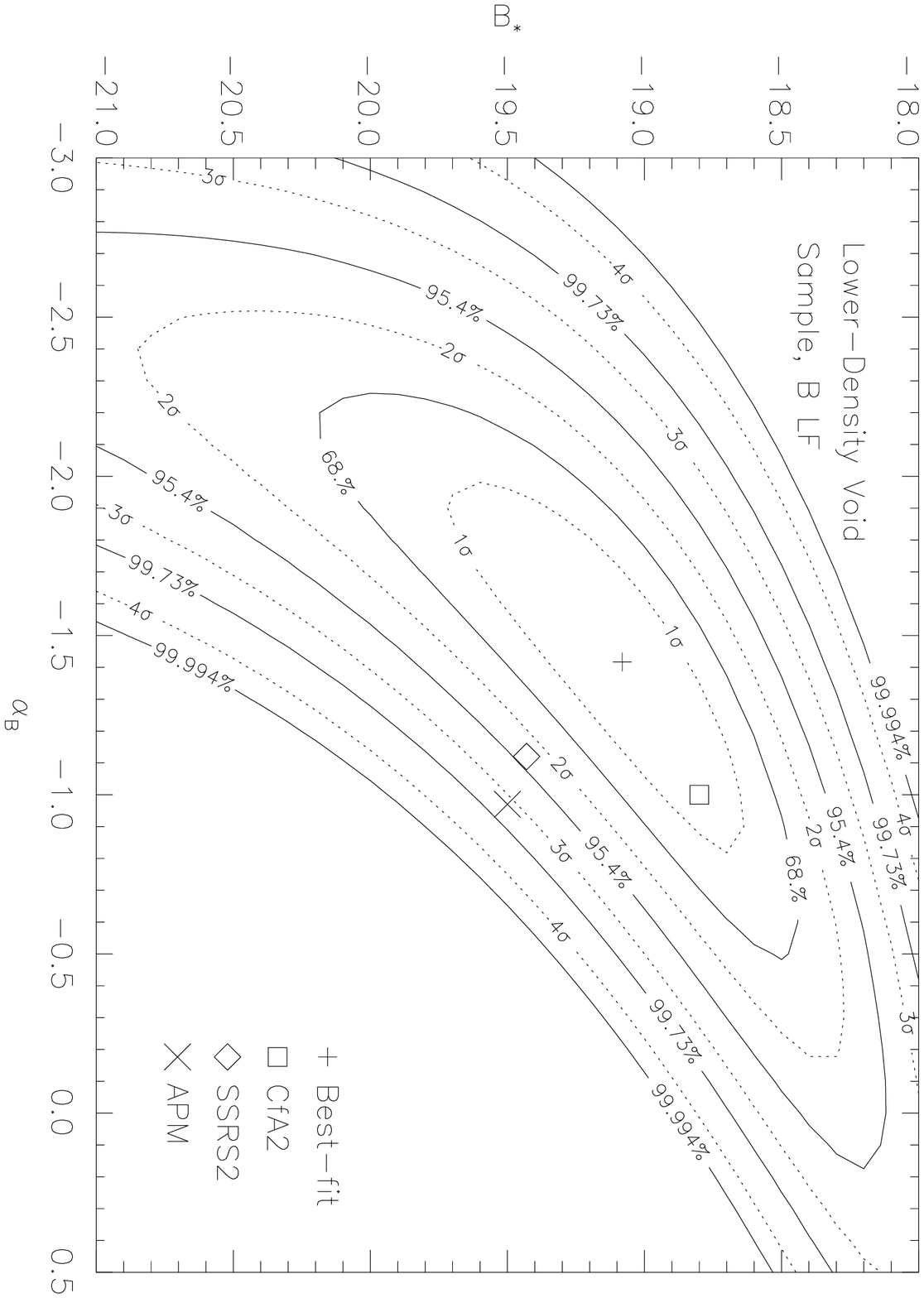}{5in}{90}{60}{60}{0}{-310}}
\vskip -0.9in
\mbox{b) \plotfiddle{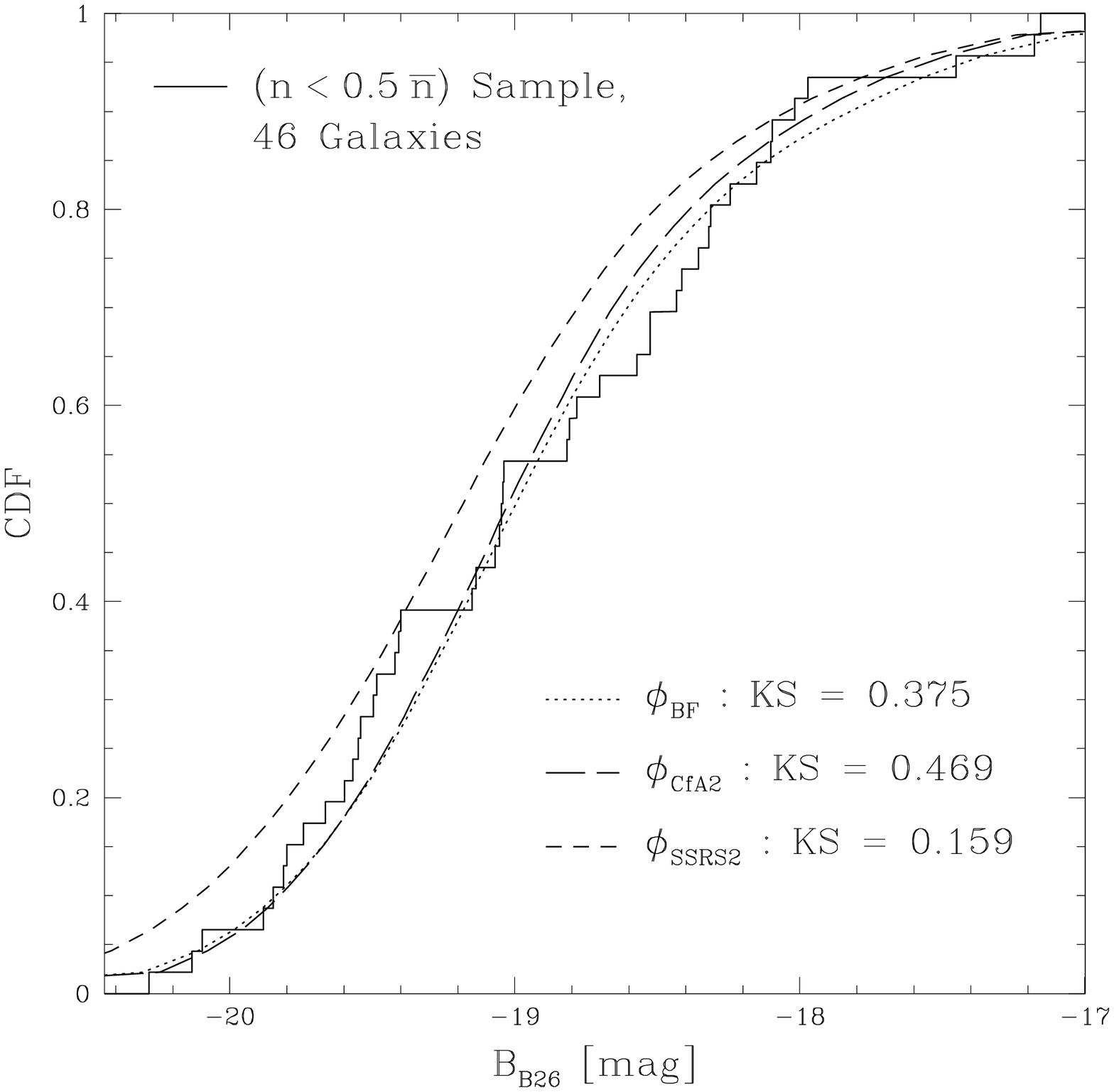}{5in}{0}{60}{60}{-430}{-405}}
\vskip 4.3in
\caption{\footnotesize  As Figure \ref{fullblffig}, but with the
$B$ magnitudes in the lower-density void sample.
\label{loloblffig}}
}
\end{figure}
\clearpage
\begin{figure}[bp]
{
\vskip -6in
\mbox{a) \plotfiddle{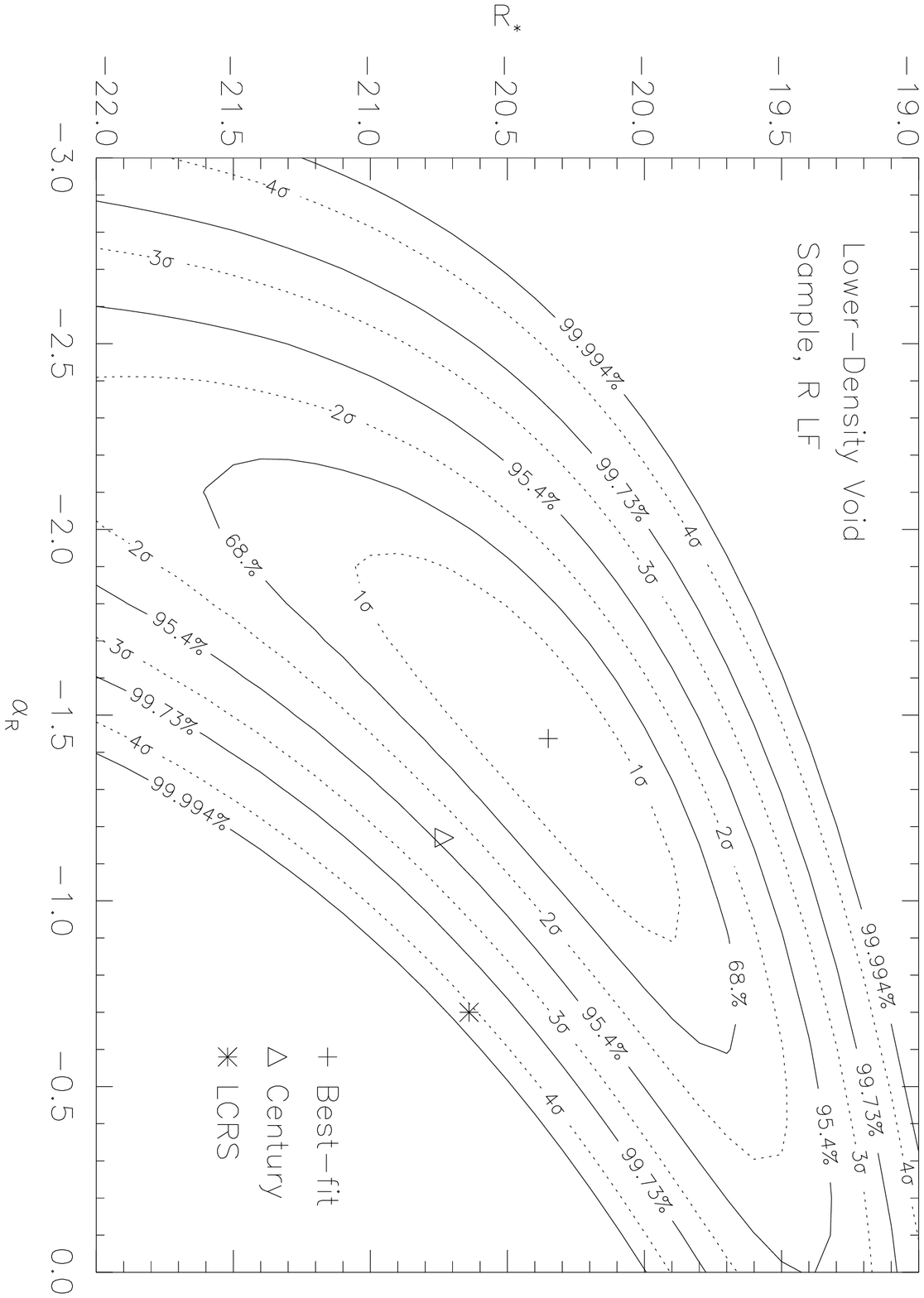}{5in}{90}{60}{60}{0}{-310}}
\vskip -0.9in
\mbox{b) \plotfiddle{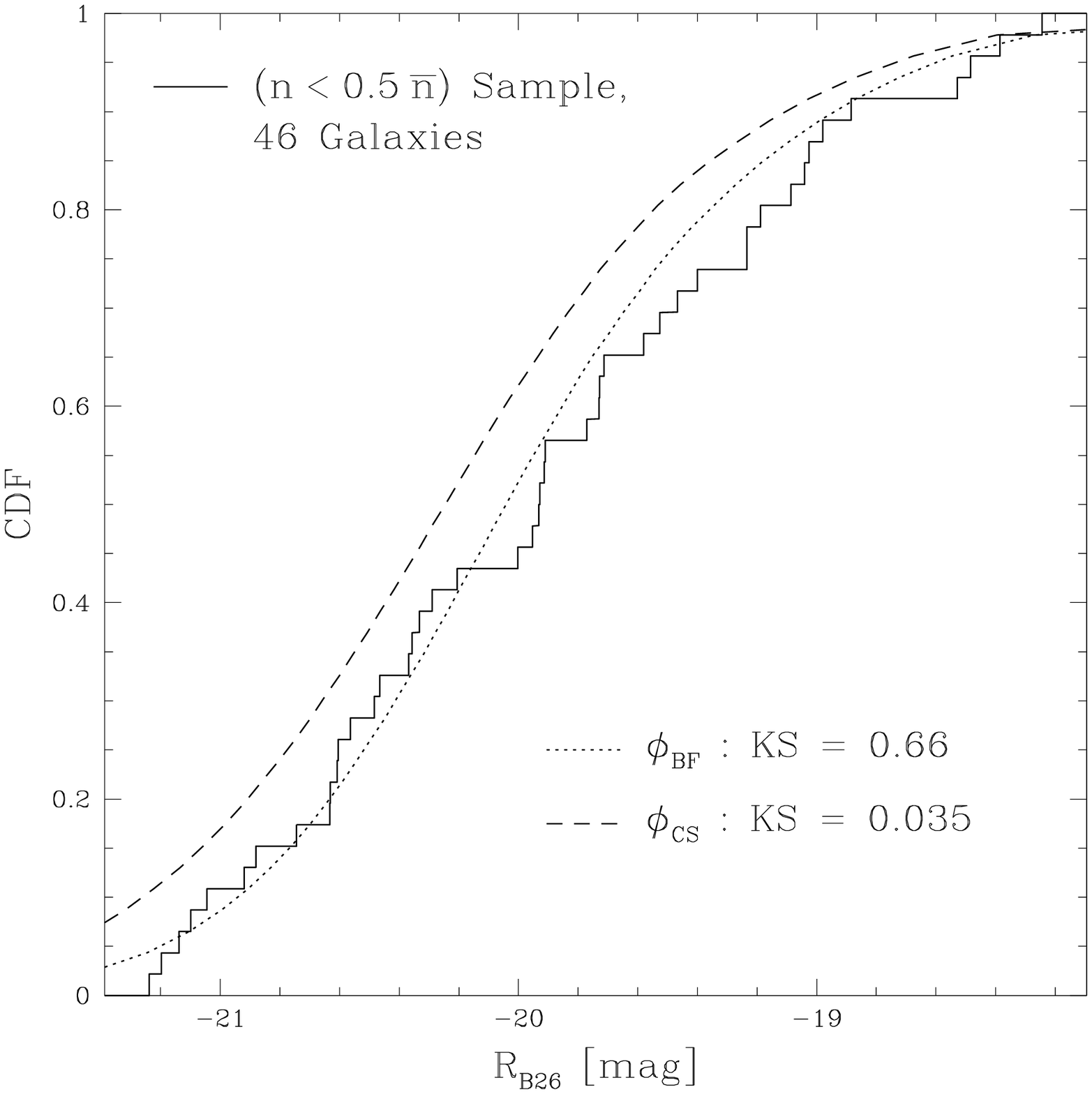}{5in}{0}{60}{60}{-430}{-405}}
\vskip 4.3in
\caption{\footnotesize  As Figure \ref{fullrlffig}, but with the
$R$ magnitudes in the lower-density void sample.
\label{lolorlffig}}
}
\end{figure}
\clearpage
\begin{figure}[bp]
{
\vskip -6in
\mbox{a) \plotfiddle{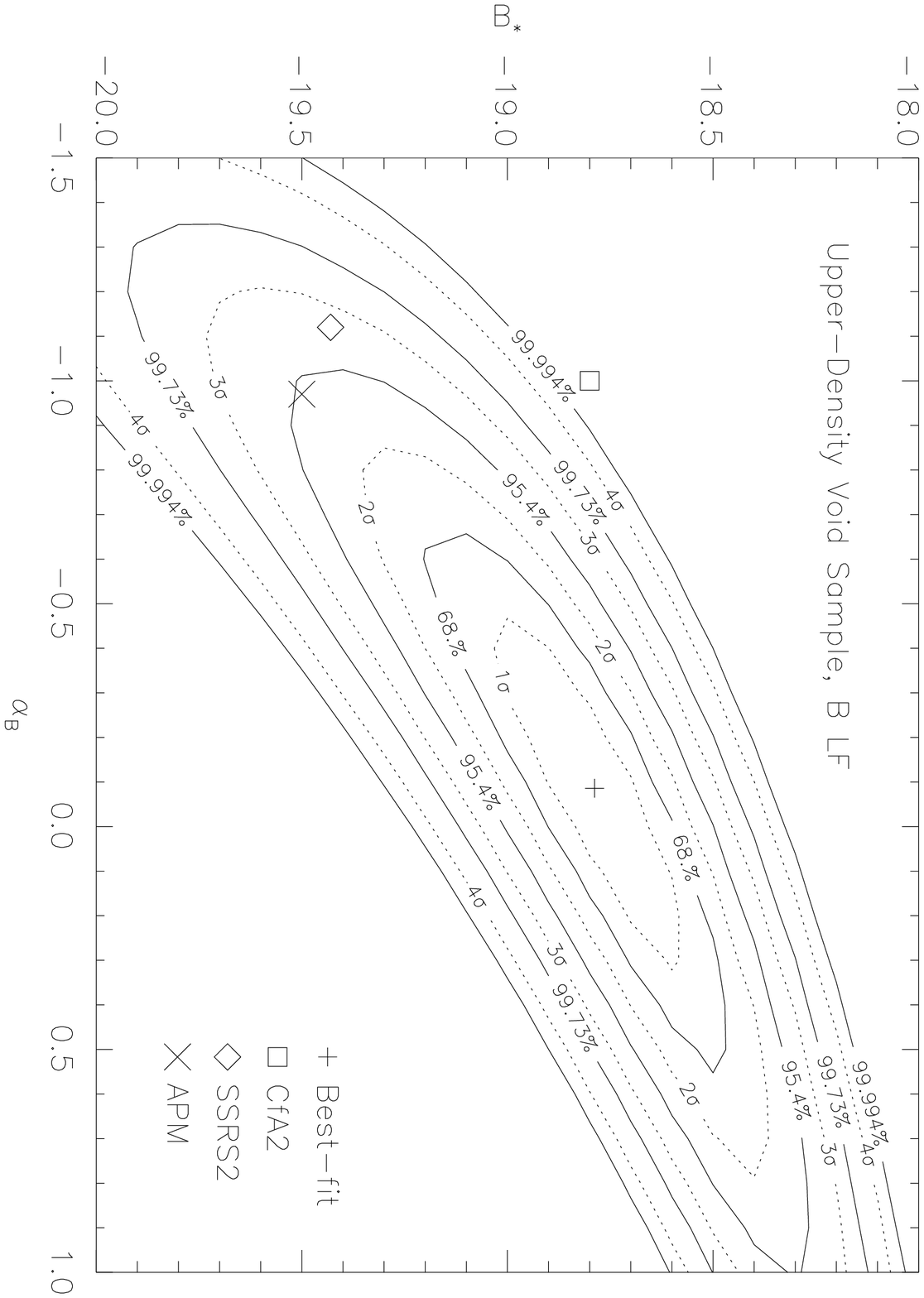}{5in}{90}{60}{60}{0}{-310}}
\vskip -0.9in
\mbox{b) \plotfiddle{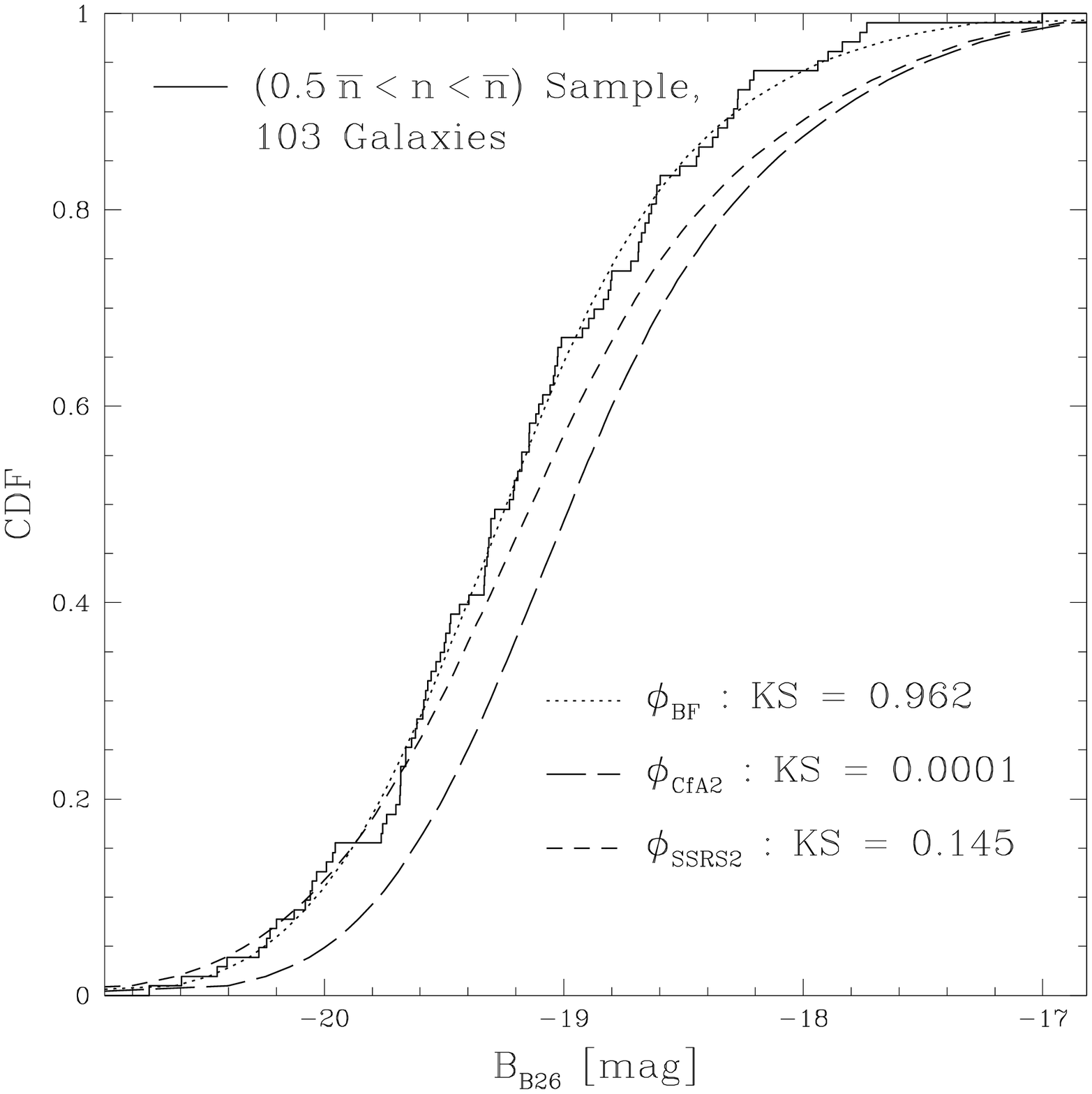}{5in}{0}{60}{60}{-430}{-405}}
\vskip 4.3in
\caption{\footnotesize  As Figure \ref{fullblffig}, but with the
$B$ magnitudes in the higher-density void sample.
\label{hiloblffig}}
}
\end{figure}
\clearpage
\begin{figure}[bp]
{
\vskip -6in
\mbox{a) \plotfiddle{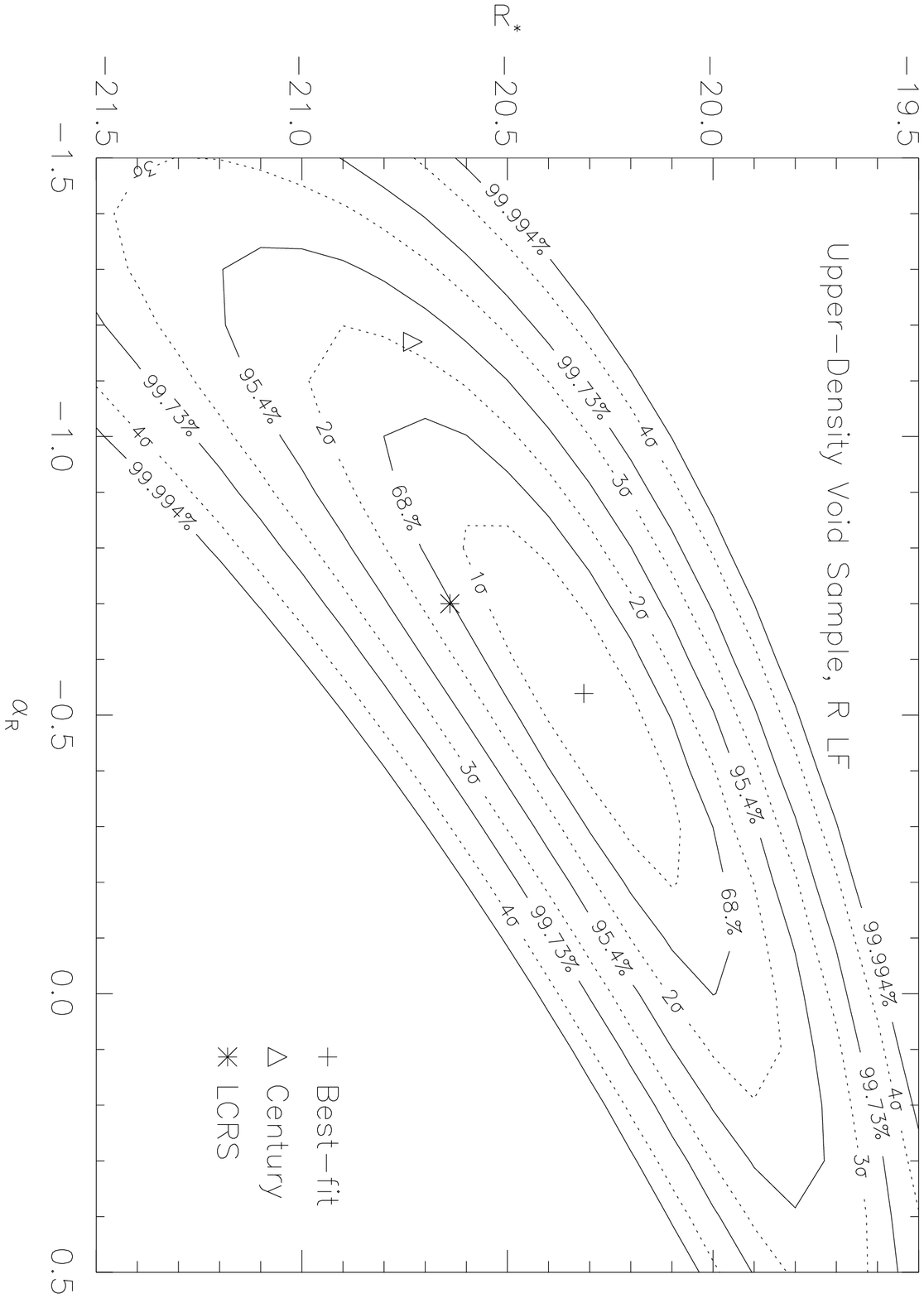}{5in}{90}{60}{60}{0}{-310}}
\vskip -0.9in
\mbox{b) \plotfiddle{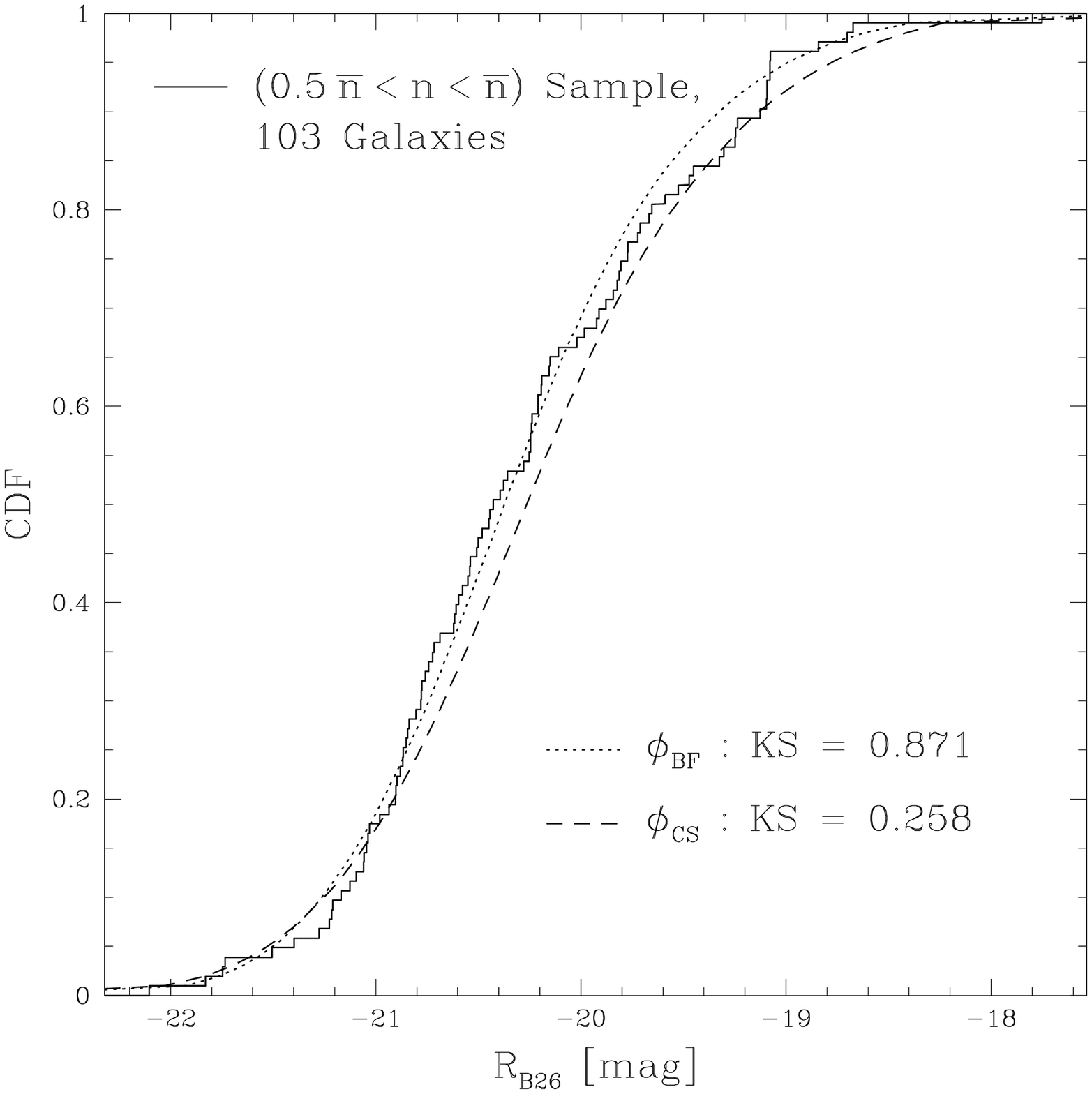}{5in}{0}{60}{60}{-430}{-405}}
\vskip 4.3in
\caption{\footnotesize  As Figure \ref{fullrlffig}, but with the
$R$ magnitudes in the higher-density void sample.
\label{hilorlffig}}
}
\end{figure}
\clearpage
\begin{figure}[bp]
{
\vskip -1in
\plotone{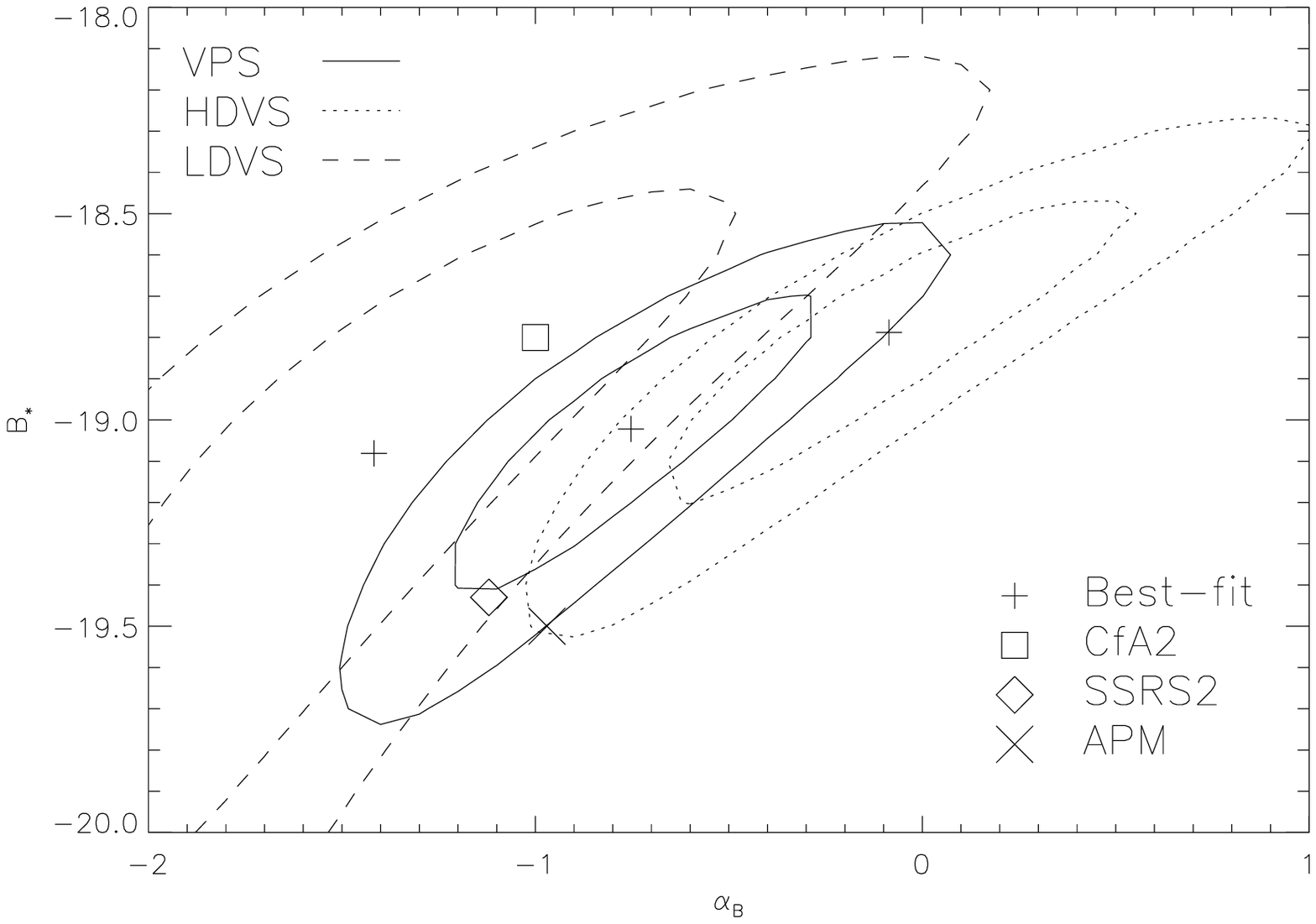}
\vskip 1pt
\plotone{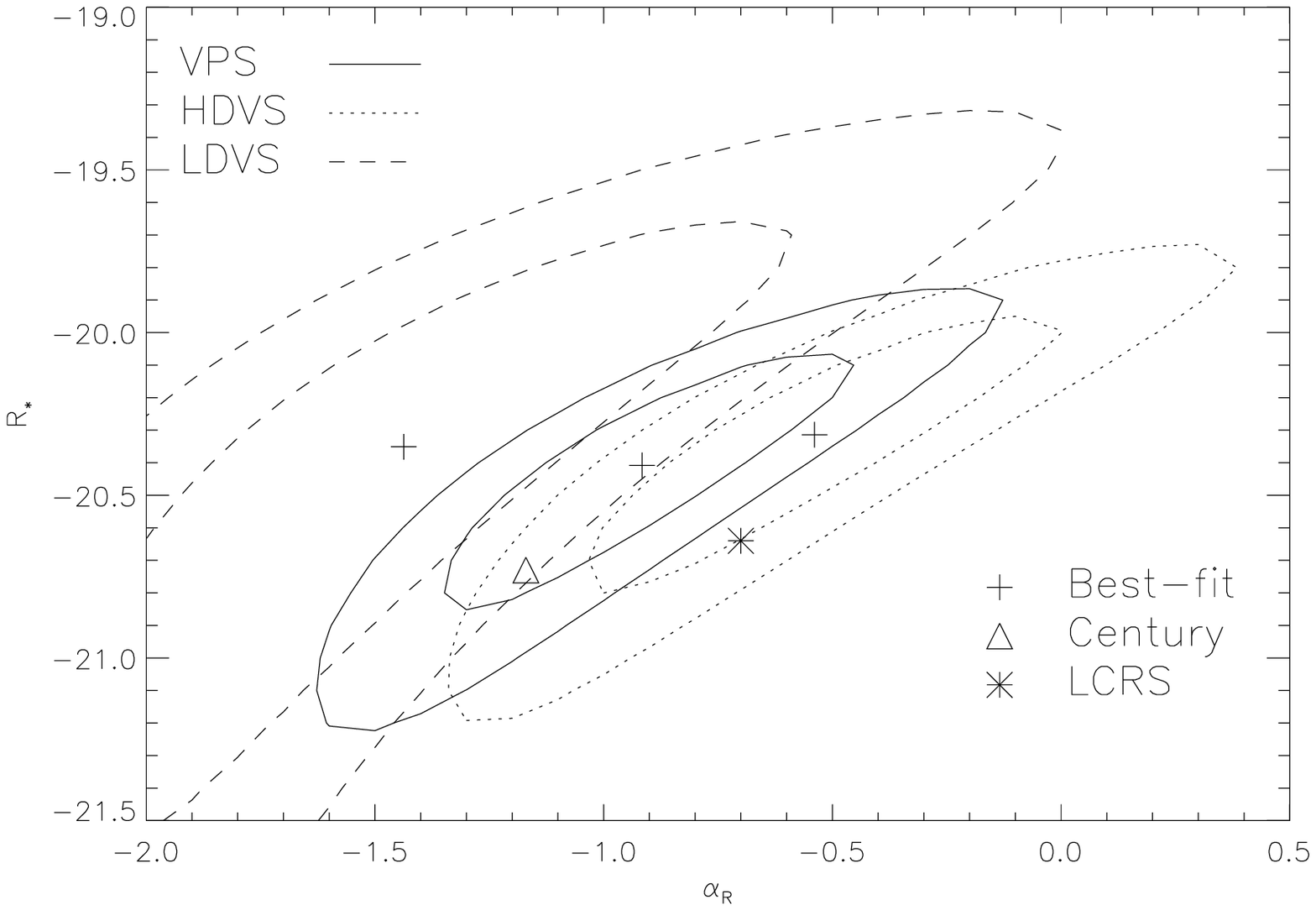}
\caption{\footnotesize  Summary of the LF-fitting results for the $B$ magnitudes
(top) and the $R$ magnitudes (bottom).  The $1\sigma$ and $2\sigma$
likelihood contours of Schechter function parameters $\alpha$ and
$M_*$ are plotted for the VPS (solid), the HDVS (dotted), and the LDVS
(dashed).  The best-fit parameters in each case are marked by a plus
symbol; typical survey LFs are marked by symbols as indicated on the
plots.
\label{lfsummfig}}
}
\end{figure}
\clearpage
\end{document}